\documentclass{ws-ijmpa}
\begin{document}

\markboth{Piero Nicolini} {Noncommutative Black Holes, The Final Appeal To Quantum Gravity}

%
\catchline{}{}{}{}{}
%

\title{NONCOMMUTATIVE BLACK HOLES, THE FINAL APPEAL TO QUANTUM GRAVITY: A REVIEW}

\author{PIERO NICOLINI}

\address{Department of Mathematics and Computer Science, Consortium of Magneto-Fluid-Dynamics,
University of Trieste  and INFN, National Institute for Nuclear Physics, Via Valerio 12\\
Trieste, I-34127, Italy \\
Department of Physics, California State University Fresno, 2345 E. San Ramon Ave., \\
Fresno, CA 93740-8031, United States\\
Piero.Nicolini@cmfd.univ.trieste.it}

 \maketitle

\begin{history}
\received{11 July 2008}
\revised{24 November 2008}
\end{history}

\begin{abstract}
 We present the state of the art regarding the relation between the physics of
Quantum Black Holes and Noncommutative Geometry. We start with a review of
models proposed in the literature for describing deformations of General
Relativity in the presence of noncommutativity, seen as an effective theory of
Quantum Gravity. We study the resulting metrics, proposed to replace or at
least to improve the conventional black hole solutions of Einstein's equation.
In particular, we analyze noncommutative-inspired solutions obtained in terms
of quasi-classical noncommutative coordinates: indeed because of their
surprising new features, these solutions enable us to circumvent long standing
problems with Quantum Field Theory in Curved Space and to cure the singular
behavior of gravity at the centers of black holes. As a consequence, for the
first time, we get a complete description of what we may call the black hole
SCRAM, the shut down of the emission of thermal radiation from the black hole:
in place of the conventional scenario of runaway evaporation in the Planck
phase, we find a zero temperature final state, a stable black hole remnant,
whose size and mass are determined uniquely in terms of the noncommutative
parameter $\theta$. This result turns out to be of vital importance for the
physics of the forthcoming experiments at the LHC, where mini black hole
production is foreseen in extreme energy hadron collisions. Because of this, we
devote the final part of this review to higher dimensional solutions and their
phenomenological implications for TeV Gravity.

\end{abstract}

\ccode{PACS numbers: 04.70.Dy, 04.60.Bc, 11.10.Nx, 04.50.Gh}

\begin{flushright}
{\it Per aspera ad astra}\\ Seneca, ``Hercules furens''\\  $2^{nd}$ Act, v. 437
\end{flushright}

\section{Introduction}

In 1975 Steven Hawking showed, in a remarkable paper\cite{Hawking:1974sw}, that
black holes can evaporate, namely that they can emit thermal radiation like a
black body. This conjecture opened the window towards the mysteries of Quantum
Gravity, since, at that time, the reconciliation between Quantum Mechanics and
General Relativity was kept in a lethargic state as a matter of secondary
concern. Indeed, theoretical physicists were basically interested in the
Standard Model of particle physics and gravity was simply assumed to be too
weak with respect to the already unified fundamental interactions. Even strings
were employed as a theory of hadrons. Direct quantization of General Relativity
into a quantum field theory in which the gravitational force is carried by
gravitons led to the unpleasant appearance of infinities. This lack of
renormalization was somehow an underestimated problem or at least considered as
being too far away, only relevant at a scale 19 orders of magnitude higher than
the mainstream high energy physics at the time. Therefore Gravity was kept
classical and the major efforts were directed towards the Standard Model of
particle physics, calculating Feynman diagrams, cross sections and decay rates.
In this spirit, Hawking's conjecture about black hole evaporation was supported
by semiclassical arguments\cite{DeWitt:1974na}\cdash\cite{Birrell:1982ix} which
hold only when the energy of the emitted particle is small compared to the
black hole mass and it is possible to neglect the back reaction on the metric
during the emission process\footnote{The physical picture of Hawking radiation
is in terms of particles tunnelling through the horizon. Only recently has a
transparent derivation of it been given in terms of a quasi-classical
tunnelling calculation (see
Refs.~\refcite{Parikh:1999mf}--\refcite{Akhmedov:2008ru}).}. After realizing
that black holes can emit photons, gravitons and other elementary particles as
products of the decay, Hawking raised a natural and profound question about the
final stage of the evaporation and the possible loss of information encoded in
quantum states of matter. All of the troubles about the black hole evaporation
were basically due to a breakdown of the semiclassical description, which
requires that the black hole mass $M_{BH}\gg M_P$, where $M_P$ is Planck mass.
As the decay proceeds and the black hole mass decreases, this condition
eventually can no longer be met and a quantum theory of gravity must be
invoked. Indeed the Planck scale marks the {\it limes} between the classical
and the quantum behavior of the spacetime.

After more than $30$ years of intensive research, the scenario is now
drastically different and we have at least two plausible candidate theories of
Quantum Gravity: Superstrings\footnote{In (Super)String Theory one adopts
1-dimensional quantum objects instead of the more conventional point-like
structures\cite{Green:1987sp,Green:1987mn}. This represents the first chapter
in a program for studying the physics of extended quantum objects, which also
include 2-dimensional branes and 3-dimensional bubbles (for more details see
Refs.~\refcite{Pavsic:2001qh}--\refcite{Ansoldi:1997hz}).} and Loop Quantum
Gravity\cite{Rovelli:2004tv}. Both of them have great merits and still-open
problems, but at present both suffer from a basic limitation: the absence of
any support from experiments. For this reason they cannot be considered as
being more than theoretical speculations, even if they are both physically very
promising, aesthetically very attractive and mathematically elegant and well
defined. The main problem is that we do not yet have any experimental data
demonstrating some quantum gravitational effect. From historical,
epistemological and philosophical points of view, the long-standing inability
to find observable effects of Quantum Gravity is probably the most dramatic
crisis in theoretical physics and perhaps in all of science. What makes the
situation even more embarrassing is that Quantum Gravity is expected to be the
synthesis of two physical theories which are the most successful ever
formulated from the point of view of their capacity for reproducing
experimental results\footnote{The relative error in the General Relativity
prediction for the rate of change of the orbital period of the Hulse-Taylor
binary pulsar PSR 1913+16 is $10^{-14}$ (see Ref.~\refcite{Taylor:1989sw}).
Experimental data and Quantum Field Theory predictions of the value of the
anomalous magnetic moment of the electron agree to within a relative error of
$10^{-11}$ (see Ref.~\refcite{Kinoshita_1995}).}.

What is the role of black holes in this context? It is very likely that black holes are fated to provide the final
answer about our knowledge of Quantum Gravity and close the logical path which they started in 1975, igniting the
initial interest in this area. Indeed, in the next few months\footnote{The Large Hadron Collider (LHC) is being
built at the CERN lab and is planned to circulate the first beams in August 2008, while the first collisions are
expected in October 2008 with the first results arriving soon afterwards\cite{LHC}.} we could be very close to
having revolutionary experimental evidence for both particle phenomenology and gravitational physics: the
detection of some signature of the production of mini black holes\cite{Argyres:1998qn}\cdash\cite{Hsu:2002bd} that
are predicted to be produced as a result of high energy hadronic collisions. According to theory, decay of a mini
black hole would lead to the sudden emission of a huge quantity of particles independently of their spin, charge,
quantum numbers or interaction properties, as long as their rest mass is smaller than the black hole
temperature\cite{Casanova:2005id}. Therefore mini black holes may be excellent intermediate states for long-sought
but yet undiscovered particles, like the Higgs boson or supersymmetric particles, possibly with cross sections
much larger than for the direct processes.

For these reasons, the starting-up of the LHC, or of some other future
experiments\footnote{Formation and detection of mini black holes is expected in
both future hadron colliders, such as the Very Large Hadron Collider (VLHC) at
Fermilab, and the particle shower of Ultra High Energy Cosmic Rays (UHECR),
impacting Earth's atmosphere\cite{Feng:2001ib}.}, will open up very good
opportunities for getting information about Quantum
Gravity\cite{Feng:2001ib}\cdash\cite{Rizzo:2006uz} and the very early universe,
as well as for solving some basic questions, whose answers are too often taken
for granted.

First, do (mini) black holes exist? From velocity measurements for the whirlpool of hot gas surrounding it,
astronomers have found convincing evidence for the existence of a supermassive black hole in the center of the
giant elliptical galaxy $\mathrm{M}87$\footnote{In 1994, Hubble Space Telescope data produced an unprecedented
measurement of the mass of an unseen object at the center of $\mathrm{M}87$. Based on the kinetic energy of the
material whirling about the center, the object is estimated to be about 3 billion times the mass of our Sun and
appears to be concentrated within a volume of space smaller than that of our solar system.}. Astronomers have also
detected radio emission coming from within 30 million kilometers of the dark object SGR A*, thought to be a
colossal black hole, that lies at the center of the Milky Way\cite{Falcke:1994fz}\cdash\cite{Ghez:2003hb}.
Previously, X-ray emission from the binary star system Cygnus X-1 convinced many astronomers that it contains a
black hole, which is supported by more precise measurements which have recently become available. To this purpose
we also have to mention V404 Cygni, one of the most evident cases of stellar black hole: observations at X-ray and
optical wavelengths have shown that this is a binary system in which a late yellowish G star or maybe an early
orange-red K star revolves, every 6.47 days, around a compact companion with a diameter around $60-90$ km and a
probable mass of 8 to 15 solar masses, well above the mass limit at which a collapsed star must become a black
hole\cite{Sanwal:1996}\cdash\cite{Corbel:2008yw}. In spite of these observations, however, there are still some
ranges of mass in which the existence of black holes is unclear, in particular of black holes of less than $3$
solar masses. The relevance of these objects is connected with the possibility of observing the Hawking radiation.
Indeed, what we know for sure is that for astrophysical black holes the Hawking radiation is negligible because
their temperatures can be at most some tens of nK, far below $T_{\mathrm{CMB}}\sim 2.7$ K, the temperature of
Cosmic Microwave Background (CMB) radiation. On the other hand tidal effects are significant in the case of mini
black holes that could be very hot and very bright if their mass is sufficiently small \cite{Page:1976wx}.

As second point, we might be able to conclude that (mini) black holes really can evaporate. Indeed for the above
reasons the detection of these objects is the unique direct way\footnote{There exist alternative proposals to
observe the Hawking radiation by an ``indirect'' means in the so-called Analogue Models, namely condensed matter
systems which behave, to some extent, like gravitational ones \cite{Unruh:1980cg}: in Bose-Einstein condensates
the supersonic region is the acoustic analog of a black hole, while the flux of phonons corresponds to the Hawking
radiation. The resulting Hawking temperature is $T_H\sim 10$ nK as compared with a condensate temperature $T_c\sim
\ 100$ nK (see Ref.~\refcite{Balbinot:2006ua} for a review and Refs.~\refcite{Balbinot:2007de}--\refcite{Carusotto:2008ep} for recent results). \\
Another interesting proposal is that concerning the possibility of experimentally detecting the Unruh effect in
particle storage rings in terms of the Sokolov--Ternov effect\cite{Akhmedov:2007xu,Akhmedov:2006nd}. This could be
the simplest prototype of vacuum polarization effects in curved space. } to have experimental evidence of the
Hawking conjecture, one of the most important predictions of Quantum Field Theory in Curved Spacetime and of the
associated semiclassical gravity.

Thirdly, we could find out what it is the fate of a radiating black hole. If
mini black holes can be created in high energy particle collisions, the black
holes produced will pass through a number of phases before completing their
evaporation.  After a loss during the first two phases of their ``hair'' (i.e.
the associated long-range fields) and of their angular momentum (the
``balding'' and spin-down phases respectively), the picture of the evaporation
will be described by the Schwarzschild phase, in which the resulting
spherically symmetric black hole loses energy by the emission of Hawking
radiation, with a gradual decrease in its mass and an increase in its
temperature. Since the Schwarzschild geometry has a curvature singularity at
the origin, there would be a divergent behavior of the Hawking temperature if
the black hole were to shrink to the origin as a result of losing mass by
thermal emission. However, we do not expect that this divergent behavior will,
in fact, take place since in the vicinity of the origin, the evaporating black
hole will be dramatically disturbed by strong quantum gravitational
fluctuations of the spacetime manifold. In other words the black hole will
undergo a Plank phase of the evaporation during which a theory of Quantum
Gravity must be used. Observations of the final stages of black hole
evaporation could provide the profile of the temperature as a function of the
mass of the black hole and hence let us pick out, for the first time, the
correct quantum gravitational theory.

All of this promising and very fascinating program is, on the other hand,
subject to a severe and unavoidable constraint: the inferred existence of large
spatial extra dimensions. This hypothesis is currently considered to be the
unique viable solution to the long-standing hierarchy problem, namely the
presence of two fundamental scales in nature, the electroweak scale and the
Planck scale, separated by $16$ orders of
magnitude\cite{ArkaniHamed:1998rs}\cdash\cite{Randall:1999vf}. The main point
in this potential resolution of the problem is that extra dimensions can be
assumed to be as large as around a millimeter, if we suppose that Standard
Model fields are constrained in a four dimensional sub-manifold of the higher
dimensional spacetime, and that only gravity can probe the additional
transverse dimensions. In connection with this, recent experiments involving
direct measurement of Newtonian gravity put limits on the size of extra
dimensions at length scales below $1$ mm.\cite{Hoyle:2004cw,Kapner:2006si} In
this type of scenario, we can identify a new scale $M_\ast$, derived from the
Planck scale through the following relation
 \begin{equation}
M_\ast^{(2+n)}=M_{P}^{2}/R^{n}
\end{equation}
 where $R$ is the mean size of each of the $n$ extra dimensions. If $R$ is large
enough with respect the Planck length $\ell_P$, then $M_\ast$ will be much
smaller than $M_P$ and we obtain a unique fundamental scale $M_\ast \sim 1 $
TeV for both the electroweak and gravitational interactions. Conversely, in the
absence of any large extra dimensions, Quantum Gravity is lost forever: it is
very likely that at the current rate of technological progress, mankind will
probably become extinct before any experimental evidence of Quantum Gravity
would become accessible. Giving a possible solution of the hierarchy problem is
not the only key consequence of the presence of large extra dimensions: there
are also further important consequences which we will summarize briefly here.
First, any black hole smaller than the size of the extra dimensions can be
considered, to a good approximation, as being totally submerged in a $4+n$
dimensional isotropic spacetime, with one time dimension and $3+n$ spatial
ones.  This allows one to use the higher-dimensional Schwarzschild solution to
describe at least the neutral non-rotating phase of the black hole's life. As a
result, the cross section for creation of mini black holes due to parton
collisions significantly increases in the presence of large extra dimensions.
Indeed, the corresponding Schwarzschild radii $r_H$ become of the order of
$10^{-4}$ fm, which is large with respect to the parton impact parameter $b$.
Therefore we can estimate the black hole production cross section by the
geometrical approximation $\sigma\sim\pi r_H^2\sim 400$ pb and so, at the
estimated luminosity for the LHC ($L\sim 10^{33}$ cm$^{-2}$s$^{-1}$) we find
that roughly ten black holes would be created per second\cite{Bleicher:2007hw}.
Another consequence of the introduction of large extra dimensions is that the
mini black holes would be colder and thus live longer than their
four-dimensional analogs, and so would be more easily detectable once created.
Indeed, one typically finds mini black hole temperatures of the order of $100$
GeV and lifetimes of the order of $10^{-26}$ s, that are interpreted as those
of resonances.

At this point, the game seems to be over: we have candidate theories of Quantum
Gravity and forthcoming experiments which are potentially able to check them.
We should then try to find some easily testable theoretical predictions and
wait for their experimental confirmation. The problem is that the final state
of black hole evaporation cannot be efficiently described by means of the
aforementioned theories of Quantum Gravity. Indeed (Super)String Theory
provides a quantum description of black holes only for a few cases, namely for
the extremal (and near-extremal) charged black hole
models\cite{Strominger:1996sh,Callan:1996dv}. This is rather unsatisfactory,
since the Planck phase, during the terminal stage of the evaporation, occurs in
a neutral regime due to the rapid discharge of the black hole in the very
initial stages. On the other hand, Loop Quantum Gravity suffers from the
absence of a clear semiclassical limit. What is missing is a systematic way of
computing scattering amplitudes and cross sections by perturbative techniques,
a fact that is the basic limitation for obtaining significant quantitative
results\cite{Rovelli:2004tv}. Therefore, in the absence of a full quantum
description of all of the significant black hole evaporation phases, one uses
effective theories to describe the quantum gravitational behavior, at least in
some regimes. The most common effective tool is Quantum Field Theory in Curved
Space, which works efficiently at least until the black hole quantum back
reaction destroys the hypothesis of a fixed background
spacetime\cite{Padmanabhan:2003gd}. Very recently, stimulated by the need for
going beyond the rough semiclassical approximation of Quantum Field Theory in
Curved Space, significant new approaches have been proposed based on
Noncommutative Geometry arguments. Indeed there is a long-held belief that
Quantum Gravity should have an uncertainty principle which prevents one from
measuring positions to accuracies better than that given by the Planck length:
the momentum and energy required to make such a measurement would themselves
modify the geometry at these
scales\cite{DeWitt:1967ub}\cdash\cite{Calmet:2004mp}. Therefore, one might wish
to describe these effects, at least effectively, by a model theory having a new
sort of uncertainty principle among the coordinates. In the same way as happens
with coordinates and momenta in conventional quantum theory, the uncertainty
would come from a noncommutative relation and so one is led to examine the
possibility that position measurements might fail to commute, postulating the
existence of a noncommutative manifold
 \begin{equation}
 \left[ \textbf{x}^i,\textbf{x}^j\right]  \neq 0.
\end{equation}
 A feature of a Noncommutative Geometry would be the presence of quantum
fluctuations able to remove the infinities which usually appear and cause
the bad short-distance behavior of field theories, including gravity.

The purpose of this review is to explore the current status of the physics of
Quantum Black Holes from the viewpoint of Noncommutative Geometry, used as an
effective tool for modelling the extreme energy quantum gravitational effects
of the final phase of the evaporation, which are plagued by singularities at a
semiclassical level. In Section 2, we start by reviewing the most popular
Noncommutative Geometry models existing in the literature and their role in
comparison with General Relativity. In Section 3, we discuss the four
dimensional noncommutative Schwarzschild solution, from both geometrical and
thermodynamical points of view. In Section 4, we study how noncommutativity
affects the Maxwell field and present the four dimensional noncommutative
Reissner-Nordstr\"{o}m solution, providing a detailed analysis of both the
Hawking and Schwinger pair production mechanisms.  In Section 5, we consider
evaporation in the extra-dimensional scenario, for both neutral and charged
solutions, reviewing its phenomenological consequences in view of a possible
experimental detection at the LHC. The final Section is devoted to future
perspectives.

\section{Models of Fuzzy Geometry}

Noncommutativity is an old idea in theoretical physics, dating back to the
early times of quantum theory and originally postulated as a way to improve the
renormalizability properties of a theory at short distances or even to make it
finite\cite{Pauli:1985}\cdash\cite{Yang:1947ud}. On the other hand, the
development of efficient renormalizion techniques in field theories, decreased
the interest in noncommutativity until recent times when the quantization of a
unrenormalizable theory like gravity was studied extensively.  In the 1980s,
noncommutativity was largely developed by mathematicians in order to
characterize a Euclidean manifold in terms of the spectrum of the Dirac
operator on the manifold, providing a generalization of differential geometry,
which became known as Noncommutative
Geometry\cite{Connes:1985}\cdash\cite{Connes:2000by}.

In the high energy physics community, the interest about noncommutativity was revitalized when, in the Theory of
Open Strings, it has been shown that target spacetime coordinates become noncommuting operators on a $D$-brane in
the presence of a constant Neveu-Schwarz B field\cite{Witten:1985cc,Seiberg:1999vs}.  As a result, the open
strings, at least in the low energy limit when matter decouples from gravity, induced a quantum field theory on
noncommutative spacetime, often called Noncommutative Field Theory and characterized by the transparent presence
of a noncommutative coordinate algebra. In some sense, this physically motivated approach is somehow against the
spirit of the mathematicians' community approach\cite{Connes:1985}\cdash\cite{Connes:2000by}, which was formulated
in a coordinate free language in terms of diffeomorphism invariants.  Also in the framework of Loop Quantum
Gravity, the resulting quantum geometry exhibits a noncommutative behavior: indeed operators that measure areas
and volumes of regions of a spatial sub-manifold, strictly speaking fail to
commute\cite{Thiemann:2001yy}\cdash\cite{Ashtekar:2007px}.

Independently of its interpretation, namely as a fundamental picture of quantum spacetime or as a formalism
capturing some aspects of String Theory and/or Loop Quantum Gravity, Noncommutative Geometry is currently employed
to implement the ``fuzziness'' of spacetime by means of
\begin{equation}
\left[\, \mathbf{x}^i\ , \mathbf{x}^j\, \right]= i \, \theta^{ij} \label{ncx}
\end{equation}
where, in the simplest case, $\theta^{ij}$ is an anti-symmetric, real, $D\times D$ ($D$ is the dimension of
spacetime) matrix, which has dimension of a length squared. Such matrix determines the fundamental cell
discretization of spacetime much in the same way as the Planck constant $\hbar$ discretizes the phase space.  As a
consequence, the resulting geometry is ``pointless'', since the notion of point is no longer meaningful because of
the uncertainty
\begin{equation}
\Delta x^i\ \Delta x^j \ge\frac{1}{2}\left|\theta^{ij}\right|
\end{equation}
induced by the noncommutative behavior of coordinates.  Therefore one could interpret this loss of resolution as
the emergence of a natural effective ultraviolet cut off, regulating not only gravity but any quantum field
theory. If we define $\theta$ to be the average magnitude of an element of $\theta^{ij}$, physically
$\frac{1}{\sqrt{\theta}}$ corresponds to the energy scale beyond which the conventional differential spacetime
manifold turns out to be a noncommutative one. To this purpose, there is a general consensus about the appearance
of noncommutative phenomenology at intermediate energies between the scale of the Standard Model of particle
physics and the Planck scale. For this reason, the inclusion of noncommutativity in field theory in flat space can
provide lower bounds about the energy threshold beyond which a particle moves and interacts in a distorted
spacetime.  The presence of noncommutative effects in flat space is, on the other hand, a feature which merits
further inspection: the defining relation (\ref{ncx}) provides a subtle process of quantization of the spacetime
manifold ${\cal M}$, we define in terms of a class of equivalent atlases, without regard of any sort of field,
function or tensor defined over it.  Only after having constructed this groundwork we can think to introduce
various kinds of structure over the manifold, even when ${\cal M}$ is subject to a noncommutative behavior. Among
the possible structures, we can independently consider matter fields $\psi_\alpha$ propagating over ${\cal M}$ and
any kind of tensorial quantity, including a symmetric $(0,2)$ non degenerate tensor, called metric tensor
$g_{\mu\nu}$ describing the curvature properties of ${\cal M}$.  For this reason, the quantization of the geometry
encoded in (\ref{ncx}) can be considered a genuine background independent procedure: indeed the metric tensor
works as a mathematical apparatus superimposed on ${\cal M}$ and measures the gravitational interaction on the
underlying spacetime manifold, much in the same way as the Newtonian gravitational field $\phi$ in the
pre-relativistic physics. As a result, the quantum mechanical fluctuations of $g_{\mu\nu}$ are the response of the
corresponding quantum fluctuations of ${\cal M}$ governed by the noncommutative relation (\ref{ncx}). Therefore,
the quantum behavior of ${\cal M}$ is an independent fact of metric tensor and can be considered even when only
matter fields $\psi_\alpha$ are allowed to propagate on ${\cal M}$. The manifold ${\cal M}$, its fluctuations and
curvature look similar to the case of a surface, that at short distances (high energies) appears rough, even if
nothing is said whether the surface is still flat or already curved.

Noncommutativity could be also able to describe a possible Lorentz violation: the granular structure of space
would mean that different wave lengths of light could travel at different speeds and play a role in the unexpected
energy threshold of cosmic rays. However, the violation of Lorentz invariance may or may not occur in
Noncommutative Geometry: indeed there exist many formulations, based on different ways of implementing non-local
deformations in field theories\footnote{For a review about this topic see Refs.
\refcite{Douglas:2001ba}--\refcite{Szabo:2001kg}.}, starting from a noncommutative relation and preserving Lorentz
symmetry
\begin{equation}
\left[\, \mathbf{x}^\mu\ , \mathbf{x}^\nu\,\right]= i \, \theta^{\mu\nu} \label{ncxL}
\end{equation}
where now $\theta^{\mu\nu}$ is an anti-symmetric second-rank
tensor\cite{Doplicher:1994zv}\cdash\cite{Smailagic:2004yy}.

The most popular approach to Noncommutative Geometry is founded on the replacement of the point-wise
multiplication of fields in the Lagrangian by a non-local Groenewold-Moyal
$\star$-product\cite{Groenewold:1946kp,Moyal:1949sk}. This technique, largely inspired by the foundations of
quantum mechanics\cite{Weyl:1931,Wigner:1932eb}, provides a systematic way to describe noncommutative spaces and
to study field theories propagating thereon.   Since physically, it turns out that $\theta$ is a free parameter,
which we can imagine increasing from zero to go from the commutative to the noncommutative regime, a deformation
of $C\left( {\cal M} \right)$, the bounded continuous function on ${\cal M}$, is an algebra ${\cal A}_\theta$ with
the same elements and addition law but different multiplication law, called $\star$-product to distinguish it from
the original point-wise multiplication of functions.  The notation has a further virtue, since let us somehow
circumvent the ordering conditions in ${\cal A}_\theta$. Indeed, if we want to specify monomials or any kind of
product of noncommutative coordinates, we have to give a ordering prescription. On the other hand, we can
establish an isomorphism between the noncommutative algebra ${\cal A}_\theta$ and the conventional algebra of
functions $C\left( \Re^D \right)$, employing a linear map $S$, called {\it symbol} of the operator. Therefore we
represent the multiplication law in ${\cal A}_\theta$ in term of the $\star$-product of symbols
\begin{equation}
\hat{f}\hat{g}\equiv S^{-1}\left[S [\hat{f}]\star S[\hat{g}]\right]
\end{equation}
where $\hat{f},\hat{g}\in {\cal A}_\theta$. There could be many valid definitions of $S$, corresponding to
different choices of ordering prescription for $S^{-1}$.  A convenient and standard choice for $S^{-1}$ is the
Weyl ordered symbol ${\cal W}$, largely used in the early times of quantum
mechanics\cite{Weyl:1931,Wigner:1932eb}. As a result we map the basis monomials in $C\left( {\cal M} \right)$ onto
the symmetrically ordered basis elements of ${\cal A}_\theta$, namely $x^\alpha x^\beta\rightarrow
\frac{1}{2}\left({\bf x}^\alpha {\bf x}^\beta + {\bf x}^\beta {\bf x}^\alpha\right)$.  The extension of ${\cal W}$
to the algebra isomorphism, induced the Groenewold-Moyal $\star$-product $ {\cal W}\left[f\star
g\right]\equiv{\cal W}[f] \ {\cal W}[g]=\hat{f}\ \hat{g}$, where $f,g \in C\left( {\cal M} \right)$ and $
\hat{f},\hat{g}\in {\cal A}_\theta$. The representation turns out to be \begin{equation} f(x)\star g(x)=
\sum_{k=0}^\infty \frac{1}{k!}\left( \frac{i}{2}\right)^k \theta^{\alpha_1 \beta_1}\cdots\theta^{\alpha_k
\beta_k}\
\partial_{\alpha_1} \cdots\partial_{\alpha_k}f(x)\
\partial_{\beta_1}\cdots\partial_{\beta_k}g (x) \label{rsp}
\end{equation}
and the $\star$-product can be shown to be associative and noncommutative. Therefore, the spacetime
noncommutativity may be encoded through ordinary products in the noncommutative ${\cal A}_\theta$ of Weyl
operators, or equivalently through the deformation of the product of the commutative algebra of functions on
spacetime $C\left( {\cal M} \right)$ to the noncommutative $\star$-product.

To transform a conventional field theory into a noncommutative
one, we may write the field action in terms of Weyl operator
${\cal W}[\phi]$, corresponding to the field $\phi (x)$; then we
can map the action back to coordinate space to get
\begin{equation} S=\int d^D x \left[
\frac{1}{2}\left(\partial_\mu\phi\right)\star\left(\partial^\mu\phi
\right) -\frac{m^2}{2}\left( \phi\star\phi\right)
+\frac{\lambda}{24} \left(\phi\star\phi\star\phi\star\phi\right)
\right].
\end{equation}
In spite of its mathematical exactitude, the $\star$-deformed field theory suffers nontrivial limitations. The
Feynman rules, obtained directly from the above action, lead to unchanged propagators, which implies that the
$\star$-deformed field theory is identical to the ordinary field theory at the level of free fields, a fact
somehow physically counterintuitive or at least unexpected.  On the other hand, the only modifications, concerning
vertex contributions, are even responsible for the non-unitarity of the theory and for UV/IR mixing. In other
words UV divergences are not cured but rather are accompanied by surprisingly emergent IR ones.

The key point of the origin of these malicious features, is
basically related to perturbative expansion in $\theta$ of the
above action, which is the unique viable procedure to extract
phenomenologically testable results from the formal field
theoretical apparatus. As a result any truncation at a desired
order in the noncommutative parameter basically destroys the
non-locality encoded in (\ref{ncxL}) and gives rise to a local
field theory, plagued by spurious momentum-dependent interactions.
In other words the desired non-locality induced by the
noncommutative manifold is connected to the presence of infinite
derivative terms in the product of functions. For this reason any
sort of expansion in $\theta$ provides a local field theory, which
has nothing to do, even remotely, with the original non-local
field theory. Only recently, have there been attempts to restore
unitarity\cite{Bahns:2002vm} and to overcome the problem of UV/IR
mixing\cite{Bahns:2003vb,Grosse:2004yu}, even if the restriction
of noncommutative corrections only to interaction terms remains a
non intuitive feature. Against this background, the most efficient
way out is that of a radical change of perspective about the
employment of noncommutative coordinates and the resulting algebra
${\cal A}_\theta$.

Indeed, to proceed in this direction, we have to invoke the lesson
of quantum mechanics: any quantum operator $A$ is subject to
fluctuations due to the fuzziness of the quantum phase space,
induced by the Heisenberg (non) commutation relations. Therefore,
the operator eigenvalues are the unique physically admissible
outcomes of a measure of the observable represented by $A$ and the
mean value $\left<A\right>$ of these eigenvalues provides what is
more reminiscent of the corresponding classical observable. Indeed
mean values, being statistical mixtures of eigenvalues, do take
into account the quantum fluctuations of the observable and evolve
according to laws which resemble the classical dynamical
equations.  In this spirit, to extract some physically meaningful
content from the noncommutative algebra ${\cal A}_\theta$, one has
to consider the set of continuous real functions, which
corresponds the mean values $\left<\hat{f}\right>$ of each element
$\hat{f}\in {\cal A}_\theta$. This procedure lets us work with
conventional multiplication laws because mean values preserve the
non-local character encoded in the noncommutative nature of the
algebra ${\cal A}_\theta$.  Thus, the problem is to find proper
noncommutative states to calculate mean values
$\left<\hat{f}\right>$: starting from a noncommutative relation
(\ref{ncxL}), there are no coordinate eigenstates for the
coordinate operators and no coordinate representation can be
defined. Pioneering works\cite{Glauber:1963tx} in quantum optics
and referred to phase space, suggest that coherent states,
properly defined as eigenstates of ladder operators built from
noncommutative coordinate operators only, are the closest to the
sharp coordinate states, which we can define for noncommutative
coordinates.  In other words, coordinate coherent states are the
minimal uncertainty states over the noncommutative manifold and
let us calculate the aforementioned mean values.

As an example of the action of mean values on the noncommutative
algebra ${\cal A}_\theta$, let us consider the simplest
noncommutative manifold, i.e. the noncommutative
plane\cite{Cho:1999sg,Smailagic:2003rp}
 embedded in a $2+1$ dimensional spacetime, in
which only spatial coordinates are noncommutative
\begin{equation}
\left[\, \mathbf{x}^i\ , \mathbf{x}^j\, \right]= i \, \theta\epsilon^{ij} \label{nc2d}
\end{equation}
with $i,j=1,2$, while $\theta^{ij}=\theta\epsilon^{ij}$ can be
written in terms of a unique noncommutative parameter $\theta$ and
the anti-symmetric tensor $\epsilon^{ij}$.  The relation
(\ref{nc2d}) tells us that we cannot build eigenstates
$\left.|x^1,x^2\right>$ and therefore one is led to consider the
operators ${\bf z}$ and ${\bf z}^\dagger$ in order to define a
convenient set of states
\begin{eqnarray}
{\bf z}\equiv\frac{1}{\sqrt{2}}\left(\mathbf{x}^1+i\mathbf{x}^2\right)\\
{\bf z}^\dagger\equiv\frac{1}{\sqrt{2}}\left(\mathbf{x}^1-i\mathbf{x}^2\right)
\end{eqnarray}
satisfying the relation $\left[\, {\bf z} \ , {\bf z}^\dagger\, \right]=  \theta$.  The advantage of shifting the
noncommutative character from $\mathbf{x}^1$, $\mathbf{x}^2$ to ${\bf z}$  and ${\bf z}^\dagger$ consists in the
possibility of having eigenstates
\begin{eqnarray}
{\bf z}\left.|z\right>=z\left.|z\right>\\
\left<z|\right. {\bf z}^\dagger=\left<z|\right. z^*
\end{eqnarray}
where $\left.|z\right>\equiv
\exp\left(-\frac{zz^*}{2\theta}\right)\exp\left(-\frac{z}{\theta}z^\dagger\right)\left.|0\right>$,
with $\left.|0\right>$  the vacuum state annihilated by ${\bf z}$
and $z$ the complex eigenvalues. These coordinate coherent states
$\left.|z\right>$ allow us to associate to any quantum operator
$F\left(\mathbf{x}^1, \mathbf{x}^2\right)$ the corresponding
function $F(z)\equiv \left<z|\right. F\left(\mathbf{x}^1,
\mathbf{x}^2\right) \left.|z\right>$ representing its mean value
over the fluctuating manifold, in terms of quasi classical
coordinates or simply {\it quasi coordinates}, namely mean values
$x^1$, $x^2$ of coordinate operators
\begin{eqnarray}
\left<z|\right. \mathbf{x}^1 \left.|z\right>=\sqrt{2} \Re z\equiv x^1\\
\left<z|\right. \mathbf{x}^2 \left.|z\right>=\sqrt{2} \Im z\equiv x^2 .
\end{eqnarray}
These definitions are the prelude to the quantum field theory on the noncommutative plane
\begin{equation}
\phi (t, z) =\sum_{E, p}\left[{\bf a}_p^\dagger \ e^{-iEt}\left<z|\exp(ip_j{\bf x}^j)|z\right> +h.c.\right]
\end{equation}
where ${\bf a}_p$, ${\bf a}_p^\dagger$ are the conventional lowering/rising operators acting on Fock states with
definite energy and momentum.  The novelty of the above definition lies in $\left<z|\exp(ip_j{\bf x}^j)|z\right>$,
which introduced a gaussian damping factor $e^{-\theta p^2/4}$. In other words, the noncommutativity of
coordinates manifests as a modification of the integration measure in momentum space, once ordinary finite
functions like $F(z)$ are represented by their inverse Fourier transform. An illuminating feature of this
noncommutative behavior is given by the form of the momentum space free propagator \begin{equation}
G(E,\vec{p}^2)=\frac{e^{-\theta p^2/4}}{-E^2+\vec{p}^2+m^2}
\end{equation}
which elegantly exhibits the sought ultraviolet behavior.  As a
consequence the corresponding Green's function equation exhibits a
modified source term
\begin{equation}
(\triangle +m^2)\ G_\theta (\vec{x}-\vec{y}) =\rho_\theta (\vec{x}-\vec{y})
\end{equation}
namely the conventional Dirac delta has been replaced by a Gaussian distribution, the signature of the pointless
geometry.

The nice feature of this approach is that the noncommutative
fluctuations of the manifold produce an ultraviolet damping factor
in the Fourier transform of field functions defined over it, even
at the level of a free field, independently of their tensorial
nature and preserving the conventional point-wise product among
them. Important progress in this area has been made in order to
extend the above two dimensional toy model to a full
noncommutative field formulation, in which all of the coordinates
are noncommutative: as a consequence the resulting field theory is
UV finite, free from any IR contribution, while unitarity and
Lorentz invariance are recovered if one imposes a single
noncommutative parameter $\theta$ in the theory, namely
\begin{equation}
\left[\, \mathbf{x}^\mu\ , \mathbf{x}^\nu\,\right]= i \,
\theta\,\sigma^{\mu\nu} \label{ncxLINV}
\end{equation}
where $\sigma^{\mu\nu}$ is an antisymmetric dimensionless tensor
determined in Ref.~ \refcite{Smailagic:2004yy}.

The effective quantum description of the physics of radiating
black holes still needs a lot of work. First of all, one has to
extend the flat space noncommutative formulation in order to
include the gravitational interaction: indeed the modification of
gravity due to the introduction of (\ref{ncx}) followed two
distinct paths, which are mainly based on the two formalisms which
we have just introduced. Therefore, we will present the resulting
noncommutative equivalent of a black hole and we will try to
understand the role of $\theta$ versus the Plank phase of the
evaporation in both the above approaches.

\section{Black Hole Solutions in Noncommutative Gravity}

The formulation of a full consistent noncommutative version of General Relativity is a business of primary
interest and currently the subject of a vast literature. Indeed once a model of Noncommutative Geometry is
assumed, we would like to know the deformation of the gravitational field equations due to the fuzziness of the
manifold. At present, in spite of the promising work in this
field\cite{Chamseddine:1992yx}\cdash\cite{Calmet:2006iz}, we are still far from a widely acknowledged theory of
Noncommutative Gravity, which, in some sense, is expected to provide the transition from the smooth differential
to the stringy (or loopy) picture of spacetime.  In this review, we shall not enter into the debate about the
accuracy of the proposed models of Noncommutative Gravity in literature, but we will present the formulation which
concretely led to new metrics, obtained by solving, exactly or approximately, the noncommutative version of
Einstein's equation, in order to study the physically reliable effect in the physics of the resulting black holes.
To this purpose, a key point of all noncommutative formulations will be the definition of the equivalent of the
line element: indeed, any differential displacement connecting two infinitesimally close spacetime events appears
not well defined in a pointless geometry. Furthermore, in presence of the noncommutative fuzziness, we do expect
some sort of removal of singularities which occur in the conventional differential spacetime manifold.  The latter
point can be really considered as the quality certificate which justifies the employment of any Noncommutative
Geometry model and discriminate the efficiency of the mathematical machineries.

\subsection{The noncommutative equivalent of the Schwarzschild metric}

We start with the noncommutative corrections to the Schwarzschild metric, proceeding along the line suggested by
Ref.~\refcite{Chamseddine:2000si}, namely employing a deformation of Einstein's equation induced by gauging the
noncommutative $ISO(3,1)$ group\footnote{The $ISO(3,1)$ group is the full Poincar\'e group, including both the
elements of $SO(3,1)$ and spacetime translations.}.  More concretely, one has to consider the gauge field strength
of the noncommutative gauge group $ISO(4,1)$ and the contraction to the group $ISO(3,1)$ to determine the deformed
vierbein in terms of the undeformed ones.  After this stage, one can construct the deformed curvature scalar
\begin{equation}
\int d^4 x\sqrt{\hat{e}}\star\hat{e}^\mu_{\star a}\star\hat{R}^{ab}_{\mu\nu}\star(e^\nu_{\star
b})^\dagger\star\sqrt{\hat{e}}^\dagger
\end{equation}
where $\hat{e}_\mu^a$ is the deformed vierbein, while $\hat{e}=det(\hat{e}^a_\mu)$ and the inverse vierbein
$\hat{e}^\mu_{\star a}$ is given by $\hat{e}^\mu_{\star a}\star\hat{e}_\mu^b=\delta_a^b$.  Here the
$\star$-product has been extended in order to be invariant under diffeomorphism transformations. The explicit form
of the above quantities can be obtained as an expansion in $\theta$
\begin{eqnarray}
\hat{e}_\mu^a &=& e_\mu^a +i\theta^{\nu\rho}e_{\mu\nu\rho}^a
+\theta^{\nu\rho}\theta^{\kappa\sigma}e^a_{\mu\nu\rho\kappa\sigma}+ O(\theta^3)\\
 \hat{e}^\mu_{\star a} &=& e^\mu_{ a}
+i\theta^{\nu\rho}e^\mu_{a\nu\rho} +\theta^{\nu\rho}\theta^{\kappa\sigma}e^\mu_{a\nu\rho\kappa\sigma}+ O(\theta^3)
\label{Vierbein}
\end{eqnarray}
where $e_{\mu\nu\rho}^a$ and $e^a_{\mu\nu\rho\kappa\sigma}$ can be written in terms of $e_\mu^a$,
$\theta^{\mu\nu}$ and the spin connection $\omega_\mu^{ab}.$   As a result, in Ref.~\refcite{Chaichian:2007we},
the corresponding deformed metric has been introduced, assuming
\begin{equation}
\hat{g}_{\mu\nu}=\frac{1}{2}\eta_{ab}\left(\hat{e}^a_\mu\star\hat{e}^{b\dagger}_\nu
+\hat{e}^b_\mu\star\hat{e}^{a\dagger}_\nu\right) \label{Metric}
\end{equation}
where $\hat{e}^a_\mu$ are written in terms of the conventional Schwarzschild vierbein by means of an expansion in
$\theta$.  Therefore the resulting deformed Schwarzschild metric $\hat{g}_{\mu\nu}$  has the following nonzero
components
\begin{eqnarray}\hat{g}_{00}&=&g_{00}-\frac{\alpha(8r-11\alpha)}{16r^4}\theta^2+O(\theta^4)\\
\hat{g}_{11}&=&g_{11}-\frac{\alpha(4r-3\alpha)}{16r^2(r-\alpha)^2}\theta^2+O(\theta^4)\\
\hat{g}_{22}&=&g_{22}-\frac{2r^2-17\alpha r +17\alpha^2}{32r(r-\alpha)}\theta^2+O(\theta^4)\\
\hat{g}_{33}&=&g_{33}-\frac{(r^2+\alpha r -\alpha^2)\cos\psi-\alpha(2r-\alpha)}{16r(r-\alpha)}\theta^2+O(\theta^4)
\label{CHASCHW}
\end{eqnarray}
where $\alpha=2GM/c^2$ and $g_{\mu\nu}$ is the known Schwarzschild metric.  As one can see, the singular behavior
at the origin of the Schwarzschild manifold is unaffected by the noncommutative corrections, but it is rather
accompanied by even worse $1/r^4$ terms.


Another kind of deformation of gravity is based on the requirement that $\theta^{\mu\nu}$ is a covariantly
constant tensor in curved space and that the symmetries of spacetime reduce to volume-preserving diffeomorphism
which also preserve $\theta^{\mu\nu}$.  The consequences of such proposal have been analyzed by means of an
extension of both Moyal and Kontsevich product\cite{Kontsevich:1997vb}\footnote{The employment of the Kontsevich
product let us consider the case in which $\theta^{\mu\nu}$ is no longer constant, but satisfies the Jacobi
identity.} to curved space and lead to the noncommutative contribution to the metric at the linearized
level\cite{Harikumar:2006xf}
\begin{eqnarray}
g_{00}&=&1 -\frac{2GM}{r}\left(1+3G\vec{\theta}^2-G(\vec{n}\cdot\vec{\theta})^2\right)\\
g_{0i}&=&0,\\
g_{ij}&=&-\delta^{ij}-\frac{2GM}{r}\left[n^in^j+G\delta^{ij}\left( \vec{\theta}^2-(\vec{n}\cdot\vec{\theta})^2
\right) +G\left(\theta^in^j+\theta^jn^i\right)\vec{n}\cdot\vec{\theta}\right]
\end{eqnarray}
where $n^i=x^i/r$ is the radial unit vector, while $\theta^j$ is defined in such a way that
$\theta^{ij}=\varepsilon^{ijk}\theta^k$. As a result, the Newtonian potential has a noncommutative contribution
proportional to $\vec{\theta}^2$, which can be incorporated into $G$ to give rise to an effective Newton constant,
even if no regularization mechanism occurs at the origin. Furthermore, the Newtonian potential also acquires an
even worse angular dependent term $(\vec{n}\cdot\vec{\theta})^2$, which goes like $1/r^2$, a contribution which
appears physically inconsistent.

Recently, another approach\cite{Kobakhidze:2007jn} has been proposed to calculate noncommutative long-distance
correction to the classical geometry, on the ground of semiclassical arguments, namely considering only the
corrections due to the interaction of linearized graviton field with noncommutative matter energy momentum tensor
\begin{equation}
{\cal L}_{int}=\frac{1}{2}\int d^4 x \ h_{\mu\nu}\ T^{\mu\nu}_{NC}
\end{equation}
where
\begin{equation}T^{\mu\nu}_{NC}=\frac{1}{2}\left(\partial^\mu\phi\star\partial^\nu\phi+\partial^\nu\phi\star\partial^\mu\phi\right)-
\frac{1}{2}\ \eta^{\mu\nu}\left(\partial_\alpha\phi\star\partial^\alpha\phi-m^2\phi\star\phi\right).
\end{equation}
The graviton field, for a nearly static source, can be written as the Fourier integral
\begin{equation}
h_{\mu\nu}(x)=-16\pi G\int\frac{d^3 q}{(2\pi)^3}e^{i\vec{q}{r}}\frac{1}{\vec{q}^2}\left(T_{\mu\nu}^{
NC}(q)-\frac{1}{2}\eta_{\mu\nu}T_{NC}(q)\right)
\end{equation}
where $T_{\mu\nu}^{ NC}(q)=\left<p_2\left|:T_{\mu\nu}^{ NC}(x):\right|p_1\right>$ and $q_\mu=(p_{2}-p_{1 })_\mu$.
As usual, the actual calculation is performed truncating the representation of the $\star$-product at a given
order, this time to retain terms with four derivative at the most
\begin{equation}
T^{\mu\nu}_{NC}\approx
T^{\mu\nu}_{0}+\eta^{\mu\nu}\frac{m^2}{16}\theta^{\alpha\beta}\theta^{\rho\sigma}\partial_\alpha
\partial_\beta\phi\partial_\rho\partial_\sigma\phi,
\end{equation}
where $T^{\mu\nu}_{0}$ is the conventional commutative free scalar field energy momentum tensor. The resulting
metric reads
\begin{eqnarray}
g_{00}&=&\left[1-2\frac{Gm}{r}+2\frac{G^2m^2}{r^2}-2\frac{G^3m^3}{r^3}+...\right]+\delta h_{00}^{NC}\\
g_{ij}&=&\left[-\delta_{ij}\left(1+2\frac{Gm}{r}\right)-\frac{G^2m^2}{r^2}\left(\delta_{ij}+\frac{r_i
r_j}{r^2}\right)+2\frac{G^3m^3}{r^3}\frac{r_ir_j}{r^2}+...\right]+\delta h_{ij}^{NC}\\
g_{0i}&=&0
\end{eqnarray}
where the noncommutative corrections coming from the four derivatives term in the energy momentum tensor are
\begin{eqnarray}
\delta h^{NC}_{00}&=&\frac{Gm^3\theta^{0i}\theta^{0j}}{4\pi}\left[-\frac{\delta_{ij}}{r^3}+\frac{3r_ir_j}{r^5}\right]
\label{Kob1}\\
\delta h_{km}^{NC}&=&-\delta_{km}\delta h_{00}^{NC}\\
\delta h_{0k}^{NC}&=&0.
\end{eqnarray}
As we can see, a post-post-Newtonian term appears as a result of
the noncommutativity, which, at least for the Solar system, would
lead to $\sqrt{\theta}>10^{-8}$ cm, an unlikely lower bound to
have significant correction dominating the post-post Newtonian one
of the classical theory.  A second term in (\ref{Kob1}) reveals a
violation of the Lorentz invariance, even though there are no
signals for this feature throughout the formulation and the
calculations. Finally the above results are not capable of probing
the geometry at the origin and smear out the singularity. This
problem seems to persist also in the effective metric emerging in
the matrix model framework for noncommutative $U(n)$ gauge theory:
even if any conclusion would be premature, preliminary results
fatally provide a badly behaving geometry at the
origin\cite{Steinacker:2007dq}. As we shall see, the absence of
any regularization mechanism is due to the employment of an
actually local field theory, rather than to the asymptotic nature
of the corrections as it might seem to be at first sight.

Employing a mere ``substitution'' of the radial coordinate with $\hat{r}\equiv
\left(\sum_i\hat{x}_i^2\right)^{1/2}$, where $\hat{x}_i$'s satisfy the noncommutative relation (\ref{ncx}),
another line element has been recently proposed as possible equivalent of the Schwarzschild
one\cite{Nasseri:2005ji,Nasseri:2005yr}
\begin{equation}
ds^2=\left(1-\frac{2M}{\hat{r}}\right)dt^2-\frac{d\hat{r}d\hat{r}}{\left(1-\frac{2M}{\hat{r}}\right)}-\hat{r}\hat{r}
\left(d\psi^2+\sin^2{\psi}d\phi^2\right).
\end{equation}
Writing this line element in terms of the commutative coordinates,
$x_i=\hat{x}_i+\frac{1}{2}\theta_{ij}\hat{p}_j$, with $\hat{p}_i=p_i$, one can determine the corrections due to
the above substitution,  by a perturbative expansion only
\begin{equation}
g_{00}\simeq\left(1-\frac{2M}{r}\right)-\frac{GM}{2r^3}\left[\vec{L}\vec{\theta}-\frac{1}{8}
\left(p^2\theta^2-(\vec{p}\cdot\vec{\theta})\right)\right]+{\cal O}(\theta^3)
\end{equation}
where $L_k=\varepsilon_{ijk}x_ip_j$, with $p_i$ given by
$[x_i,p_j]=i\delta_{ij}$ and $\vec{\theta}$ by
$\theta_{ij}=\frac{1}{2}\epsilon_{ijk}\theta_k$. This result is
subject to some concern: the noncommutative behavior of spacetime
is an exclusive feature of the radial coordinate, through the
``substitution'' $r \rightarrow \hat{r}$, a fact in contradiction
with the covariant nature of the commutative coordinates; there
are some ambiguities in the notation since the differential
displacement $ds^2$ is written in terms of hermitian operators
$x^i$ and $p^i$ rather than functions and nowhere is it defined,
$d\hat{r}$ a symbol which appears somehow inconsistent; the line
element is no longer a solution of any gravitational field
equation, which could support this procedure; the linear and
angular momentum terms appear inconsistent, leading $1/r^3$
corrections, which make worse the behavior of the geometry at the
origin.

\subsubsection{Towards the noncommutative inspired Schwarzschild black hole}

The analysis which we have made up to now put a severe limitation
about the possibility of employing the above metrics for all
practical purposes in the field of particle phenomenology, where
the Planck phase of the resulting black hole is expected to
happen. As we have already highlighted, any perturbative expansion
in $\theta$ cannot any longer govern the nonlocal character of the
noncommutative manifold and cannot cure the singular behavior of
conventional black hole solutions, leading to weakly reliable
physical conclusions about the final stage of the evaporation.
Against this background, a new perspective has been proposed in
2005, in order to take seriously into account the role of the
noncommutative quasi coordinates, obtained by averaging position
operators. Indeed in Ref.~\refcite{Gruppuso:2005yw}, the
modification of the Newtonian potential coming from the $h_{00}$
component of the graviton field
\begin{equation}
h_{00}(x)=-8\pi G\int d^4y \ G(x,y)\ T_{00}(y),
\end{equation}
has been for the first time addressed, considering gaussian smearing of both the energy momentum tensor and the
propagator in momentum space.  Contrary to what we have already discussed about Ref.~\refcite{Kobakhidze:2007jn},
the calculation has been performed exactly, namely without any sort of expansion in the noncommutative parameter
$\theta$. The final result leads to the following form of the potential $\phi(r)$
\begin{equation}
\phi(r)=-G\ \frac{M}{r}\ erf\left(\frac{r}{2\sqrt{\theta}}\right) \label{ncNewton}
\end{equation}
where the error function is defined as
$erf(x)\equiv\frac{2}{\sqrt{\pi}}\int_0^x dt\ e^{-t^2}$.  The
asymptotic behaviors of $\phi(r)$ reveal how noncommutativity
works: indeed at infinity, namely for $r\gg\sqrt{\theta}$, the new
solution matches the conventional $1/r$ form, while for small
distances, namely for $r\ll\sqrt{\theta}$, the appearance of the
noncommutative parameter $\theta$ regularizes the divergence in
such a way that $\phi(0)=-G\frac{M}{\sqrt{\pi\theta}}$. For the
first time $\theta$ has here played the role of minimum length,
the sought short distance cut-off descending from the intrinsic
non-locality of a field propagating on a noncommutative manifold.

This result opened the road to further investigations on the
nature of the gravitational interaction, when a noncommutative
manifold is considered.  Indeed the form (\ref{ncNewton}) has been
exploited to propose the $2D$ noncommutative equivalent of the
Schwarzschild black hole\cite{Nicolini:2005de} and to determine
and solve the linearized noncommutative version of the Einstein
equation\cite{Nicolini:2005gy,Nicolini:2005zi}. Contrary to what
was found in Refs.~
\refcite{Harikumar:2006xf,Kobakhidze:2007jn,Steinacker:2007dq},
the weak field regime provides important conclusions about the
role of noncommutativity in General Relativity and black hole
evaporation. Indeed, noncommutativity, being an intrinsic property
of the manifold, does not depend on the curvature; therefore if
any effect is produced by noncommutativity it must appear also in
the weak field regime. Second, in the considered case of static,
spherically symmetric geometry, the expression for the temperature
does not depend on the weak field expansion and the new
thermodynamic behavior of the black hole can be studied, even
before the complete Einstein field equation has been analyzed. The
adoption of quasi coordinates in the spirit of Ref.~
\refcite{Smailagic:2004yy} leads to a modified form of the
linearized Einstein equation
\begin{eqnarray}
&&\vec\nabla^2\, g_{00} =
8\pi\,G \,\rho_\theta\left(\,\vec{x}\,\right)\ ,\nonumber\\
&&\rho_\theta\left(\,\vec{x}\,\right)= \frac{M}{\left(\,2\pi\theta\,\right)^{3/2}}\,
\exp\left(-\vec{x}^{\,2}/4\theta\,\right) \label{poisson}
\end{eqnarray}
providing an effective smearing of the conventional Dirac $\delta$
source term, in agreement with the presence of a minimum length
$\sqrt{\theta}$ throughout the manifold.  On the other hand, the
geometrical part, coming from the Einstein tensor is formally left
unchanged, even if one has to keep in mind that this tensor is now
written in terms of the aforementioned quasi coordinates.

\begin{figure}[ht!]
\begin{center}
\includegraphics[width=5.5cm,angle=270]{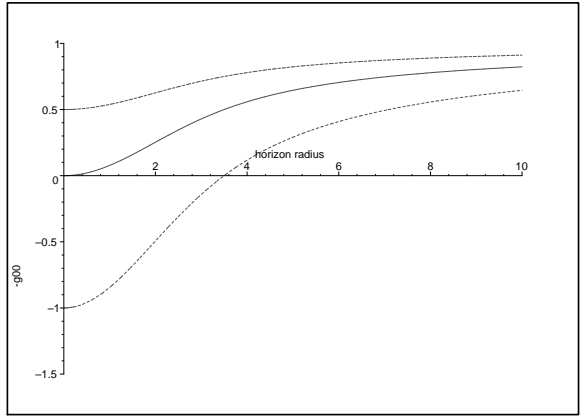}
\caption{\label{fg00} {\it The linearized solution.}  The function
$g_{00}$ vs the radial distance $r/\sqrt{\theta}$ for some value
of the mass $M /\sqrt{\theta} M_P^2$. The dashed line, corresponds
to a mass $M=0.5 M_0$ for which there is no event horizon. The
dotted line corresponds to a mass $M=2 M_0$, which describes a
black hole, which is regular at its center $r=0$. The solid curve
is the borderline case, namely the case of $M=M_0$ in which the
horizon radius $r_H$ is shrunk to the origin.}
\end{center}
\end{figure}

Therefore one can obtain, without any perturbative expansion in $\theta$, the noncommutative equivalent of the
linearized Schwarzschild line element
\begin{equation}
ds^2 = \left( 1- \frac{2MG}{\sqrt{\pi}\ r}\gamma\left(1/2\ , r^2/4\theta\, \right) \right) dt^2 - dl^2
\end{equation}
where $\gamma$ is the lower incomplete Gamma function, with the definition
\begin{equation}
\gamma\left(1/2\ , r^2/4\theta\, \right)\equiv \int_0^{r^2/4\theta} dt\, t^{-1/2} e^{-t}
\end{equation}

\begin{figure}[ht!]
\begin{center}
\includegraphics[width=5.5cm,angle=270]{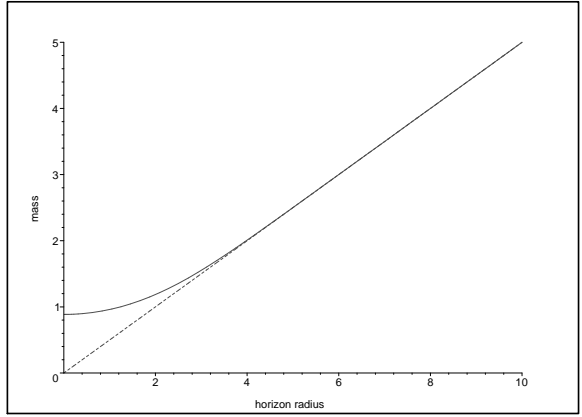}
\caption{\label{fig1} {\it The linearized solution.} The mass $M/\sqrt{\theta}M_P^2$ vs horizon
$r_H/\sqrt{\theta}$ relation. In the commutative case, dashed line, the mass is the linear function
$M=r_H/2\ell_P^2$ vanishing at the origin, while in the noncommutative case, solid line, $M\left(\, r_H\to
0\,\right)\to M_0=0.5\sqrt{\pi\theta}M_P^2$, i.e. for $M< M_0$ there is no event horizon.}
\end{center}
\end{figure}

\begin{figure}[ht!]
\begin{center}
 \includegraphics[width=5.5cm, angle=270 ]{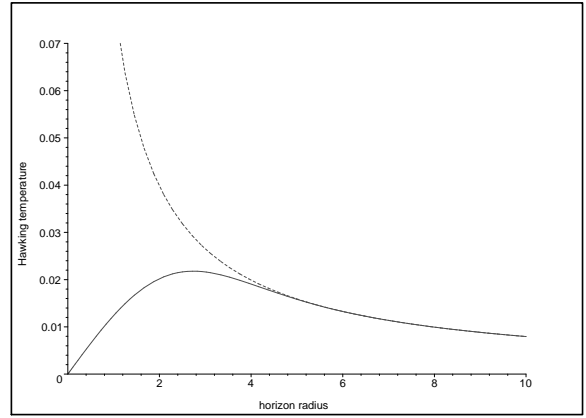}
\end{center}
\caption{\label{fig2} {\it The linearized solution.} The Hawking
temperature $T_H\sqrt{\theta}$ as a function of the horizon radius
$r_H/\sqrt{\theta}$. In the noncommutative case, solid curve, one
can see that the temperature reaches a maximum value $T_H^{Max.}=
2.18\times 10^{-2}/\sqrt{\theta}$ for $r_H= 2.74\sqrt\theta$, and
then decreases to zero as $r_H\to 0$ (SCRAM phase). The
commutative, divergent behavior (Planck phase), dashed curve, is
cured.}
\end{figure}

Under the condition $r>\sqrt{\theta}\sim \ell_P$, the above
metric, for its intrinsic nonlocal character, anticipates all the
desired features of the noncommutativity, which emerge in the
solution of the full Einstein equation: on the geometrical side,
the quantum fluctuations of the noncommutative spacetime manifold
cure the curvature singularity at the origin; on the
thermodynamical side, the resulting black hole undergoes, after a
temperature peak, a slowdown of the Hawking emission and
experiences a SCRAM phase\footnote{The term SCRAM, probably the
backronym for ``Safety Control Rod Axe Man'' or ``Super Critical
Reactor Axe Man'', refers to an emergency shutdown of a
thermonuclear reactor. The term has been extended to cover
shutdowns of other complex operations or systems in an unstable
state, but is also has a ``standard'' meaning ``go away quickly'',
in particular when we address children or animals. } in place of
the Planck phase, shrinking to an absolute zero stable relic.
Indeed from the horizon equation (see Fig.~\ref{fg00}), one learns
that noncommutativity implies a minimum non-zero mass $M_0$ in
order to have an event horizon (see Fig.~\ref{fig1}). Another
important aspect is that the above solution is stable, unaffected
by any nonlinear perturbation due to back reaction effects.
Indeed, preventing a proper Planck phase, noncommutativity keeps
the black hole at a state too cool to have a thermal energy
comparable with the black hole mass, namely $T/M < 2.5\times
10^{-2}\ell_P^2/\theta$.

\subsection{The full noncommutative inspired Einstein's equation}

Even if in a simplified form, the above linearized model showed,
for the first time, the noncommutativity nicely at work in its
regularization business.  The need to have more reliable results
about the fate of a radiating black hole, namely for
$r\sim\sqrt{\theta}$, struck at the time when the spherically
symmetric, neutral solution of the full noncommutative inspired
Einstein equation was first found, a solution which has been the
subject of the seminal paper\cite{Nicolini:2005vd} in this
research area and successfully improved the conventional
Schwarzschild geometry. As we shall see, the solution,
intentionally called ``noncommutative inspired'', is subject to a
tricky, dual interpretation, often in balance between the
classical and the quantum formalism. Indeed the calculations,
which we are going to perform, have the appearance of those of
classical General Relativity, even if their physical content is
purely quantum gravitational. The reason of this can be found in
the employment of the noncommutative quasi coordinates, that give
rise to a quantum formalism in terms of noncommutative deformed
classical functions.  Therefore, it will be natural that some
features of the solution have their explanation only in terms of
Quantum Gravity.

We shall start from some basic considerations about the known
Schwarzschild solution, which is often considered a ``vacuum
solution'', since it solves the homogeneous equation
$R_{\mu\nu}=0$.  Even if, for computational purposes the latter
homogeneous equation provides the solution sought, strictly
speaking any ``vacuum solution''
 has to be found only in the absence of matter.
 Indeed, as elegantly shown in Ref.\refcite{Balasin:1993fn},
  the above homogeneous equation $R_{\mu\nu}=0$ works computationally
 because one implicitly assumes a source term, the energy momentum tensor, concentrated on a region, the origin,
  excluded from the domain $\mathbb{R}\times\left(\mathbb{R}^3  \setminus \left\{0\right\}\right)$ of the solving line element.
  As a result, one is led to the physically inconsistent
  situation in which curvature is generated by a zero energy momentum tensor, while, properly speaking, only
  the Minkowski spacetime can be considered a ``vacuum solution'' of  Einstein's equation.
    These considerations are necessary in view of the noncommutative equivalent of the Schwarzschild solution,
  because in any pointless geometry one cannot employ the shortcut of the homogeneous equation, which is based on
  the assumption that matter is concentrated in a single point. Indeed we do expect that the energy momentum
  tensor is diffused throughout a region of linear size governed by the noncommutative parameter $\theta$.

\subsubsection{The noncommutative inspired Schwarzschild solution}

  We shall start from the noncommutative version of the Einstein equation in terms of quasi
  classical coordinates
  \begin{equation}
  G^{\mu\nu}_\theta=\frac{8\pi G}{c^2}\ T^{\mu\nu}_\theta
\end{equation}
in which the only relevant modifications occur in the source term, while $G^{\mu\nu}_\theta$ is formally left
unchanged.  Indeed, considering the energy density distribution of a static, spherically symmetric, noncommutative
diffused, particle-like gravitational source, one gets
\begin{equation}
\rho_\theta\left(\,r\,\right)= \frac{M}{\left(\,4\pi\theta\,\right)^{3/2}}\, \exp\left(-r^2/4\theta\,\right)
 \end{equation}
 which replaces the conventional Dirac $\delta$ distribution\cite{DeBenedictis:2007bm}
 and enters the temporal component of the energy momentum
 tensor
 \begin{equation}T^{\,\,\,\, 0}_{\theta\, 0 }=-\rho_\theta\left(\, r
\,\right).
\end{equation}
The covariant conservation $T^{\mu\nu}_\theta\,;\,\nu=0$ and the
additional ``Schwarzschild like'' condition $g_{00}=-g_{rr}^{-1}$,
completely specify the form of the energy momentum tensor
\begin{equation} T_{\theta\,\mu}^{\,\,\,\,\nu} = \left(
\begin{array}{cccc}
-\rho_\theta &  &  &\\
 & p_r &  &\\
 &  & p_\perp & \\
 & & & p_\perp
\end{array} \right)\label{t00}
\end{equation}
with $p_r= -\rho_\theta$ and $p_\perp = -\rho_\theta
-\frac{r}{2}\,\partial_r\rho_\theta\left(\,r\,\right)\label{perp}$. The above energy momentum tensor is rather
unusual because differs from the conventional perfect fluid tensor, which exhibits isotropic pressure terms: on
the other hand it is easy to show that $p_r$ and $p_\perp$ are different only within few $\sqrt{\theta}$ from the
origin and the perfect fluid condition is reestablished for larger distances. Plugging the above
$T^{\mu\nu}_\theta$ into the resulting Einstein equation, we find the line element\footnote{We are using the
convenient units $G=c=1$.}:
\begin{equation}
 ds^2 =-\left(\, 1- \frac{4M}{r\sqrt{\pi}}\, \gamma(3/2 \ ,
r^2/4\theta\,)\, \right)\, dt^2 + \left(\, 1-\frac{4M}{r\sqrt{\pi}}\, \gamma(\, 3/2\ , r^2/4 \theta\, )\,
\right)^{-1}\, dr^2 + r^2\, d\Omega^2 \label{ncs}
\end{equation}
where $d\Omega^2=d\vartheta^2 +\sin^2\vartheta\, d\phi^2$ and $\gamma\left(3/2 \ , r^2/4\theta\, \right)$ is the
lower incomplete Gamma function
\begin{equation}
\gamma\left(3/2\ , r^2/4\theta\, \right)\equiv \int_0^{r^2/4\theta} dt\, t^{1/2} e^{-t}.
\end{equation}
The above metric describes a self-gravitating, anisotropic
fluid-type matter, whose non vanishing radial pressure is a
consequence of the quantum fluctuation of the spacetime manifold
and balances the inward gravitational pull, preventing the matter
collapsing into a point. This is one of the first physically
relevant effects due to noncommutativity: indeed on purely
classical grounds, one would be led to believe in a matter
collapse, the contrary of what happens when the intrinsic quantum
gravitational nature of the solution is taken into account. As
expected, the relevance of the noncommutative correction occurs in
a neighborhood of the origin i.e. $r\lesssim\theta$, where both
the radial and the tangential pressures together with the matter
density are regularized by $\theta$ and share a finite value,
$M/(4\pi\theta)^{3/2}$. Asymptotically far away from the origin,
the density and the pressures virtually vanish, reproducing the
regime which supported the name of ``vacuum solution''. To have a
complete understanding of the further surprising features, it is
worthwhile to analyze both the geometrical and the thermodynamical
behavior of the solution.

\subsubsection{Horizons, curvature and energy conditions}

Let us start from the study of the horizon equation
$-g_{00}(r_H)=g^{rr}(r_H)=0$.

\begin{figure}[ht]
\begin{center}
\includegraphics[width=5.5cm,angle=270]{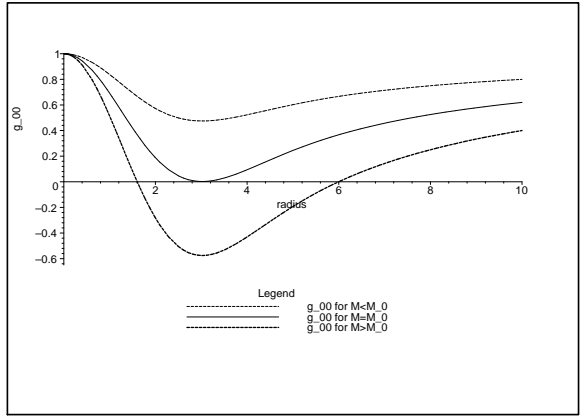}
\caption{\label{hor} {\it The noncommutative Schwarzschild solution.} The function $-g_{00}=g_{rr}^{-1}$ vs
$r/\sqrt{\theta}$, for various values of $M/\sqrt\theta$. Intercepts on the horizontal axis give radii of the
horizons. The lighter curve corresponds to $M=1.00 \sqrt\theta $ (no horizon), the darker curve below corresponds
to $M=M_0\approx1.90\, \sqrt\theta $ (one degenerate horizon at $r_H= r_0\approx 3.0\, \sqrt\theta$) and finally
the lowest curve corresponds to $M= 3.02\, \sqrt\theta $ (two horizons at $r_H=r_-\approx 1.60\sqrt{\theta}$ and
$r_H=r_+\approx 6.00\sqrt{\theta}$).}
\end{center}
\end{figure}

Fig. \ref{hor} illustrates the role of noncommutativity and the
new behavior with respect to the conventional Schwarzschild
solution: instead of a single event horizon at $r_H=2M$, the
intersection of the function $g_{00}$ with the $r$ axis shows that
there is a minimum mass $M_0\approx 1.90\sqrt{\theta}$ below which
no event horizon occurs. For $M>M_0$ the line element exhibits two
distinct horizons, while for the borderline case, $M=M_0$, we find
a unique degenerate horizon at $r_H=r_0\approx 3.0\sqrt{\theta}$,
corresponding to the extremal black hole.

As signature of consistency of this noncommutative solution, we find that the curvature singularity at the origin
is cured. Indeed the short distance behavior of the Ricci scalar is given by
\begin{equation}
 R\left(\, 0\,\right)=\frac{4M}{\sqrt\pi\, \theta^{3/2}}.
  \label{ricci0}
\end{equation}
Therefore for $r\ll\sqrt{\theta}$, the curvature is constant and
positive. Instead of the curvature singularity one finds a
deSitter core, whose cosmological constant is governed by
noncommutativity, namely $\Lambda=M/3 \sqrt\pi\ \theta^{3/2}$. The
presence of such a deSitter core contributes to shedding light on
the origin of the finite pressure terms, which we discussed above.
In analogy to the appearance of cosmological constant terms
induced by quantum vacuum fluctuations, when described in terms of
a fluid type energy momentum tensor\cite{Zel'dovich:1968zz}, the
previous pressure terms are the fluid type picture of the quantum
noncommutative fluctuations, which determine a regular deSitter
core at the black hole center.  The above result is in full
agreement with previous attempts to match an outer Schwarzschild
geometry with an inner deSitter core either through
time-like\cite{Aurilia:1984cm,Aurilia:1987cp,Aurilia:1983ih} and
space-like shells\cite{Frolov:1989pf,Frolov:1988vj}, attempts
which gave rise to the vast literature about regular black holes
(for a recent review see Ref.~\refcite{Ansoldi:2008jw} and the
references therein). The novelty and the virtue of the above
solution lies in the fact that the singularity in not cured
placing by hand a regular core at the origin, but it is the
noncommutative parameter $\theta$ that provides a smooth
transition between the inner regular deSitter geometry and the
outer Schwarzschild one, without invoking any matching condition.
In other words, contrary to other approaches, the regularity of
the spacetime is no longer an artifact, but it directly descends
from the noncommutative defining relation (\ref{ncxL}).

\begin{figure}[ht]
\begin{center}
\includegraphics[width=5.5cm,angle=270]{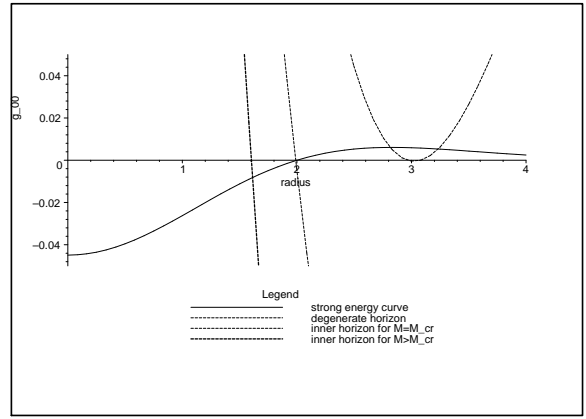}
\caption{\label{SEC} {\it The noncommutative Schwarzschild
solution.} The solid curve is the function $(\rho_\theta
+p_r+2p_\perp) \times \theta$ vs $r/\sqrt{\theta}$. Dashed lines
corresponds to $-g_{00}$ for various value of the mass. The
intercepts of $(\rho_\theta +p_r+2p_\perp)$ with the $r$-axis mark
the border, at $r=2\sqrt{\theta}$, between the classical and
quantum description of the spacetime manifold. We observe that for
the extremal case, $M=M_0$, the degenerate horizon $r_0>
2\sqrt{\theta}$ is in the region where the strong energy condition
is satisfied. Finally there is the curve of the critical case,
namely $-g_{00}$ for a value of the mass $M=M_{cr}\approx
2.35\sqrt{\theta}$, such that the inner horizon coincides with
$2\sqrt{\theta}$. This means that for $M<M_{cr}$ the classical
picture can still be employed to describe the black hole, except
in a limited region in the vicinity of the origin, where the
classical singular behavior non longer exists.}
\end{center}
\end{figure}

A very tricky matter regards the energy conditions of the
solution. Again, for a correct interpretation of the results, one
has to keep in mind the noncommutative origin and the quasi
classical procedure, which we have employed up to now. Indeed if
we interpret the metric in Eq. (\ref{ncs}) merely as a solution of
classical Einstein's equation, one would have some concern for the
violation of the strong energy condition
\begin{equation}
\rho_\theta +p_r+2p_\perp \ge 0
\end{equation}
even if the weak energy conditions, $\rho_\theta +p_r\ge 0$ and
$\rho_\theta +p_\perp\ge 0$ are always satisfied. Indeed
spherically symmetric, asymptotically Schwarzschild, regular
solutions of this kind have known global structure and are
characterized in a precise way: in all generality if they have an
(outer) event horizon, they must have an (inner) Cauchy horizon
and they fail to be globally hyperbolic. To this purpose, also a
non-vanishing matter density distribution beyond the outer horizon
is not a novelty and can be considered as a matter of secondary
concern. Indeed black holes may be dirty, namely they may be
surrounded by some kind of matter, admitting the equilibrium as
possible
configuration\cite{Letelier:1979ej}\cdash\cite{Bronnikov:2008ia}.
On the other hand, as recently underlined\cite{Ansoldi:2008jw}, an
interesting feature of the noncommutative inspired solution is
that the violation of the strong energy condition could take place
outside the inner Cauchy horizon, implying a runaway extension of
the region where gravity switches to a repulsive quantum
interaction. To shed light on this matter it is convenient to note
that
\begin{equation}
\rho_\theta +p_r+2p_\perp \sim e^{r^2/4\theta}\left(\frac{r^2}{2\theta}-2\right)
\end{equation}
which implies a strong energy condition violation for
$r<2\sqrt{\theta}$. This means that, in spite of the nonlinear
nature of gravity, the classical description of matter and energy
breaks down only in a region where we supposed, since the
beginning, that quantum effects would have been dominant. In other
words in the core around the origin, namely within
$2\sqrt{\theta}$ from the black hole center, gravity is actually
described by Noncommutative Geometry rather than by General
Relativity. For this reason, all the classical arguments cease to
be valid at small scales. Anyway, it is interesting to study the
extension of the above core around the origin with respect to the
position of horizons: to this purpose it is convenient to define
$M_{cr}$, the value of the mass parameter for which the region of
strong energy violation is confined behind the inner Cauchy
horizon. In other words, we have that black holes total with mass
$M_0<M<M_{cr}\approx 2.35\sqrt{\theta}$ the breakdown of the
classical picture of the solution occurs only in a limited region
around the origin, just to cure the singularity thanks to the
quantum fluctuations. (see Fig. \ref{SEC}). A related source of
concern would be the potential blue shift instability at the inner
horizon. Indeed an observer crossing the inner horizon would
experience an arbitrarily large blue shift of any incoming
radiation and see the entire history of the exterior region in a
finite lapse of his own proper time, as he approaches the horizon.
This suggested that any small perturbation would disrupt the
horizon and develop a curvature singularity. Previous analyses
have shown that the inner horizon of the Reissner-Nordstr\"{o}m
geometry\cite{McNamara:1978}\cdash\cite{Markovic:1994gy}, the past
horizon of the Schwarzschild white
hole\cite{Eardley:1974}\cdash\cite{Eardley:1975kp} and those of
the Schwarzschild wormhole\cite{Redmount:1985} are unstable. Even
if we do not have yet a definitive answer about this matter, the
noncommutative inspired black hole appears to be quite different
from the above cases: again the blue shift instability has to be
addressed taking into account, that the onset of a curvature
singularity is meaningful only for a classical differential
manifold. In a noncommutative background, the propagation of any
field, representing the perturbation, is subject to the presence
of a natural ultraviolet cut-off: therefore no observer could
experience an infinite amount of energy, approaching the Cauchy
horizon.  As a consequence of this, we can conclude that the
conventional arguments about the instability of Cauchy horizons
should be reviewed, when noncommutative effects are included.

For the special case of absence of event horizons, i.e. $M < M_0
$, the strong energy condition violation again occurs only in a
neighborhood of the origin. In particular, we have an everywhere
regular spacetime and no naked singularity appears. The resulting
regular gravitational system bears a strong resemblance to the
gravastar, the hypothetical gravitational vacuum star, whose model
is currently at the center of a hot scientific
debate\cite{Mazur:2001fv}. The new feature, introduced by
noncommutativity, is that the occurrence of a gravastar is subject
to the upper bound on its mass $M < M_0 $ . For this reason we
would suggest to call it ``mini-gravastar'', to distinguish it
from its more conventional forerunner. Finally, at the opposite
side, i.e. the large mass regime $M\gg M_0$, the solution
reproduces the standard Schwarzschild geometry, because the inner
horizon shrinks to the origin, while the outer horizon coincides
with $2M$, up to negligible exponentially suppressed corrections.
The last analysis can be equivalently made also by studying the
line element (\ref{ncs}) at large distances, i.e. $r\gg
\sqrt\theta$, a scale at which Noncommutative Geometry turns into
General Relativity and effectively describes the gravitational
field in terms of a smooth classical manifold.

\subsubsection{The black hole thermodynamics}

\begin{figure}[ht]
\begin{center}
\includegraphics[width=5.5cm,angle=270]{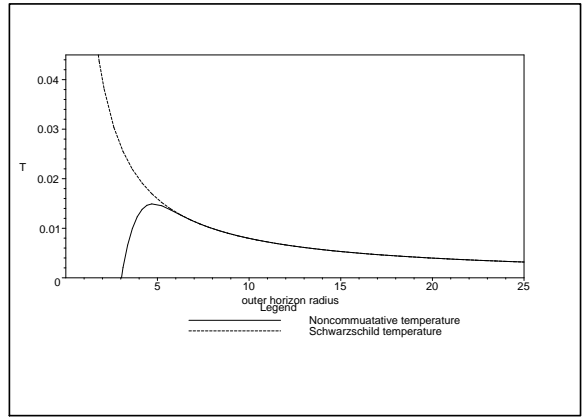}
\caption{\label{NCT} {\it The noncommutative Schwarzschild solution.} The Hawking temperature
$T\times\sqrt{\theta}$ vs the outer horizon radius $r_+/\sqrt{\theta}$. The dashed curve refers to the
conventional result for the Schwarzschild black hole, with a divergent Planck phase in the final stage of the
evaporation. The solid curve is the new behavior of the temperature due to noncommutative effects: for $r_+
\gtrsim 6\,\sqrt\theta$, the temperature follows the conventional behavior, while  it reaches a maximum $T_H\simeq
0.015\times 1/\sqrt{\theta}$, at $r_+=r_{max}\approx 4.76\sqrt{\theta}$, corresponding to a mass $M=M_{max}\simeq
2.4 \times\sqrt{\theta}$. Finally there is a SCRAM phase, leading to $T_H=0$ for $r_+=r_0=3.02\sqrt{\theta}$,
corresponding to a black hole remnant, the extremal black hole configuration.}
\end{center}
\end{figure}

The thermodynamics of the noncommutative inspired black hole is
even more interesting.  In agreement to what has already been seen
in the linearized case, the behavior of the temperature reflects
the regularity of the manifold. Indeed, also looking at the
temperature profile, we can conclude that the black hole is
essentially the Schwarzschild one for large distances, where the
conventional result $\sim 1/M$ still holds. The exciting novelty
appears, when, during the evaporation, the outer horizon $r_+$
shrinks to $\sim 6\,\sqrt\theta$: the Hawking temperature deviates
from the conventional behavior and reaches a maximum at
$r_+=r_{max}\approx 4.76\sqrt{\theta}$, after which the black hole
cools down, with a relevant slow down of thermal emission. Thus,
the SCRAM is over at $r_+=r_0\approx 3.02\sqrt{\theta}$, when the
black hole reaches the extremal configuration with mass
$M_0\approx 1.90\sqrt{\theta}$, the temperature is zero and the
Hawking emission breaks off. Analytically one can determine
\begin{equation}
T_H = \frac{1}{4\pi\,r_+}\left[\, 1
 -\frac{r^3_+}{4\,\theta^{3/2}}\,
  \frac{e^{-r^2_+/4\theta}}{\gamma\left(\, 3/2\ ; r^2_+/4\theta \right)}
\,\right] \label{thnc}
\end{equation}
where the mass $M$ is written in terms of $r_+$ from the horizon
equation.  Again for $r_+<r_0$ we cannot speak of an event horizon
and no temperature can be defined.  To this purpose, the final
zero temperature configuration can be considered a black hole
relic, namely the remnant of the black hole evaporation via
Hawking emission. Such a relic could provide a sort of solution to
the long standing problem of the information
paradox\cite{Hawking:2005kf}: indeed, due to occurrence of the
cool SCRAM phase in place of the hot Planck phase, the initial
information is confined and preserved into the black hole relic,
even after the end of the evaporation.  A detailed analysis of
spin-2 quantum amplitudes led to the conclusion that unitarity is
preserved during the life of the black hole\cite{DiGrezia:2006rw}.
From Fig. \ref{NCT} it follows that the black hole is ``colder''
with respect to what one would imagine: indeed, even if the
maximal temperature could reach the astonishing value $\sim
2.1\times 10^{30}$K, it is not able to prime any back reaction
effect. This is one of the key points of the SCRAM phase, which
lets us employ the solution obtained during all of the black
hole's life: in fact to have significant back reaction, one should
have $E_{max}\sim M$, where $E_{max}$ is the peak of the thermal
energy, a condition which leads to $\sqrt{\theta}\lesssim
10^{-34}$ cm, an unphysical constraint that can never be met.

\begin{figure}[ht]
\begin{center}
\includegraphics[width=5.5cm,angle=270]{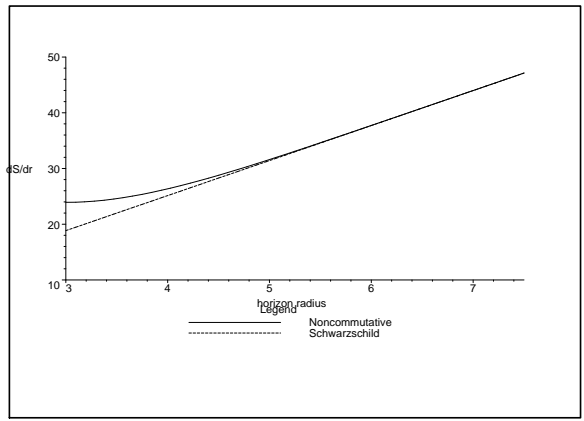}
\caption{\label{EntropyDER} {\it The noncommutative Schwarzschild solution.} The function $dS_H/dr_+ \times
\sqrt{\theta}$ vs the horizon radius $r_+/\sqrt{\theta}$. Deviations from the conventional linear behavior occur
around the $r_+\approx 4.76\sqrt{\theta}$, corresponding to the temperature peak and the beginning of the SCRAM
phase.}
\end{center}
\end{figure}

Regarding the remaining thermodynamical quantities, several
authors\cite{Nozari:2006bi}\cdash\cite{Banerjee:2008du} have
studied the effects of noncommutativity on the laws regulating
black holes. In particular, much effort has been recently devoted
to the entropy area law, employing the second law of
thermodynamics
\begin{equation}
dS_H= \frac{1}{T_H}\, \frac{d M (r_+)}{d r_+}\
 d r_+
\label{dsh}
\end{equation}
where the function $M(r_+)= \Gamma\left(3/2\right)\ r_+/2\gamma\left(3/2;\ r_+^{2}/4\theta\right)$ is given by the
horizon equation $-g_{00}(r_+)=g^{rr}(r_+)=0$. Once integrated from the extremal horizon $r_0$ to a generic outer
horizon $r_+$, the above formula provides that, once integrated from the extremal horizon $r_0$ to a generic outer
horizon $r_+$, provides
\begin{eqnarray}
\Delta S_H&=& \frac{1}{4G_\theta(r)}\left(A_+-A_0\right)+\delta S_H \label{Entropy}
\end{eqnarray}
 where $A_+=4\pi r_+^2$, $A_0=4\pi r_0^2$,  while the $G_\theta (r)$ is an effective Newton constant\footnote{The line element (\ref{ncs}) can be written as
\begin{equation} ds^2=-\left(1-2MG_\theta/r\right)dt^2+ \left(1-2MG_\theta/r\right)^{-1}dr^2 +r^2 d\Omega^2
\end{equation}resolving the long standing problem of an asymptotically safe gravitational coupling\cite{weinberg}.
Contrary to what was found in Ref.~\refcite{Bonanno:2000ep}, the
short distance behavior of the incomplete gamma function implies
$G_\theta\sim r^3$ for small $r$, curing the curvature singularity
without any artificial introduction of an {\it ad hoc} smoothing
function $d(r)$. } given by
\begin{equation}
G\longrightarrow G_\theta\left(\, r\, \right)\equiv G \frac{2}{\sqrt{\pi}} \, \gamma\left(\, \frac{3}{2}\ ,
\frac{r^2}{4\theta}\,\right) \label{geff}
\end{equation}
and $\delta S_H$ is a small correcting term. The first term
corresponds to the noncommutative version of the famous
Bekenstein-Hawking area law, while further noncommutative
corrections are always exponentially small. Indeed, since
$r_0>\sqrt{\theta}$, we can approximate
$\gamma(3/2,r^2/4\theta)\approx \sqrt{\pi}/2$ and find
\begin{equation}
\delta S_H\approx\frac{1}{2\sqrt{\theta}\ G}\left(r_0^3e^{-r_0^2/4\theta}-r_+^{3}e^{-r_+^2/4\theta}\right).
\end{equation}
 We conclude that the area law is maintained up to exponentially small corrections.
In support of the above conclusion, there is Fig.
\ref{EntropyDER}: we can see that the conventional area-entropy
relation holds until $r_+\sim r_{max}\approx 4.76\sqrt{\theta}$,
where the maximal temperature takes place and the SCRAM phase
starts. Therefore noncommutative corrections are dominant in the
region $r_0\lesssim r_+\lesssim r_{max}$. The behavior of the
entropy remains qualitatively equivalent to that of the
Schwarzschild case at least with respect to the Third Law of
Thermodynamics: the entropy near to absolute zero is governed only
by the temperature and tends to a constant minimum value
independently of the other parameters. In particular we have
$\Delta S_H=0$, for the extremal black hole relic at $r_+=r_0$
(see
Refs.~\refcite{Myung:2006mz,Spallucci:2008ez,Banerjee:2008du}).

\begin{figure}[ht!]
\begin{center}
\includegraphics[width=5.5cm,angle=270]{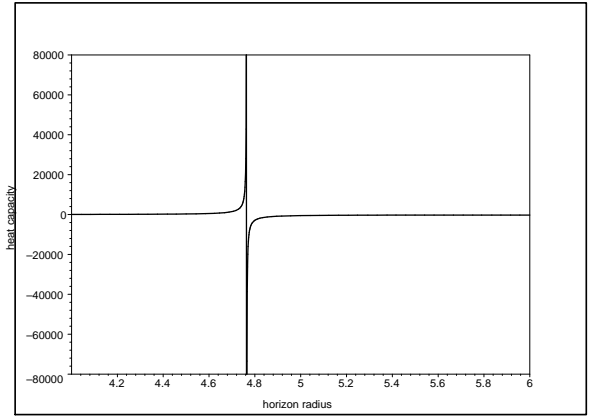}
\caption{\label{heat} {\it The noncommutative Schwarzschild solution.} The heat capacity $C\times\theta$ vs
$r_+/\sqrt{\theta}$. }
\end{center}
\end{figure}

A further inspection into the role of noncommutativity can be made
by looking at the black hole heat capacity
 $C\equiv dM/dT_H$, that can be written as
\begin{equation}
C(r_+)=T_H \left(\frac{dS_H}{dr_+}\right)\left(\frac{dT_H}{dr_+}\right)^{-1}.
\end{equation}
This relation highlights the black hole exchanges of energy with
the environment. Any zero for the temperature corresponds to
$C=0$, leading to the conclusion that the black hole relic is a
thermodynamically stable object, that cannot exchange energy and
evaporate.  On the other hand, any zero in $dT_H/dr_+$ produces a
pole and marks a change of sign of the heat capacity, switching
the black hole to a $C>0$ stable system, namely the very beginning
of the SCRAM phase.  The occurrence of a divergent behavior of the
heat capacity should not be worrying, since it is only an artifact
in the definition of $C$.  To this purpose, in support to the
consistency of the procedure, it is worthwhile to see that, even
at critical points of $C$ and $T_H$, the entropy
\begin{equation}
S_H=C\left(T_H\right)\left( \frac{dT_H}{T_H}\right) \label{entemp}
\end{equation}
is always a positive finite function, giving correct black hole
thermodynamics. Therefore the infinite discontinuity of the heat
capacity could signal the presence of some sort of phase
transition between a Schwarzschild unstable regime and a SCRAM
stable regime\cite{Pavon:1991kh,Myung:2007ti}. Finally we
introduce the black hole free energy
\begin{equation}
F(r_+)=M(r_+)-T_H(r_+)\ S_H(r_+)
\end{equation}
whose variations reveal, for $\delta F>0$ spontaneous processes or, for $\delta F<0$, disfavored processes or even
equilibrium states, for $\delta F=0$. Therefore it is worthwhile to study the behavior of $dF/dr_+$, which is zero
when either $T_H$ or $dT_H/dr_+$ are zero.  A deeper inspection can be made looking at second derivatives in the
critical points: to this purpose, numerical estimates\cite{Myung:2006mz} confirm that, as expected, the free
energy has a minimum at $r_+=r_{max}\approx 4.76\sqrt{\theta}$. As a consequence, $dF/dr_+$ is positive for
$r_0\le r_+\le r_{max}$, because of the positive heat capacity which characterize the SCRAM phase. This means that
the evaporation turns into a disadvantageous process, slowing down in the vicinity of the zero temperature
configuration.

\subsubsection{Black hole lifetime and detection}

Another important aspect of the black hole thermodynamics is the
time scale of the evaporation. On the basis of the above
considerations, we can assume that this process takes place
essentially in two phases, the Schwarzschild and the SCRAM phase,
 before and after the temperature peak respectively.  The formula
governing the black hole evaporation rate is
\begin{equation}
 \frac{dM}{dt}=-A_+\, \Phi
  \end{equation}
where $\Phi$ is the flux of thermal energy emitted by the
evaporating black hole.  Even if, in the absence of significant
back reaction, the integration of the above equation holds for all
of life of the black hole, it requires much attention in
particular in the vicinity of the final zero temperature state.
Indeed from a thermodynamical point of view, approaching the
absolute zero would take an infinite time. On the other hand the
temperature breaks off beforehand, namely when the black hole
reaches the thermal equilibrium with the environment at
$T_H=T_{CMB}\approx 2.7$K. For these reasons, we do expect a
finite black hole life. It is also interesting to have a
preliminary estimate of this life, just to fix the time scale. To
this purpose a rough calculation of the time it takes the black
hole to shrink from an initial configuration to $r_+=r_{max}$, can
be made by means of
\begin{equation}
\frac{dM}{dt}\sim -\ 4\pi r_+^2\ T_H^4
\end{equation}
Therefore the life of the black hole is almost in agreement with the conventional results, namely
\begin{equation}
Life\sim 10^{59}\left(\frac{M_{in}-M_{max}}{M_\odot}\right)^3 Gyr,
\end{equation}
where $M_{in}$ is the black hole initial mass, $M_{max}\approx 2.4 \sqrt{\theta}M_P/\ell_P$ is the mass at the
temperature peak and $M_\odot$ is the solar mass. As expected, the evaporation produces a significant lowering of
the black hole mass only for very tiny initial masses. For this reason, our results would be in agreement with
earlier proposal of the {\it elementary black holes}\cite{Holzhey:1991bx}, namely very small black holes, whose
mass would be of the order of the Planck mass, emerging as residue of the evaporation. In our scenario, the
stability of such small objects\cite{Markov:1965}\cdash\cite{Hawking:1971vc}, previously called {\it
maximos}\footnote{Other names are {\it Friedmons}, {\it Cornucopions}, {\it Planckons} and {\it Informons}, even
if there are slight but nontrivial differences among them\cite{Hossenfelder:2004af}. }, is regained since the
evaporation no longer occurs at zero temperature. The nature of maximos is rather striking: they have a very small
cross section $\sim 10^{-66}$ cm$^2$ and a large mass $\sim 10^{28}$ eV making their direct observation difficult.
On the other hand, due to their small interaction with ordinary matter, maximos are thought to act like dark
matter and be a considerable part of the matter in the Universe\cite{Markov:1981db,Barrow:1992hq}. In our
scenario, a possible solution to the problem of their observation would be the detection of their thermal emission
before its breaking off at the zero temperature, but again one has to deal with unreachable energy scales
($\lesssim 1.5\times 10^{26}$ eV). Regarding the thermodynamics, the properties of such black hole remnants has
been extensively studied\cite{Bowick:1988xh}\cdash\cite{Alexeyev:2002tg}, but on a modern perspective, their role
can be revitalized adopting additional space-like dimensions, with consequent lowering of energy
scales\cite{Koch:2005ks}.

Regarding the applications of the noncommutative inspired solution, we mention the description of the dual effect
of noncommutativity in both the matter and geometry sector, adopting the $r-t$ section of metric (\ref{ncs}) in
the world sheet of the 2D bosonic Polyakov string, governed by a noncommutative anomaly induced effective action.
The vacuum expectation value of the energy momentum tensor of the resulting 2D scalar field theory in the
Boulware, Hartle-Hawking and Unruh vacua has been explicitly calculated: as expected the standard short distance
divergences were regularized by the presence of quantum coordinate fluctuations\cite{Spallucci:2006zj}, improving
previously known results\cite{Zaslavskii:1991eb}\cdash\cite{Grumiller:2007ju}.

Finally on the more phenomenological side, we know for certain
that noncommutative effects are not visible until the electroweak
scale, therefore we can safely assume $\sqrt{\theta}<10^{-16}$cm.
On the opposite limit, the noncommutative correction to the
planets perihelion precession of the solar system has been
recently evaluated\cite{Nozari:2007rh}. This could lead to a
stringent lower limit on $\sqrt{\theta}$ on the grounds of
astronomical data. Anyway, we can safely consider
$\sqrt{\theta}\gtrsim 10^{-33}$cm. This striking difference of
possible scales, is another feature of the still unsolved
hierarchy problem. To this purpose, it is clear that with the
neutral four dimensional noncommutative inspired solution, we are
only scratching the surface of a gold mine. Far more important
consequences of noncommutativity will emerge from the
extradimensional solutions, in which we may consider as unique
scale $1/\sqrt{\theta}\sim M_\ast\sim 1$ TeV. A detailed analysis
of the extradimensional solutions will be the subject of one of
the next chapters.

\section{Noncommutative Charged Black Holes}

A logic step forward in this review is to consider the natural extension of the above neutral solutions in order
giving the resulting black hole ``Abelian hair'', represented by a long range electric field. Apart from a matter
of completeness, this generalization is motivated by the understanding of the role of noncommutativity in those
phases preceding the Schwarzschild phase: indeed it is vital to know whether the neutralization of the hole still
occurs in short times with respect its total mass loss, leading to a neutral zero temperature remnant
configuration.  On the other hand this program requires a preliminary inspection of noncommutative electrodynamics
and the analysis of an emerging new decaying channel, i.e. the Schwinger pair production, that, together with the
Hawking emission, contributes to the black hole mass loss. As we shall see, the determination of noncommutative
charged solutions is crucial also in view of the possible detection of some evidence for mini black holes as a
result of the forthcoming extreme energy experiments at LHC: it is very likely that the initial hadronic charge
will be distributed among the fragments after the collision, fragments which could include a mini black hole.
Therefore, a charged mini black hole could be produced surrounded by a cloud of pairs, repelling those particles
of the its own sign. In other words, the presence of many electrons near one of the hadronic fragments would be a
potential signal for the presence of a charged mini black hole.

We start along the lines suggested in Ref.~\refcite{Chamseddine:2000si} to have the extension of the result
obtained in Ref.~\refcite{Chaichian:2007we}. The procedure presented in
Ref.~\refcite{Mukherjee:2007fa,Chaichian:2007dr} is based on the quest for a solution of the following equation
\begin{equation}
\tilde{R}^\nu_\mu -\frac{1}{2}\tilde{R}\delta^\nu_\mu=-T^\nu_\mu.
\end{equation}
Here $T^\nu_\mu$ is the electromagnetic energy momentum tensor, while the curvature tensors are obtained by
contraction of the deSitter gauge group SO$(4,1)$ into the ISO$(3,1)$ Poincar\'e group, setting the torsion to
zero and employing the Seiberg-Witten map\cite{Seiberg:1999vs} to get the noncommutative character of the theory.
As a result the representation of the noncommutative deformed vielbein is obtained by expanding the
$\star$-product and it is found to be coincident with that already seen in (\ref{Vierbein}) for the Schwarzschild
case. As a result, one can compute the resulting deformed metric by the formula (\ref{Metric}) to obtain the
following non-vanishing components
\begin{eqnarray}
\hat{g}_{00}&=&g_{00}-\frac{1}{r^6}\left[Mr^3-\frac{11M^2+9Q^2}{4}r^2-\frac{17MQ^2}{4}r-\frac{7Q^4}{2}\right]\theta^2+O(\theta^4)\\
\hat{g}_{11}&=&g_{11}+\frac{\left[-2Mr^3+3(M^2+Q^2)r^2-6MQ^2r+2Q^4\right]}{4r^2(r-2Mr+Q^2)^2}\theta^2+O(\theta^4)\\
\hat{g}_{22}&=&g_{22}+\frac{1}{16}\left[1-\frac{15M}{r}+\frac{26Q^2}{r^2}+\frac{4(Mr-Q^2)^2}{r^2(r^2-2Mr+Q^2)}\right]\theta^2+O(\theta^4)\\
\hat{g}_{33}&=&g_{33}+\frac{1}{16}\left[\frac{4r^2(M^2-Mr)+8Q^2(r^2-2Mr)+8Q^4}{r^2(r^2-2Mr+Q^2)^2}\sin^2\psi+\cos^2\psi\right]\theta^2+\nonumber
\\&&+O(\theta^4).
\end{eqnarray}
The above results match the previous neutral solution (\ref{CHASCHW}) found in Ref.~\refcite{Chaichian:2007we}.
Contrary to what one would expect, the electromagnetic energy momentum tensor is kept in the ordinary form, namely
no $\star$-deformation has been considered for the electromagnetic field. From a physical point of view, this fact
appears unclear since the energy content of the electromagnetic field is expected to be intrinsically related to
the fluctuation of the noncommutative manifold.  Apart from these general comments, the major concern about the
above metric is the still persistent inability to cure the curvature singularity, which, this time, affects both
the gravitational field and the electromagnetic field, giving rise to even worse $1/r^6$ terms.  This is another
malicious feature of the expansion of the $\star$-product in $\theta$, indeed a procedure without any outlet to
reliable physical effects.

In the same way, another proposal for a noncommutative Reissner-Nordstr\"{o}m solution exhibits a bad short
distance behavior. By a mere substitution of the radial coordinate in terms of its noncommutative equivalent
$r\to\hat{r}=\sum_i(\hat{x}^i)^2$, the following line element has been introduced in Ref.~\refcite{Nasseri:2005rb}
\begin{equation}
ds^2=\left(1-\frac{2M}{\hat{r}}+\frac{Q^2}{\hat{r}\hat{r}}\right)dt^2-\frac{d\hat{r}d\hat{r}}{\left(1-\frac{2M}{\hat{r}}+\frac{Q^2}{\hat{r}\hat{r}}\right)}-\hat{r}\hat{r}
\left(d\psi^2+\sin^2{\psi}d\phi^2\right).
\end{equation}
As we have already seen for the Schwarzschild case\cite{Nasseri:2005ji,Nasseri:2005yr}, again the notation appears
unclear: once $\hat{r}$ is written in terms of the matrix $\theta^{ij}$ and the conventional position operators
$x^i$ and momenta $p_i$, $ds^2$ is still far from what we mean by line element. Despite this, allowing a sort of
classical interpretation for the operators, an expansion over the noncommutative parameter has been performed,
giving rise to
\begin{equation}
g_{00}\simeq\left(1-\frac{2M}{r}+\frac{Q^2}{r^2}\right)+\left(\frac{Q^2}{2r^4}-\frac{M}{2r^3}\right)\left[\vec{L}\vec{\theta}-\frac{1}{8}
\left(p^2\theta^2-(\vec{p}\cdot\vec{\theta})\right)\right]+{\cal O}(\theta^3),
\end{equation}
where $L_k=\varepsilon_{ijk}x_ip_j$ and $\theta_{ij}=\frac{1}{2}\epsilon_{ijk}\theta_k$.  Apart from the
ambiguities, the proposed line element fails the main goal of any noncommutative geometry approach and exhibits,
by the presence of the charge, an even worse $1/r^4$ term, with an inconsistent spherical symmetry breaking.

\subsection{The noncommutative inspired Reissner-Nordstr\"{o}m solution}

We are now ready to generalize the results already seen in Ref.~\refcite{Nicolini:2005vd} and study the
noncommutative inspired charged black hole\cite{Ansoldi:2006vg}, a geometry which is going to solve all of the
inconsistencies of the above line element in a single stroke.  We shall start from the new field, the Maxwell
field $F^{\mu\nu}$, studying its behavior within the framework of noncommutative quasi coordinates.  In agreement
with what we have already seen for the linearized solution in Ref.~\refcite{Nicolini:2005zi}, the Poisson equation
governing the electrostatic potential exhibits an enlarged source term instead of the conventional Dirac $\delta$.
Therefore, when a pointlike charge is considered in a noncommutative background, the effective current density is
given by
\begin{equation}
J^\mu\left(\,x\,\right)= \rho_{el.}\left(\, x \,\right)\,\delta^\mu_0
\end{equation}
where, as expected, $\rho_{el.}$ has a Gaussian profile
\begin{equation}
\rho_{el.}=\frac{{\cal Q}}{\left(\,4\pi\theta\,\right)^{3/2}}\, \exp\left(-{\vec x}^2/4\theta\,\right).
\end{equation}
As a consequence the electric field $E(r)$ turns out to be an everywhere regular function
\begin{equation}
E\left(\ r\,\right)=\frac{2{\cal Q}}{\sqrt\pi\, r^2}\, \gamma\left(\, \frac{3}{2}\ ; \frac{r^2}{4\theta}\,\right)
\label{clmb}
\end{equation}
giving rise to a nonsingular Maxwell field $F^{\mu\nu}=\delta^{0[\, \mu\,\vert}  \delta^{r\,\vert\, \nu \,]}\,
E\left(\, r\,\right) $.  With the above ingredients, we are ready to consider the system describing quasi
classically both the electromagnetic field and the gravitational field, a system which resembles the conventional
 Einstein-Maxwell system
\begin{eqnarray}
&& R^\mu{}_\nu-\frac{1}{2}\, \delta^\mu{}_\nu\, R = 8\pi\,\left(\, T^\mu_\nu\vert_{matt.} + T^\mu_\nu\vert_{el.}
\,\right)
\label{einst}\\
&& \frac{1}{\sqrt{-g}}\, \partial_\mu\,\left(\, \sqrt{-g}\, F^{\mu\nu}\, \right)= J^\nu. \label{max}
\end{eqnarray}
Here $T^\mu_\nu\vert_{matt.}$ is the same as in the neutral case, describing the energy content of the matter,
while $T^\mu_\nu\vert_{el.}$, even if  formally the usual one, takes into account, through the form of
$F^{\mu\nu}$, the smearing effect to which the propagation of the electromagnetic field is subject when
noncommutativity is considered. Thus, looking for a line element of the form $ ds^2= -f(r)\, dt^2 + f^{-1}(r)\,
dr^2 + r^2 d\Omega^2 $, one finds
\begin{equation}
 f(r)= 1 -\frac{4{\cal M}}{r\sqrt\pi}\ \gamma\left(\, 3/2\ , r^2/4\theta\,\right) + \frac{Q^2}{\pi\, r^2}\ F(r)\ ,
\end{equation}
where ${\cal M}$ is the ``bare'' mass parameter, while
\begin{equation}
F(r)\equiv  \gamma ^2\left(\,1/2\ ,r^2 /4\theta\,\right) -
 \frac{r}{\sqrt{2\theta }}\gamma\left(\, 1/2\ ,r^2 /2\theta\,\right)\ ,
 \label{rncbh}
\end{equation}
and the charge $Q={\cal Q}$ in natural units\footnote{Otherwise one has $Q={\cal Q}\ G^{1/2}/c^2$}. To have a
deeper understanding of the above solution, it is convenient to introduce the total mass-energy of the
electro-gravitational system
\begin{equation}
M =\oint_\Sigma d\sigma^\mu\left(\, T_\mu^0\vert_{matt.} + T_\mu^0\vert_{el.}  \,\right) \label{mtot}
\end{equation}
where $\Sigma$ is a $t=\mathrm{const.}$, closed three-surface. In terms of $M$ the solution reads
 \begin{equation}
  f(r)= 1
-\frac{4 M}{\sqrt\pi\, r} \gamma\left(\, 3/2\ , r^2/4\theta\,\right)
 + \frac{Q^2}{\pi\, r^2}\left[\, F\left(\, r\,\right)+\sqrt{\frac{2}{\theta}}
 \, r\, \gamma\left(\, 3/2\ , r^2/4\theta\,\right)\,\right]
 \label{ncrn}
  \end{equation}
which exhibits the expected asymptotic behavior, reproducing the Reissner-Nordstr\"{o}m geometry at large
distances.

\begin{figure}[ht!]
\begin{center}
\includegraphics[width=9.5cm,angle=0]{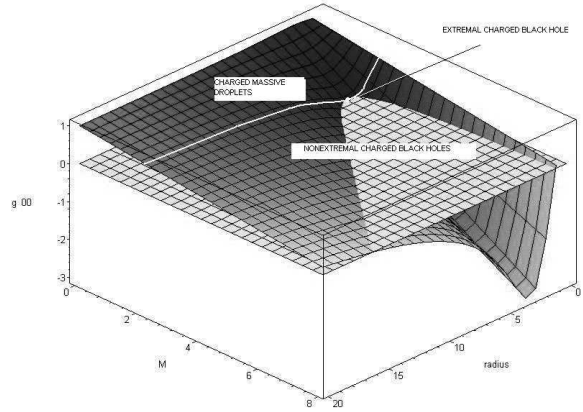}\caption{\label{g001} \textit{The noncommutative Reissner-Nordstr\"{o}m
solution.} The function $f$ is plotted versus $M$ and $r$, for a charge, $Q=1$ in $\sqrt\theta$ units.  The
intersection of the $f=0$ plane (light-grey) with $f=f(r,M)$ surface (dark-grey) gives the ``horizon curve'' whose
minimum (white-dot) gives the extremal black hole. The portion of the surface below the plane $f=0$ represents
  the spacelike region between the inner and outer horizons.}
\end{center}
\end{figure}

Our analysis proceeds with the study of the horizon equation $f=0$, that can be efficiently visualized in terms of
the plot in Fig. \ref{g001}. We can see that in analogy with the noncommutative Schwarzschild case, there are
three possible cases, this time depending on the values of both $ M$ and $Q$: therefore the line element
(\ref{ncrn}) describes either a two-horizon charged black hole or an extremal single horizon charged black hole or
a charged gravitational object without any horizon.  A key aspect of the solution is given by the fact that in all
such cases, the conventional curvature singularity at the origin is smeared out by the noncommutative fluctuations
of the spacetime manifold. Indeed the short distance behavior of the metric is of the deSitter type
\begin{equation}
 g_{00}= 1 -\frac{{\cal M}}{3\sqrt{\pi}\, \theta^{3/2}}\, r^2
   +O\left(\, r^4\,\right)
 \label{desitter}
 \end{equation}
whose effective cosmological constant $\Lambda_{eff.}={\cal M}/\sqrt{\pi}\, \theta^{3/2}$ is governed only by the
``bare'' mass because the electric field contributes with subleading $O\left(\, r^4\,\right)$ terms, due to its
linear behavior at short distance, i.e. $E(r)\sim Qr/6\sqrt{\pi}\ \theta^{3/2}$.

\begin{figure}[ht!]
\begin{center}
\includegraphics[width=7.5cm,angle=0]{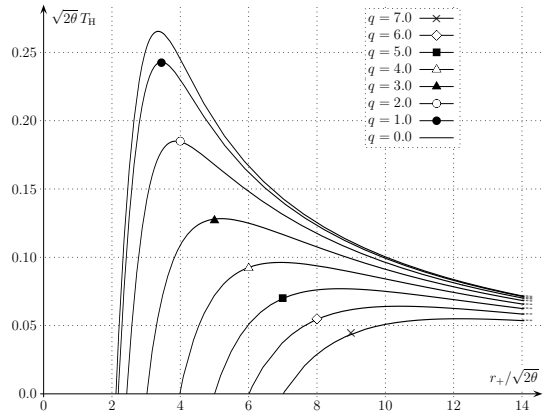}
\caption{\label{temp} \textit{The noncommutative Reissner-Nordstr\"{o}m solution}. The Hawking temperature
$T_H\times\sqrt{2\theta}$ as a function of
 $r_+/\sqrt{2\theta}$, for different values of $Q$ in $\sqrt{\theta}$ units.  The temperature drops to zero
 even in the case $Q=0$ as a result of coordinate uncertainty. The peak
 temperature drops with increasing $Q$.
}
\end{center}
\end{figure}

\subsubsection{Black hole temperature, Hawking and Schwinger mechanisms}

The regularity of the geometry has its thermodynamic equivalent in the finiteness of the black hole temperature,
which is given by
\begin{eqnarray}
 4\pi \, T_H &&= \frac{1}{r_+}
 \left[ {1 - \frac{{r^3_+ \exp ( - r^2_+ /4\theta )}}{{4\theta ^{3/2}
 \gamma\left(\,3/2\ , r^2_+ /4\theta\,\right)}}} \right]+ \nonumber\\
 &&-\frac{{4Q^2 }}{{\pi r^3_+ }}
 \left[ {\gamma ^2\left(\,3/2\ ,r^2_+ /4\theta \,\right) +
 \frac{{r^3_+ \exp ( - r^2_+ /4\theta )}}
 {{16\,\theta ^{3/2} \gamma\left(\,3/2\ , r^2_+ /4\theta\,\right)}}
 F(r_+)} \right]
\end{eqnarray}
where $r_+$ is the outer horizon.  The above formula contains the expression for the temperature already seen in
the neutral Schwarzschild case, with an additional term due to the presence of the charge and responsible for a
global lowering of the profile of $T_H$. At this point, one would be tempted to say that the final configuration
of the evaporation is an extremal black hole, with a nonzero residual charge, a radius and mass larger than
$3.02\sqrt{\theta}$ and $1.90\sqrt{\theta}$, the values, which we obtained for the neutral remnant.  Against this
scenario, there is the fact that, conventionally, the electric charge of a black hole is negligible, since the
neutralization phase occurs in very short times\cite{Novikov:1989sz,Wald:1984rg}. Furthermore, pairs of charges
can be produced quantum mechanically in the black hole's surroundings, reducing even faster the black hole
electric charge to zero.  The rate of production of pairs of particles in now ruled not only by the Hawking
emission, but also via the Schwinger mechanism\cite{Schwinger:1951nm} for the presence of the black hole electric
field. Indeed if the electrostatic potential energy on the black hole outer horizon $r_+$ is sufficiently high, at
least locally it is possible to have a uniform electric field exceeding the threshold electric field $E_{th}=\pi\,
m_e^2/e$ and priming the pair production\cite{Nikishov:1969tt}\cdash\cite{Preparata:2002wx}.  As a result, we
expect that the charged black hole balding phase will occur in a very short time, leading to a Schwarzschild phase
and finally to a SCRAM phase terminating with the formation of a zero temperature stable black hole relic, which
we already described by means of the noncommutative inspired Schwarzschild solution.  For this reason, in Fig.
\ref{temp}, the region of the plot near to the intersections with the $r_+$-axis, does not correspond to a
physically realizable situation, since charged remnants do not develop. As a consequence, the black hole back
reaction due to the Hawking emission mechanism, is still negligible: indeed the balding phase is so quick that, at
a given $r_+$ the black hole is always colder and heavier than its neutral equivalent. In support of the above
reasoning, it is sufficient to require that the electric field at $r_+$ exceeds the threshold field, namely
$E(r_+)>E_{th}=\pi\, m_e^2/e$. Writing the black hole total charge as $Ze$, i.e. an integer multiple of the
elementary charge, one finds that at any possible value for $r_+$, just a single electron charge, namely having
$Z=1$, is sufficient to have pair creation in the black hole surroundings. Indeed the threshold field, once
written in terms of $\theta$ units, is $E_{th}\sim m_e^2 \theta << 1$ and therefore the black hole electric field
is so strong that the Schwinger mechanism can occur even far away from the black hole outer horizon. In view of
this, it is useful to introduce the concept of {\it dyadosphere}, representing the spherical region where the
electric field reaches the threshold value and produces pairs: then, in our case, the dyadosphere radius $r_{ds}$
can be determined by
\begin{equation}
\frac{r_{ds}^2}{\gamma\left(\,\frac{3}{2}\ ;\frac{r^2_{ds}}{4\theta}\,\right)}= \frac{2}{\pi^{3/2}}\frac{Z e^2}{
m^2_e}.
\end{equation}
As a first approximation, we find
\begin{equation}
r_{ds}^2\simeq \frac{2Z e^2}{\pi^{3/2} m^2_e}\, \gamma\left(\,\frac{3}{2}\ ;\frac{Z e^2  }{4\pi \theta m^2_e
}\,\right) \label{rdyado}
\end{equation}
which lets us estimate the dyadosphere radius as being comparable with the electron Compton wave length
$r_{ds}\sim 1/m_e\gg \sqrt{\theta}$. From the large extension of the dyadosphere with respect to the
noncommutative scale $\sqrt{\theta}$, we conclude that the black hole described by the line element (\ref{ncrn})
is extremely unstable under Schwinger pair production, a mechanism which dominates the early life of the black
hole until a neutral phase is reached.  An important note regards the hot debate in the literature about the
occurrence of a dyadosphere for astrophysical black holes\cite{Page:2006cm}\cdash\cite{Page:2006zk}. Without
entering this debate, it is necessary to show that the criticisms against dyadosphere formation do not apply to
objects like the noncommutative black holes, which are governed by the scale $\sqrt{\theta}$: indeed  we find that
in our case the condition for the existence of the dyadosphere
\begin{equation}
\frac{E}{E_c}=\frac{e\ {\cal Q}}{r^{2}\ m_e^2}\frac{2}{\sqrt{\pi}}\, \gamma\left(\, \frac{3}{2}\ ;
\frac{r^2}{4\theta}\,\right)\ll\frac{m_p \ M G}{r^{2}\ m_e^{2}}\sim \left(\frac{m_p}{m_e}\right)\frac{M}{m_e} \sim
4\times 10^9
\end{equation}
is always met, where $m_p$ is the proton mass. Thus we can conclude that the electric field can reach the
threshold value without the electrostatic repulsion overcoming the gravitational attraction among hadronic charged
matter.  For the above reason, it is legitimate to follow the lines suggested in
Refs.~\refcite{Preparata:1998rz,Preparata:2002wx}, to have an estimate of the discharge time via the Schwinger
mechanism.  To this purpose it is worthwhile to introduce the surface charge density $\sigma (r)$, in order to
divide the dyadosphere into thin spherical shells of thickness $1/2m_e$.  Calculating the number of pairs
produced, per second inside such a spherical shell because of the presence of the electric field $E(r)=4\pi\sigma
(r)$, one finds that the discharge time is
\begin{equation}
 \Delta \tau = \frac{\theta\, m_e}{\alpha_{em}}\,
\left(\frac{2\pi}{m_e c\sqrt{\theta}}\right)^{2} \frac{s -1}{s^{2}} \exp\left(\,\frac{\pi}{s}\,\right)
\label{meanlifem}
\end{equation}
where $\alpha_{em}=1/137 $ is the fine structure constant and $s=\sigma/\sigma_c$, with the critical surface
density $\sigma_c= m_e^2/4\pi e$ being obtained when $E=E_{th}$. An upper bound on $\Delta \tau $ can be obtained
for $s\approx 1$, i.e.  $\sigma \approx \sigma_c$, to obtain $\Delta \tau \le 1.76\times 10^{-19}$ s. This result
confirms that the black hole tends to discharge in a very short time, reaching a Schwarzschild phase.

\begin{figure}[ht!]
\begin{center}
\includegraphics[width=7.5cm,angle=0]{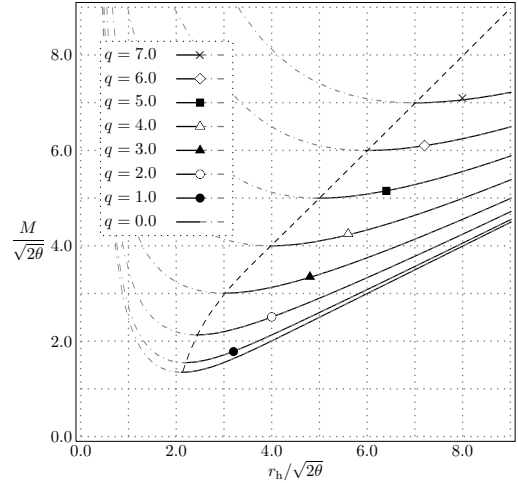}
\caption{\label{diado}\textit{The noncommutative Reissner-Nordstr\"{o}m solution} The total mass energy
$M/\sqrt{2\theta}$ is plotted as a function of $r_+/\sqrt{2\theta}$ for different $q\equiv Q/\sqrt{2\theta}$. At
the very beginning, the system is described by $Q$ and $r_+$, then it starts evolving towards its ground
  state, i.e. it tends to minimize its total mass-energy both by sliding down the $q$-curve and by jumping
  from higher to lower $q$-curves.  The
Schwarzschild black-hole is given by the lowest curve, $q=0$, which is no longer a straight line indicating that
there can be two horizons for $M> 1.90\sqrt\theta$.  The dashed curve intersects the minima of the $q$-curves
representing extremal black holes and deviating from the conventional linear behavior $M=r_+ /2$ during the SCRAM
phase because of the presence of the residual nonzero mass of the remnant. }
\end{center}
\end{figure}

\subsubsection{Charged black hole entropy}

As final note of this section, we analyze the entropy of this charged black hole, which has a further term with
respect to the neutral solution. Starting from  Ref.~\refcite{Novikov:1989sz}
\begin{equation}
dM = T_H\, dS_H + \frac{\partial M}{\partial Q}\, dQ \label{M1}
\end{equation}
where the mass is a function of both $r_+$ and $Q$, namely $M=M(r_+, Q)$ through the horizon equation. As a
consequence, we have
\begin{equation}
dM= \frac{\partial M}{\partial r_+ }\, d r_+ + \frac{\partial M}{\partial Q}\, dQ \label{M2}
\end{equation}
Comparing (\ref{M1}) and (\ref{M2}) one finds the following useful relation for the entropy
\begin{equation}
dS_H= \frac{1}{T_H}\, \frac{\partial M}{\partial r_+}.
 d r_+
\label{rndsh}
\end{equation}
Therefore we have to calculate the mass derivative with respect to $r_+$
and determine the expression for $dS_H$, that once integrated from the extremal horizon $r_0$ up to a generic
external horizon $r_+$ gives
\begin{eqnarray}
\Delta S_H=&& \frac{\pi^{3/2}}{2} \,\left[\, \frac{r_+^{2}}{\gamma\left(\, 3/2\ , r_+^2/4\theta\,\right)}
-\frac{r_e^{2}}{\gamma\left(\, 3/2\ , r_e^2/4\theta\,\right)}\,\right]
+\nonumber\\
+&& \frac{\pi^{3/2}}{2}  \int_{r_e}^{r_+} du\,u^{2} \frac{\gamma^\prime\left(\, 3/2\ ,u^2/4\theta\,
\right)}{\gamma\left(\, 3/2\ ,u^2/4\theta\,\right)^2}. \label{sbh}
\end{eqnarray}
Even if both $r_0$ and $r_+$ do depend on the charge $Q$, the above expression can be rewritten as
\begin{equation}
\Delta S_H=\frac{1}{4\ G_\theta\left(\, r\,\right)} \,\left(\, A_+  -A_e\,\right)+\delta S_H \label{unquarto2}
\end{equation}
which reproduces the celebrated relation $S_H= A_H/4$ in the same way as we have already seen for the neutral
solution in (\ref{Entropy}), after the introduction of the effective fundamental scale $G_\theta\left(\,
r\,\right)$. Again it can be shown that the correcting term $\delta S_H$ in (\ref{unquarto2}) is always
exponentially small. Thus we can conclude that the area law is maintained up to exponentially small
corrections\cite{Spallucci:2008ez}.

\section{The Extradimensional Scenario}

All of the noncommutative black hole solutions presented until now, both for the neutral and charged cases, have
been proposed in the literature as possible candidates to describe the final phase of the evaporation and the
onset of quantum gravity. Even if the regularity of a solution at the origin can be assumed as a mandatory
requirement on the theoretical side, we do not yet have a way to discriminate the correctness of a model on an
experimental basis. To this purpose, a possible way out comes from TeV-scale Quantum Gravity and the conjecture of
production of black holes at the CERN Large Hadron Collider within a few months. Therefore, it is imperative to
analyze the extensions of the proposed models to the case of large spatial extradimensions, assuming a unique mass
scale $M_{\theta}\approx M_\ast\sim 1$ TeV, in order to study the new physics coming from noncommutativity. More
in detail, even if the noncommutative mass scale $M_{\theta}$ is directly correlated to $\sqrt{\theta}$, it is
sufficient to assume only that $\sqrt{\theta}\approx 1/M_{\theta}$, without specifying the exact relationship.
Indeed, depending on the models\cite{Hewett:2000zp}\cdash\cite{Kamoshita:2002wq}, we could find that
$M_{\theta}\sim 1-10$ TeV. On the other hand, independently of noncommutativity, there is also a huge amount of
work about the possibility of observing mini black holes by detecting the products of the Hawking/Schwinger
mechanism or even revealing the presence of any black hole remnant by the study of visible and missing momenta of
the hadronic fragments after the collision\cite{Giudice:1998ck}\cdash\cite{Roy:2008we}.  All such kinds of
analysis are performed in terms of semiclassical arguments and  the conventional Schwarzschild geometry. For this
reason, on similar grounds, the fate of any evaporating black hole can only be described by means of speculative
scenarios, since no reliable prediction can be made if the black hole mass $M\sim M_\ast$.

\subsubsection{The noncommutative higher dimensional Schwarzschild solution}

In such a scheme it is natural to proceed along the lines proposed in Ref.~\refcite{Rizzo:2006zb}, in order to
combine, for the first time, noncommutative phenomenology and the properties of TeV-scale mini Schwarzschild black
holes in order to overcome the basic limitations of semiclassical approaches. In the spirit of the four
dimensional noncommutative inspired solutions\cite{Nicolini:2005vd,Ansoldi:2006vg}, we will determine a smearing
of matter distributions on length scales of order $\sqrt{\theta}$, while, as usual, the geometry sector is
formally left unchanged. Furthermore, to efficiently employ the condition of spherically symmetry, we need that
both the black hole horizon size and the noncommutative length scale $\sqrt{\theta}$ be smaller than $R$, the size
of each of the $n$ extradimensions. For later convenience, it is worthwhile to introduce $d=3+n$, the total number
of spatial dimension, i.e. $D=d+1$. Therefore the noncommutative equivalent of the higher dimensional Einstein
equation comes from the $D$ dimensional Einstein Hilbert action
\begin{equation}
S=\frac{M_\ast ^{d-1}}{2}\int \sqrt{-g}\ {\cal R}
\end{equation}
with ${\cal R}$ being the Ricci scalar in terms of noncommutative quasi coordinates, while the source term for a
massive object is determined by $\sqrt{\theta}$ in terms of a static spherically symmetric Gaussian matter
distribution
\begin{equation}
\rho\left(\,r\,\right)= \frac{M}{\left(\,4\pi\theta\,\right)^{d/2}}\, \exp\left(-r^2/4\theta\,\right).\label{uno}
\end{equation}
For the above conditions, the metric is assumed to be of the spherically symmetric form
\begin{equation}
ds^2_{(d+1)}= -f(r)\, dt^2 + f^{-1}(r)\, dr^2 + r^{2} d\Omega^2_{d-1} \label{extrads}
\end{equation}
with a further demand that $f(r)\to 1$ as $r\to \infty$, while $\Omega^2_{d-1}$ can be simply described in terms
of $d-1$ angles, $\phi_i$ where $i=1,...,n+2$.  \ In analogy to what we have seen in the four dimensional case,
two component of the energy momentum tensor $T^{MN}$ are already determined, $T_0^0=T_r^r=-\rho$, while the
remaining $d-1$ components, which are identical from the condition of spherical symmetry, can be obtained by
requiring the covariant conservation of $T^{MN}$, namely $\left. T^{MN}\right. _{;N}=0$. As a result, we find
\begin{equation}
T_i^i=-\rho-\frac{r}{d-1} \;\;\partial_r \rho
\end{equation}
for all $i=1,...,d-1$ (without summation), reproducing the the four dimensional energy momentum tensor in the
limit when $d\to 3$, i.e. $n\to 0$.  Therefore $T^{MN}$ describes a $d$-dimensional anisotropic fluid and provides
the source term of the resulting Einstein equation in the presence of noncommutative smearing effects
\begin{equation}
R_M^N=\frac{1}{M_\ast^{d-1}}\left(8\pi\ T_M^N-\frac{8\pi}{d-1}\ \delta_M^N\ T\right)
\end{equation}
where $T$ is the trace of the energy momentum tensor, $T=T_M^M$.  Plugging the line element (\ref{extrads}) into
the above equation, one finds a solution in terms of $f(r)$
\begin{equation}
f(r)= 1 -\frac{1}{M_\ast^{d-1}}\frac{ 2M }{r^{d-2} \Gamma(d/2)}  \,\, \gamma\left(\, \frac{d}{2}\ ,
\frac{r^2}{4\theta}\,\right)
\end{equation}
where
\begin{equation}
 \gamma\left(\, d/2\ , r^2/4\theta\,\right)\equiv
  \int_0^{r^2/4\theta} \frac{dt}{t}\, \, t^{d/2} \, e^{-t}
\end{equation}
and with
\begin{eqnarray}
&\Gamma\left(d/2\right)= \left(\frac{d}{2}-1\right)!\hspace{10mm}  d\,\, even&\\
&\Gamma\left(d/2\right)= \sqrt{\pi}\ \frac{(d-2)!!}{2^{(d-1)/2}}\hspace{10mm}  d\,\, odd.&
\end{eqnarray}

\subsubsection{Curvature, horizons and remnants}

The above result reproduces the four dimensional line element when  $d\to 3$ i.e. $n\to 0$, since $G=M_\ast^{-2}$,
with $M_\ast = M_P$. Also the commutative limit works, because the usual $D$-dimensional Schwarzschild solution is
reproduced when $\sqrt{\theta}/r\to 0$\footnote{The commutative limit of the above line element differs from the
usual form of the higher dimensional Schwarzschild solution in most of the literature\cite{Myers:1986un},
\begin{equation}
f(r)=1-\frac{1}{M_\ast^{d-1}}\frac{4\Gamma(d/2)}{(d-1)\ \pi^{(d-2)/2}}\frac{2M}{r^{d-2}}
\end{equation}
because in the present case the extradimensions were not compactified and therefore the constants were not matched
to the four dimensional ones. Anyway, the additional factors, being of order one, can be reabsorbed in the
definition of the fundamental scale $M_\ast$. }. On the other hand the behavior of the manifold near to the origin
is given by
 \begin{equation}
f(r)= 1 - \frac{1}{d\, \, M_\ast^{d-1}}\frac{4M }{ 2^{d-1}\pi^{(d-2)/2 }\ \theta^{d/2}}\, \, r^2 +0\left(\,
r^4\,\right) \label{desit}
\end{equation}
being of deSitter type, as expected.  Again, the $D$-dimensional curvature singularity has been dragged out by the
noncommutative fluctuations.

 \begin{figure}[ht!]
 \begin{center}
\includegraphics[width=5.5cm,angle=270]{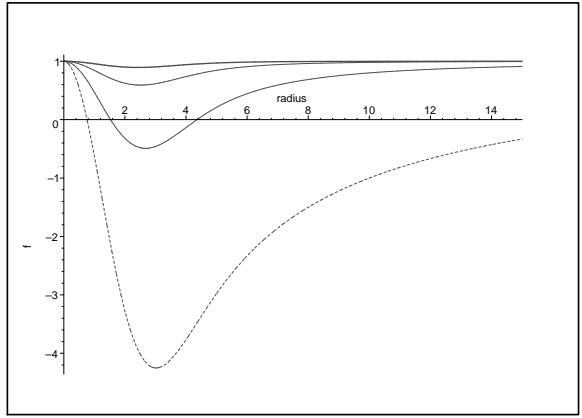}
\caption{\label{mg00} \ {\it The noncommutative extradimensional Schwarzschild solution.} The function $f$ is
poltted versus $r/\sqrt{\theta}$, for $M=10\, M_\ast $. We can observe that the curves rise with $d$ and the outer
horizon $r_+$ decreases until $d=4$. For $d\ge 5$ no black hole can be formed: the mass $M$ is so light that
cannot provide a significant gravitational disturbance, since $f\simeq 1$. }
\end{center}
\end{figure}

 \begin{figure}[ht!]
 \begin{center}
\includegraphics[width=5.5cm,angle=270]{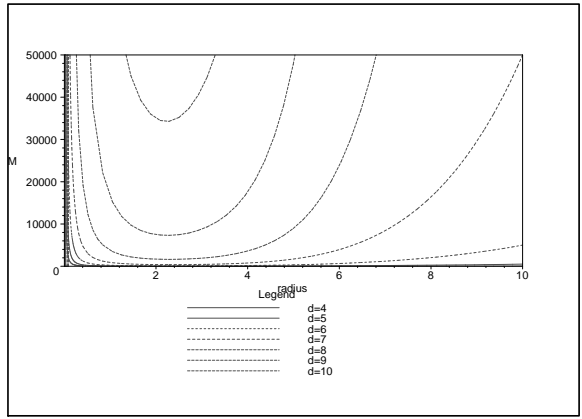}
\caption{\label{extramass} \ {\it The noncommutative extradimensional Schwarzschild solution.} The mass $M$ as a
function of $r_+$ in $\sqrt{\theta}$ units  for different values of $d$. The resulting minima of these curves
provide the remnant masses, which increase with $d$, while their radii decrease.}
\end{center}
\end{figure}

  \begin{table}[pht!]
\tbl{ Remnant masses and radii for different values of $d$, keeping $M_\ast\sim 1/\sqrt{\theta}$}
 {\begin{tabular}{@{}ccccccccc@{}} \toprule d
 & 3 &  4 &  5 &  6 &  7 &  8 &  9 &  10 \\
 \colrule $M_0$ (TeV)
 & 2.3 $\times 10^{16} $  & $6.7$  & $24$  & $94$  & $3.8\times 10^2$
 & $1.6\times 10^3$  & $7.3\times 10^3$  & $3.4\times 10^4$ \\
\colrule $r_0$ ($10^{-4}$ fm)  & $4.88 \times 10^{-16} $   & $5.29$  & $4.95$  & $4.75$  & $4.62$  &
$4.52$  & $4.46$  & $4.40$ \\
 \botrule
\end{tabular} \label{ta2}}
\end{table}

From the horizon equation $f(r)=0$, one again obtains three possibilities depending on the value of the mass $M$,
namely two horizons, one single degenerate horizon and no horizon. The plot in Fig. \ref{mg00} shows the behavior
of the outer horizon radius $r_+$, which decreases as $d$ increases for a given mass $M$. The minimum of each
curve also provides the value of the extremal radius $r_0$ which again decreases with $d$. Conversely from
Fig.~\ref{extramass}, we obtain that the remnant masses increase with $d$, even if their radii decrease.  The
values of mass and radius for the remnant are summarized in Table~\ref{ta2}. Since remnant radii are $\sim
4-5\times10^{-4}$ fm, i.e. smaller than the size of extradimensions, these remnants can be considered, to a good
approximation, as totally submerged in a $D$ dimensional isotropic spacetime, for any $d=3-10$. As further
consequence, estimating the black hole production cross section as the area of the event horizon $\sigma\sim\pi
r_+^2$, we find, for every $d$, an encouraging lower bound $\sigma\gtrsim 10$ nb, almost two orders of magnitude
larger than the conventional values. On the other hand, for $d\ge 6$ the remnant is too heavy to be produced at
the LHC and could be only detected in Ultra-High-Energy cosmic rays\cite{Cavaglia:2002si,Nagano:2000ve}.

\begin{figure}[ht!]
 \begin{center}
\includegraphics[width=5.5cm,angle=270]{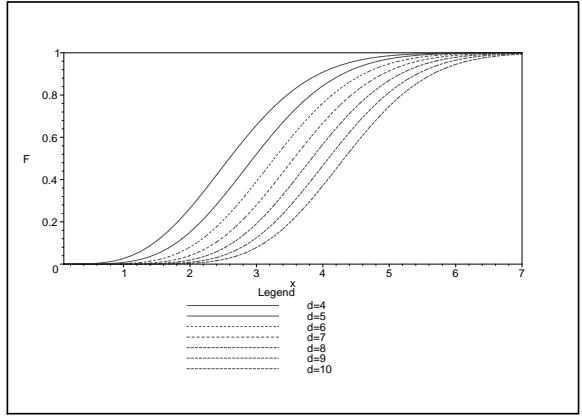}
\caption{\label{horizonradius} \ {\it The noncommutative extradimensional Schwarzschild solution.} The function
$G_d$ vs $z\equiv x/y$ for different value of $d$. The noncommutative effects are relevant only for $z \lesssim
7$, namely when the horizon size is within the noncommutative scale. The function $G_d$ is the dimensionless
effective Newton constant and realizes the ``asymptotically safe'' gravity conjecture at short distances for any
$d$. Furthermore,  we find that $G_d$ is equivalent to $x_\theta/ x$, representing the ratio between the horizon
value $x$, solution of the Eq.~(\ref{rizzouno}) and the conventional result $ \left(2m\right)^{1/(d-2)}$.}
\end{center}
\end{figure}

A more quantitative analysis can be made by the introduction of dimensionless quantities $m\equiv M/M_\ast$,
$x\equiv r_+ M_\ast$ and $y\equiv \sqrt{\theta} M_\ast$ in order to write the horizon equation in terms of
\begin{equation}
x^{d-2}=\ 2m \ G_d (z) \label{rizzozero}
\end{equation}
where
\begin{equation}
G_d(z)=\frac{1}{\Gamma(d/2)}\ \gamma \left(d/2, z^2/4 \right) \label{rizzouno}
\end{equation}
is the dimensionless effective Newton constant, with $z\equiv x/y$.  Fig.~\ref{horizonradius} shows how gravity
becomes weaker and weaker at short distances because of the presence of noncommutative fluctuations of spacetime.
Indeed the noncommutative smearing realizes the ``asymptotically safe'' gravity conjecture, eliminating the
curvature singularity without invoking any RG arguments\cite{weinberg,Bonanno:2000ep}. When $d$ is even, the
incomplete gamma function can be written in explicit form, namely $G_4 = 1 -e^{-p}\left(1+p\right)$, $G_6 = 1
-e^{-p}\left(1+p+p^2/2\right)$, $G_8 = 1 -e^{-p}\left(1+p+p^2/2+p^3/6\right)$,..., $G_d = 1
-e^{-p}\left(1+p+p^2/2+...+ p^{d-5}/(d-5)\right)$, where $p\equiv z^2/4$.  From Eq. (\ref{rizzouno}), one can
obtain that the behavior of $m$ at infinity and near to the origin is $\sim x^{d-2}$ and $\sim x^{-2}$
respectively, implying the existence of a minimum value of $m$ for some intermediate value of $x$, in agreement
with Fig. \ref{extramass}. Furthermore, in the commutative limit, i.e. $y,\theta\to 0$, one can reproduce the
usual result
\begin{equation}
x = (2m)^{1/(d-2)}
\end{equation}
a relationship that no longer occurs in the noncommutative case.  Again from Fig.~\ref{horizonradius}, we find
that, for all $d$, the noncommutative value of $x$ deviates from the usual value $ \left(2m\right)^{1/(d-2)}$ in a
region around the origin, because only there is the noncommutative scale comparable with the horizon size. As a
consequence, the existence of a minimum for the mass $m$ implies also a minimum for the horizon radius $x$ and a
physical mass threshold below which black holes do not form.

\begin{figure}[ht!]
 \begin{center}
\includegraphics[width=5.5cm,angle=270]{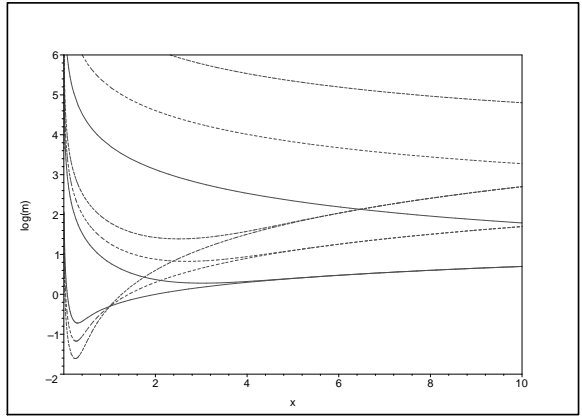}
\caption{\label{ExtraNeutralMassY} \ {\it The noncommutative extradimensional Schwarzschild solution.} The black
hole mass, $\log_{10}(m)$, as a function of $x$ for different values of $d=3$ (solid), $d=4$ (dotted) and $d=5$
(dashed). The upper (middle, lower) set of curves correspond to $y=10 (1, 0.1)$.}
\end{center}
\end{figure}

The analysis in Table~\ref{ta2} shows that black holes are too massive to be produced at LHC for large $d$,
assuming that $\sqrt{\theta}M_\ast=y=1$. More generally we can also calculate $m$ as a function of $x$, varying
$y$, the parameter governing the noncommutative scale. From Fig.~\ref{ExtraNeutralMassY}, one finds that as $y$
increases for fixed $d$ so does $m$ except where $x$ is large and we are probing the commutative regime.
Therefore, we need small values of $y$, i.e. $0.05\lesssim y \lesssim 0.2$, to obtain a range of masses accessible
to the LHC, namely $1\lesssim m \lesssim 10$, assuming $M_\ast \sim 1$ TeV.  On the other hand, having low values
for $y$ leads to the commutative regime, in which both the minimum value of the mass, $m_{min}$ and the horizon
radius where the minimum mass occurs, $x_{min}$ are algebraically zero. For this reason, it is worthwhile to
assume that $m_{min}\simeq Max\ (1, m_{min}^{a})$, where $m_{min}^a$ is the algebraic minimum value calculated
from the equation $\partial m/\partial x =0$, i.e.
\begin{equation}
G_d(p)-\frac{2\ p^{d/2} \ e^{-p}}{(d-2)\Gamma (d/2)}=0. \label{rizzomin}
\end{equation}
In the same way, a small value of $y$ implies a lowering of the value of $x$, with consequent reduction of black
hole production cross section, whose minimum $\sigma_{min}\simeq \pi x^2_{min}/M_\ast^2$ is quite small even if
not vanishing. For $y=0.05$, one finds $\sigma_{min}\approx 20$ pb, one order of magnitude below the value
expected, on qualitative arguments, for a Schwarzschild black hole in TeV gravity.  For energies far above the
black hole production threshold, the cross section significantly increases according to $\sigma\sim m^{2/(d-2)}$,
as in the commutative theory.

\begin{figure}[ht!]
 \begin{center}
\includegraphics[width=5.5cm,angle=270]{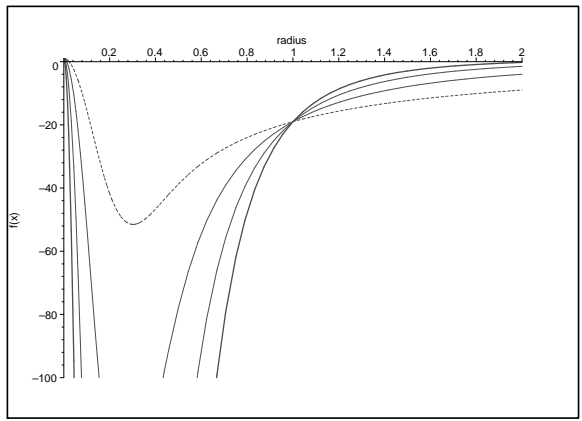}
\caption{\label{Y01m10} \ {\it The noncommutative extradimensional Schwarzschild solution.} The function $f(x)$
versus $x=rM_\ast$, for $m=10$ and $y=0.1$, namely both the black hole mass and the noncommutative scale are
heavier that the fundamental scale. Therefore nonextremal black holes form in a regime which resembles the
commutative case. The dotted curve corresponds to $d=3$, while from top to bottom on the left hand side of the
figure the solid curves are for $d=4$ to $6$. We can observe that the outer horizon radius decreases with $d$,
while the curvature near to the origin, even if finite, increases with the smallness of $y$ (contraction of the
deSitter core). The position of the minima of the above curves provides the radius of the extremal black hole
$x_0$; for $d=3$ one finds $x_3\approx 3.0 y$ as expected. }
\end{center}
\end{figure}

\begin{figure}[ht!]
 \begin{center}
\includegraphics[width=5.5cm,angle=270]{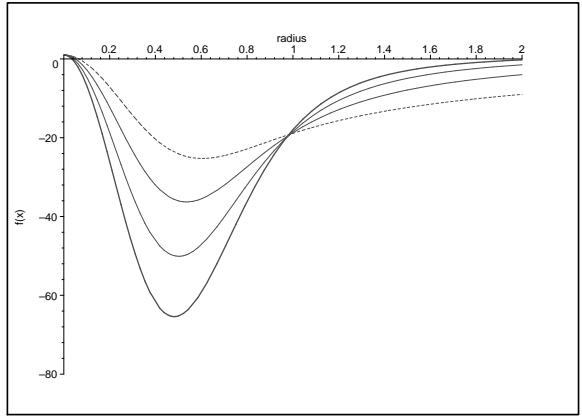}
\caption{\label{Y02m10} \ {\it The noncommutative extradimensional Schwarzschild solution.} The function $f(x)$
versus $x=rM_\ast$, for $m=10$ and $y=0.2$. Each curve rises with respect to the case $y=0.1$, since the curvature
at the origin decreases. Nonextremal black holes form with outer horizons that still occur at commutative values
$(2m)^{1/(d-2)}$. The dotted curve corresponds to $d=3$, while from top to bottom on the left hand side of the
figure the solid curves are for $d=4$ to $6$. We can observe that in agrement with Fig.~\ref{extramass}, the
extremal radius $x_0$ shrinks as $d$ increases. }
\end{center}
\end{figure}

Also the analysis of the horizon equation is somehow incomplete: the Fig. \ref{mg00} provides the profile of
$f(r)$ for $y=1$ only. Even if qualitatively we do expect a scenario with two (one or no) horizons analogous to
the case for $y$ fixed, it is worthwhile to consider the general case $f=f(x,y,m,d)$ to get quantitatively
reliable results. Indeed the parameter $y$, setting the noncommutative scale, has a crucial role: it determines
whether or not it is possible to experience the noncommutative fluctuations of the manifold at the scale $M_\ast$.
Phenomenologically we expect that $y\sim 0.1-10$, corresponding to the case of a manifold almost in a commutative
regime, since the noncommutative scale too large ($y\sim 0.1$); to the case for which the spacetime is dominated
by the relevant noncommutative fluctuations, weakly resembling to the concept of classical manifold ($y\sim 10$).
Furthermore the parameter $y$ controls the extension of the portion of the manifold subject to the noncommutative
regime: roughly speaking we can estimate this region from (\ref{desit}), calculating the extension, $L$, of the
deSitter core around the origin even if noncommutativity can still be nonnegligible even beyond it
\begin{equation}
L\sim y^{d/2}\ m^{-1/2}\ \frac{1}{M_\ast}.
\end{equation}
Thus, we find  a contraction of this core when $y<1$ for an increase of $d$ and viceversa an expansion when $y>1$.
A fact related to the extension of the deSitter core is the value of the curvature at the origin: large curvatures
implies smaller cores and viceversa. We can see this, looking at the Ricci scalar
\begin{equation}
{\cal R}\sim\frac{m}{y^d}\exp\left(-x^2/4y^2\right)
\end{equation}
which, at a given mass $m$, essentially depends only on $y$ near the origin, namely ${\cal R}\sim y^{-d}$.  The
parameter $y$ also  controls the extremal black hole.  To this purpose, we find that the extremal radius $x_{min}$
linearly depends on $y$
\begin{equation}
x_{min}(d)=  z_{0}(d) \  y
\end{equation}
where $z_0(d)$ is the root of the equation (\ref{rizzomin}), corresponding to the extremal radius for $y=1$.  On
the other hand, from (\ref{rizzozero}) one finds that the extremal mass is
\begin{equation}
m_{min}=\frac{1}{2}\  G^{-1}_d(z_0) \ \ z_0^{d-2}\ y^{d-2}.
\end{equation}
Therefore the product $z_0 y$ determines the kind of growth of $m_{min}$ with $d$.

\begin{figure}[ht!]
 \begin{center}
\includegraphics[width=5.5cm,angle=270]{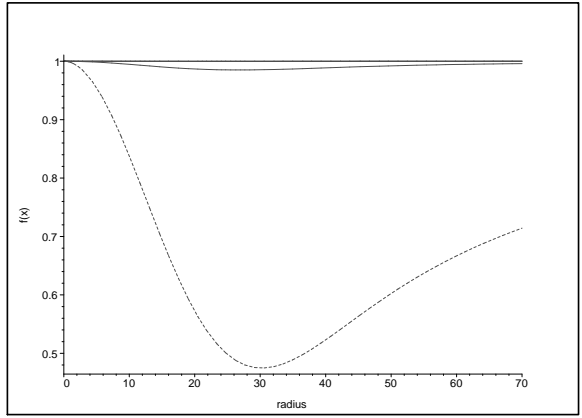}
\caption{\label{Y10m10} \ {\it The noncommutative extradimensional Schwarzschild solution.} The function $f(x)$
versus $x=rM_\ast$, for $m=10$ and $y=10$. A larger $y$ implies an extension of the region where noncommutative
fluctuations can be probed at the scale $M_\ast$, with a consequent wider smearing of the mass $m$ and a lowering
of the manifold curvature.  The smearing increases with $d$, because the mass $M$ is distributed within a larger
spacetime, leading to $f\simeq 1$ for $d\ge 4$: in other words since $y>1$ we assist an expansion of the deSitter
core, which confines all of the mass $M$ within itself for $d\ge 4$. Therefore the noncommutative effects are so
important that no black hole can form even for $d=3$.}
\end{center}
\end{figure}

\begin{figure}[ht!]
 \begin{center}
\includegraphics[width=5.5cm,angle=270]{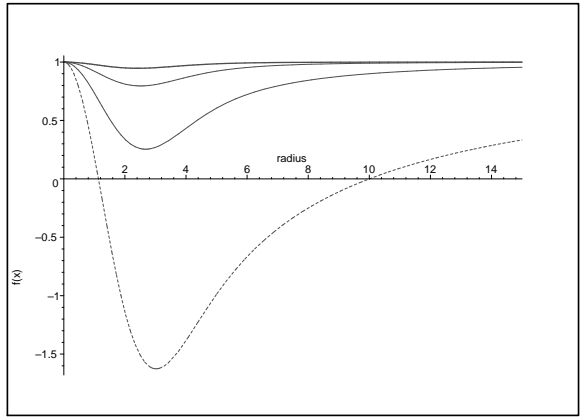}
\caption{\label{Y1m5} \ {\it The noncommutative extradimensional Schwarzschild solution.} The function $f(x)$
versus $x=rM_\ast$, for $m=5$ and $y=1$. This borderline case occurs for $y=1$, where the smearing of the mass $m$
is relevant only in the presence of extradimensions. Indeed black holes do not develop for $d\ge 4$.}
\end{center}
\end{figure}

\begin{figure}[ht!]
 \begin{center}
\includegraphics[width=5.5cm,angle=270]{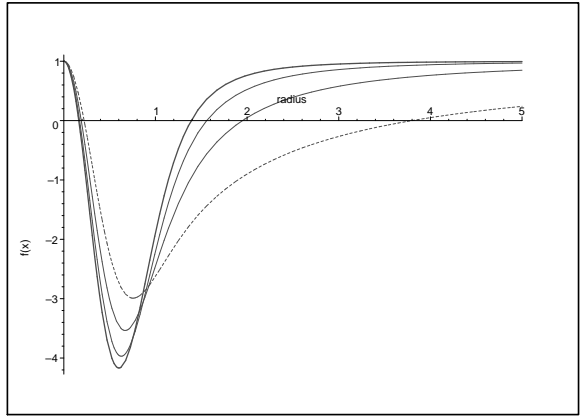}
\caption{\label{Y025m1.9} \ {\it The noncommutative extradimensional Schwarzschild solution.} The function $f(x)$
versus $x=rM_\ast$, for $m=1.9$ and $y=0.25$. The mass parameter has the value equivalent to the extremal case for
$d=3$, but the reduced value of $y$ implies a mass energy rather concentrated near the origin with a consequent
formation of two horizons. The dotted curve corresponds to $d=3$, while from top to bottom on the left hand side
of the figure the solid curve are for $d=4$ to $6$.}
\end{center}
\end{figure}

\begin{figure}[ht!]
 \begin{center}
\includegraphics[width=5.5cm,angle=270]{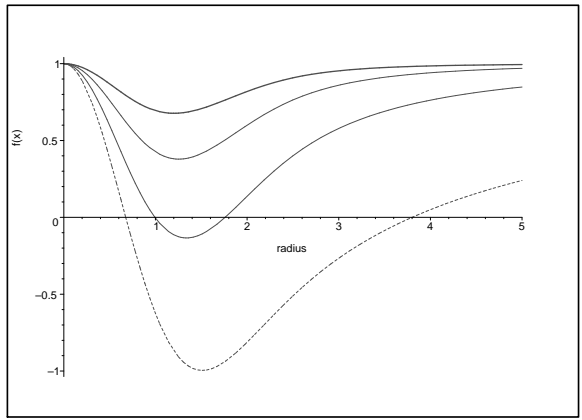}
\caption{\label{Y05m1.9} \ {\it The noncommutative extradimensional Schwarzschild solution.} The function $f(x)$
versus $x=rM_\ast$, for $m=1.9$ and $y=0.5$.  With respect the case for $y=0.25$, the noncommutative smearing
effect prevails at least for $d\ge5$, preventing the formation of black holes. In other words, the contraction of
the deSitter core is compensated by a quicker expansion of spacetime due to the additional dimensions, which let
the mass be more diffused. The dotted curve corresponds to $d=3$, while, contrary to the previous cases, from top
to bottom on the left hand side of the figure the solid curve are for $d=6$ to $4$.}
\end{center}
\end{figure}

\begin{figure}[ht!]
 \begin{center}
\includegraphics[width=5.5cm,angle=270]{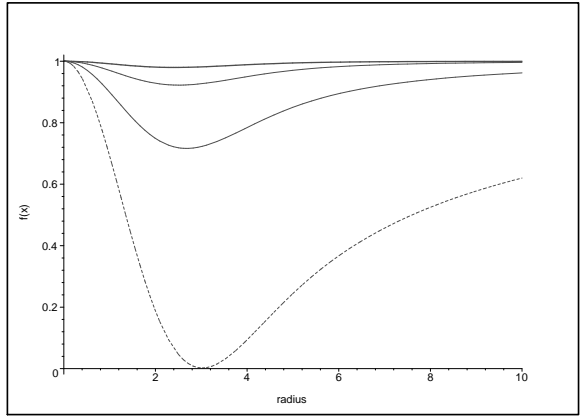}
\caption{\label{Y1m1.9} \ {\it The noncommutative extradimensional Schwarzschild solution.} The function $f(x)$
versus $x=rM_\ast$, for $m=1.9$ and $y=1$. This is a known case at least for $d=3$, for which the extremal back
hole takes place. Again the additional dimensions of spacetime determine a further smearing of the mass with
absence of black holes for $d\ge 4$ and even of curvature for the higher dimensions. The dotted curve corresponds
to $d=3$, while from top to bottom on the left hand side of the figure the solid curve are for $d=6$ to $4$. }
\end{center}
\end{figure}

\begin{figure}[ht!]
 \begin{center}
\includegraphics[width=5.5cm,angle=270]{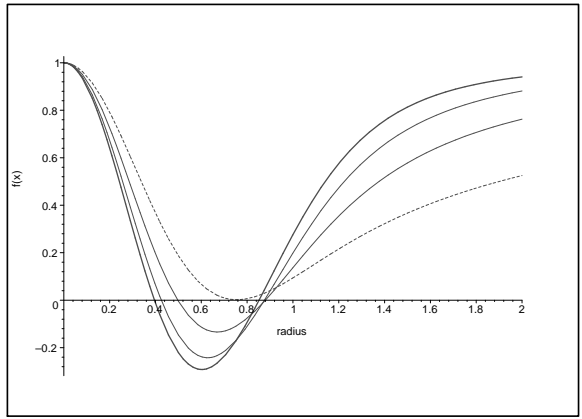}
\caption{\label{Y02m0.5} \ {\it The noncommutative extradimensional Schwarzschild solution.} The function $f(x)$
versus $x=rM_\ast$, for $m=0.475$ and $y=0.25$.  For $d=3$ we have the extremal case, while the increase of $d$
prevails on other effects, contracting the deSitter core, which shrinks with $y^{d/2}$ and reducing  and the
threshold mass to have a black hole, i.e. $m_0\sim y^{d-2}$. A smaller region influenced by noncommutative
fluctuations is accompanied by a curvature increase with $y^{-d}$. As a consequence, black holes develop for
$d>3$. The dotted curve corresponds to $d=3$, while from top to bottom on the left hand side of the figure the
solid curve are for $d=4$ to $6$. We can observe that, this time on the contrary with respect to previous cases,
the outer horizon decreases with $d$ from $d=4$.}
\end{center}
\end{figure}

\begin{figure}[ht!]
 \begin{center}
\includegraphics[width=5.5cm,angle=270]{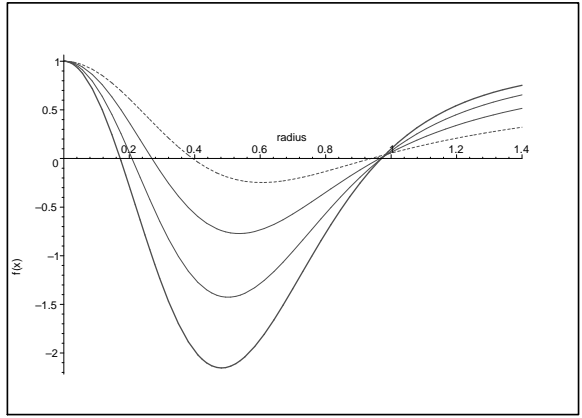}
\caption{\label{Y05m05} \ {\it The noncommutative extradimensional Schwarzschild solution.} The function $f(x)$
versus $x=rM_\ast$, for $m=0.475$ and $y=0.2$.  With respect to the case $y=0.25$, a  reduction of $y$ implies a
further contraction of the deSitter core and of the region influenced by noncommutativity, while the curvature
increases and the threshold mass decreases. Therefore black holes occur for every $d$, because the spacetime
expansion with $d$ cannot compensate the deSitter core reduction: as a consequence, the mass is more diffused as
$d$ increases and the the outer horizon increases with $d$. The dotted curve corresponds to $d=3$, while from top
to bottom on the left hand side of the figure the solid curve are for $d=4$ to $6$.}
\end{center}
\end{figure}

\begin{figure}[ht!]
 \begin{center}
\includegraphics[width=5.5cm,angle=270]{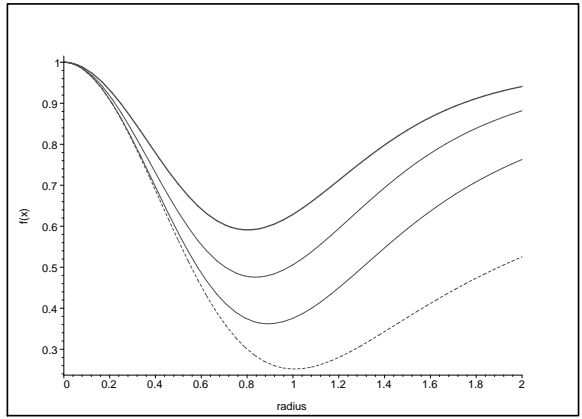}
\caption{\label{Y03m05} \ {\it The noncommutative extradimensional Schwarzschild solution.} The function $f(x)$
versus $x=rM_\ast$, for $m=0.475$ and $y=1/3$.  With respect to the case $y=0.25$, an increase of $y$ implies an
expansion of the deSitter core, an increase of threshold masses and a reduction of the curvature. Therefore, even
for the $d=3$ case black holes cannot occur. The dotted curve corresponds to $d=3$, while from top to bottom on
the left hand side of the figure the solid curve are for $d=6$ to $4$. }
\end{center}
\end{figure}

\begin{figure}[ht!]
 \begin{center}
\includegraphics[width=5.5cm,angle=270]{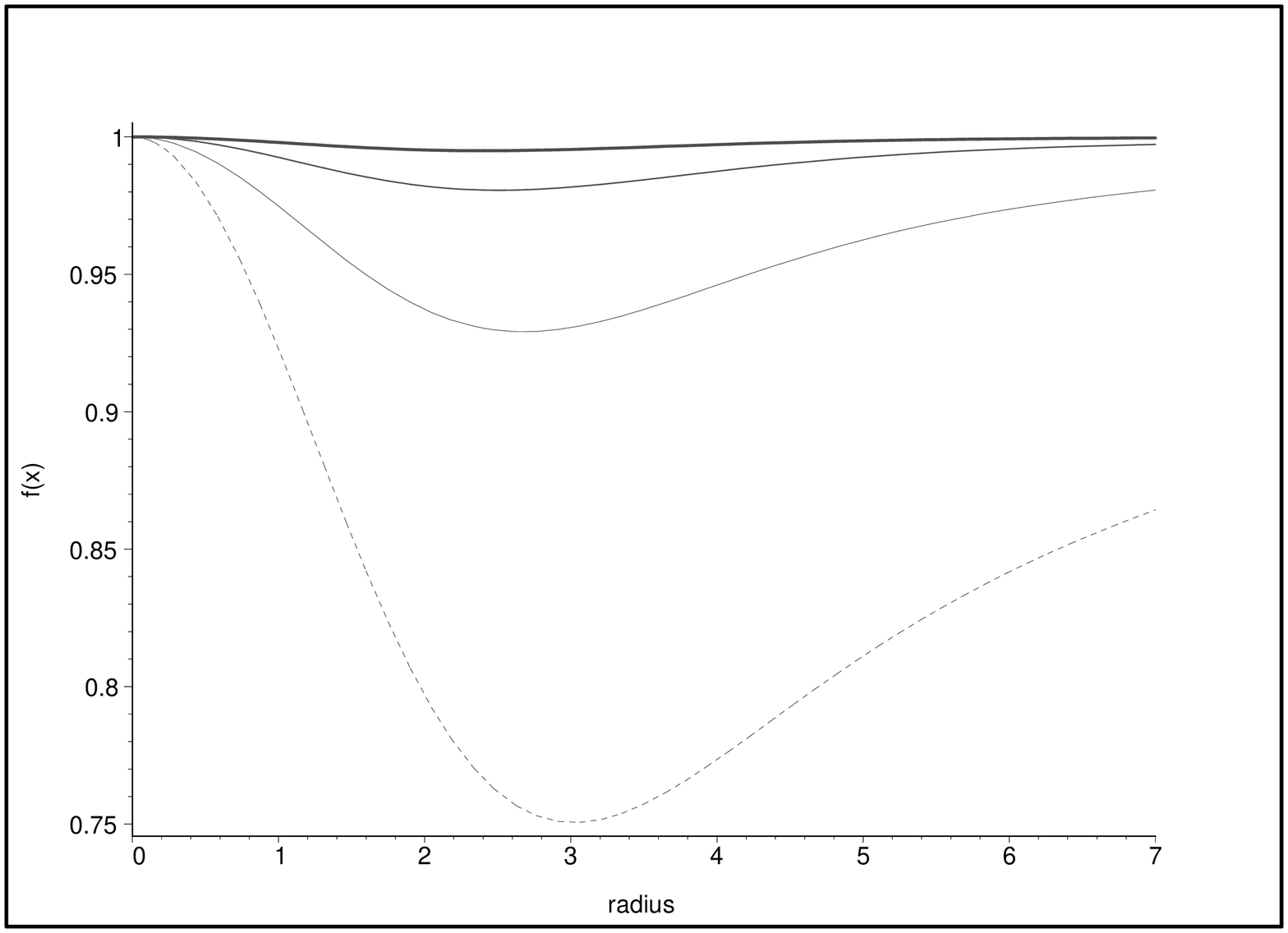}
\caption{\label{Y01m05} \ {\it The noncommutative extradimensional Schwarzschild solution.} The function $f(x)$
versus $x=rM_\ast$, for $m=0.475$ and $y=1$.  A large value of $y$ determines a manifold dominated by
noncommutativity at least in a large region. Other physical parameters turn out to be irrelevant and the $f$
behaves as expected when the diffusion of the mass increase with $d$, but no black hole occurs even for $d=3$. The
dotted curve corresponds to $d=3$, while from top to bottom on the left hand side of the figure the solid curve
are for $d=6$ to $3$.}
\end{center}
\end{figure}

With the above ingredients, we can analyze Figs.~\ref{Y01m10}--\ref{Y01m05}, which deal with function $f(x)$ for
$3<d<6$ with scale parameters $1\lesssim m \lesssim 10$ and $0.1 \lesssim y \lesssim 10$ and exhibit some common
features. For instance, the outer horizon radius generally decreases with $d$ because the mass $m$ turns out to be
further smeared along the additional dimensions. Then, a large value of $y$, i.e. $y>1$, leads to weakly curved
manifolds, because the smearing of the mass $m$ is so strong that it can disturb little the spacetime geometry. An
alternative explanation is that the deSitter core is large enough to confine almost all of the mass inside itself,
preventing the formation of a Schwarzschild region and consequent development of horizons. There are also
intermediate cases, in which the smearing is so strong to prevent the appearance of horizons for higher dimension
only, while black holes still occur for $d\lesssim 4$. An anomalous case regards
Figs.~\ref{Y02m0.5}--\ref{Y05m05}, dealing with a light mass $m\sim 0.475$ and a small $y\sim 0.25$. For $d=3$, we
have the extremal black hole, but for higher dimensions, instead of an increase of the smearing effect with
consequent absence of horizons, we obtain nonextremal black holes.  This fact can be explained by the reduction of
the minimal mass in higher dimension, being $m_{min}\sim x_{min}^{d-2}(d)$, where $x_{min}(d)<x_{min}(3)\approx
0.75$. Geometrically, we can say that a contraction of the deSitter core occurs, i.e. $L\sim y^{d/2}M_\ast^{-1}$,
with a consequent reduction of the region where noncommutativity is relevant and with an increase of curvature
near the origin, i.e. ${\cal R}\sim y^{-d}$. As a consequence black holes develop for $d>3$ only.  A slight
reduction of $y$, leads to a fatal reduction of the deSitter core also for $d=3$ with consequent formation of two
horizons (see Fig.~\ref{Y05m05}). A special feature of this near extremal case, is that the outer horizon
increases with $d$, the contrary of what one generally expects. On the other hand, a small increase of $y$ leads
to an opposite scenario in which no horizon occurs for any $d$ (See Fig.~\ref{Y03m05}). Finally in
Fig.~\ref{Y01m05}, we find a manifold dominated by noncommutative effects, while the mass $m$ is too light to
provide significant gravitational disturbances, in particular for the higher dimensional cases.

\begin{figure}[ht!]
 \begin{center}
\includegraphics[width=5.5cm,angle=270]{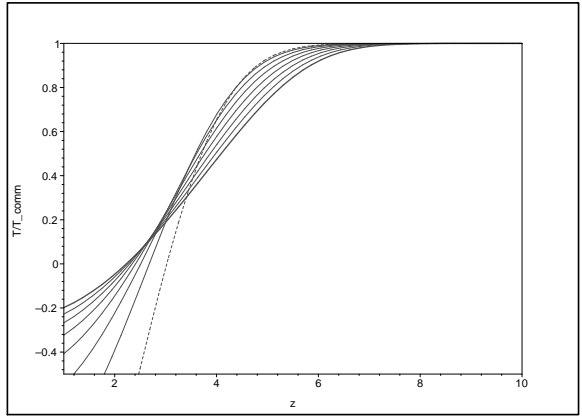}
\caption{\label{ExtraTempRatio} \ {\it The noncommutative extradimensional Schwarzschild solution.} The ratio
$T/T_{comm}$ as a function of $z$. The dotted curve corresponds to the $d=3$ case whereas, from bottom to top on
the left hand side of the figure, the solid lines corresponds to $d=4$ to $10$.}
\end{center}
\end{figure}

\subsubsection{Higher dimensional black hole thermodynamics}

The next step is about the thermodynamical analysis of the noncommutative extradimensional solution. As usual we
can calculate the black hole temperature
\begin{equation}
 T_H = \frac{d-2}{4 \pi r_+}\,\left[\, 1 -\frac{ r_+}{d-2}
 \frac{\gamma^\prime\left(\, d/2\ ,r_ +^2 /4\theta\,\right)}
 {\gamma\left(\, d/2\ , r_ +  ^2 /4\theta \,\right)}\,\right]
 \label{extratemp}
\end{equation}
Defining the dimensionless temperature $ T\equiv T_H/M_\ast$ we have
\begin{equation}
T=\frac{d-2}{4\pi x}\left[1-\frac{2}{d-2}\frac{p^{d/2}e^{-p}}{G_d(p)\Gamma(d/2)}\right] \label{dimenlesstemp}
\end{equation}
where again $p=z^2/4$ with $z=x/y$. Then it is instructive to compare this result with the commutative value
$T_{comm}=(d-2)/4\pi x$. First of all, both from (\ref{dimenlesstemp}) and from Fig.~\ref{ExtraTempRatio} we can
conclude that for large $z$ the black hole temperature coincides with the commutative result.  Noncommutative
effects appear as $z$ approaches $\sim 6$, depending on $d$: the manifold is more sensitive to noncommutativity
for higher dimensions, as we can see in the right hand side of Fig.~\ref{ExtraTempRatio}, where the lowest curve
correspond to the highest dimension.  This is in agreement with our previous arguments about the smearing effect,
which increases with $d$.  The ratio $T/T_{comm}$ rapidly decreases as $z\sim 3-4$, vanishing at the same values
where black hole mass is minimized. Again we see that $z_{min}$ decreases with $d$. For smaller $z$, we enter
negative value of the ordinate corresponding to an unphysical situation.

\begin{figure}[ht!]
\begin{center}
\includegraphics[width=5.5cm,angle=270]{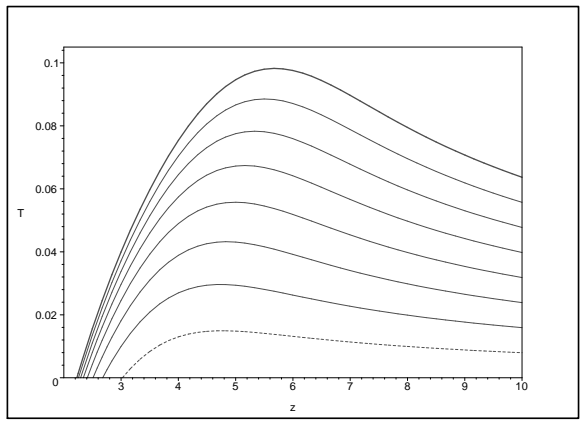}
\caption{ \label{bhtemp_max} \textit{The noncommutative extradimensional Schwarzschild solution.} The temperature
$T$ as a function of $z$, for various value of $d$. The dotted curve corresponds to the case $d=3$, while, from
the bottom to top on the right hand side of the figure, the solid lines correspond to $d=4$ to $10$.  The
temperatures increase with $d$, while the SCRAM phase leads to smaller remnant for higher $d$.}
\end{center}
\end{figure}

\begin{figure}[ht!]
\begin{center}
\includegraphics[width=5.5cm,angle=270]{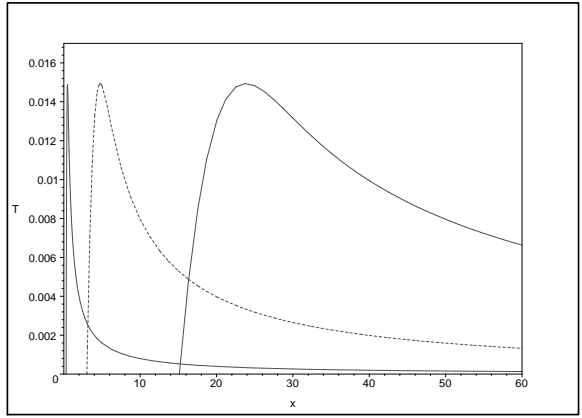}
\caption{ \label{bhtemp_maxY} \textit{The noncommutative extradimensional Schwarzschild solution.} The temperature
$T$ as a function of $x$, for $d=3$. The dotted curve corresponds to the case $y=1$, while, from the bottom to top
on the right hand side of the figure, the solid lines correspond to $y=0.1$ and $5$.  The temperatures peak is
unaffected by $y$. On the other hand smaller values of $y$ reduce the remnant radii, determining a quicker SCRAM
phase and a longer Schwarzschild phase.}
\end{center}
\end{figure}

Then, from Fig. \ref{bhtemp_max}, we conclude that the thermodynamic scenario we have already encountered with the
four dimensional solution, is confirmed in higher dimensions: after an initial Schwarzschild phase, the black hole
undergoes a SCRAM phase, ending up with a zero temperature relic. One can see also from Table \ref{ta1} that the
peak of the temperature increases with $d$, but even in the case $d=10$ it is about $98$ GeV $\left(\, \simeq
10^{15}\right.$ K$\left.\right)$ which is much lower than $M_\ast$. To this purpose, the parameter $y$ does not
play any role in enhancing the temperature peak, determining only a shift of the coordinate $x_{max}$ where the
peak takes place (see Fig. \ref{bhtemp_maxY}).  On the other hand, we still have, as reasonable lower bound for
the mass parameter, $m_{min}\sim 1$. As a result, even for the higher dimensional solutions, backreaction effects
are negligible: the dimensionless temperature is $T\ll 1\lesssim m$, as we can see from Table \ref{ta3}, where for
$d=3$ the fundamental scale is $M_\ast=M_{P}$.

\begin{table}[pht!]
 \tbl{Black hole maximal temperatures for different $d$. See Fig.(\ref{bhtemp_max})}
 {\begin{tabular}{@{}cccccccccc@{}} \toprule d
& & 3  & 4  & 5  & 6  & 7  & 8  & 9  & 10 \\
 \colrule $T_H^{max}$ (GeV)
& & $18\times 10^{16}$  & 30  & 43  & 56  & 67  & 78  & 89  & 98 \\
\colrule $T_H^{max}$ ($10^{15} K$) & & $.21\times 10^{16}$  & .35  & .50  & .65  & .78  & .91  & 1.0
& $1.1$ \\
 \botrule
\end{tabular} \label{ta1}}

\end{table}
 \begin{table}[pht!]
 \tbl{ Backreaction estimates
for different values of $d$}
 {\begin{tabular}{@{}cccccccccc@{}}  \toprule d
& & 3 &  4  & 5  & 6  & 7  & 8  & 9  & 10 \\
 \colrule $10^3 \ T/m  $
& & $< 15 $  & $< 30 $  & $< 43 $  & $< 56 $  & $ <67 $  & $<
78 $  & $< 89$  & $< 98$ \\
 \botrule
\end{tabular} \label{ta3}}
\end{table}

In analogy with what we have already seen in the four dimensional case, the stability of remnants, for any $d$, is
an obvious consequence of this extradimensional scenario. Indeed the leading term of the mass loss rate may be
written as
\begin{equation}
\frac{dm}{dt}\sim x^{d-1}T^{d+1}
\end{equation}
for bulk fields decays. The resulting black hole lifetime is then given by
\begin{equation}
Life\sim \int ^{m_{min}}_{m_{initial}}\frac{dm}{x^{d-1}(m)\ T^{d+1}(m)}
\end{equation}
with $m_{initial}$ being the starting black hole mass. Recalling that $x(m)$ never vanishes, while $T(m)\to 0$ as
$m\to m_{min}$, for all $d$ the lifetime will be driven to infinity, implying a stable relic.

\begin{figure}[ht!]
\begin{center}
\includegraphics[width=5.5cm,angle=270]{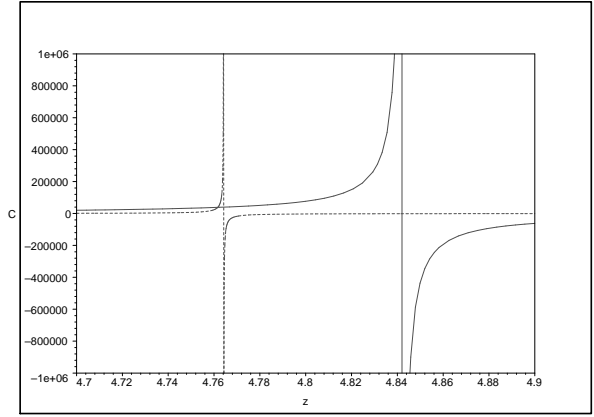}
\caption{ \label{ExtraHeat} \textit{The noncommutative extradimensional Schwarzschild solution.} The heat capacity
$C$ as a function of $z$. The dotted curve corresponds to the case $d=3$, while the solid line to $d=5$. At fixed
$y$, the noncommutative effects become relevant at larger $z$ for the higher $d$ case; indeed the temperature
peaks take place at $\sim 4.76$ and $\sim 4.84$ respectively, determining the heat capacity singularity. On the
other hand, from the slope of the curves we get that the SCRAM phase, characterized by positive $C$, is quicker
for $d=3$, leading to a larger remnant.}
\end{center}
\end{figure}

The unexpected temperature behavior in the SCRAM phase suggests the study of the higher dimensional black hole
heat capacity
\begin{equation}
C=\frac{\partial m}{\partial T}=\frac{\partial m}{\partial x}\left(\frac{\partial T}{\partial x}\right)^{-1}
\label{extraC}
\end{equation}
which turns out to be
\begin{equation}
C=-\frac{2\pi
x^{d-1}}{G_d(p)}\left[1+\frac{\frac{4pH_d(p)}{d-2}\left(1+\frac{H_d(p)}{p}-\frac{d}{2p}\right)}{1-\frac{2
H_d(p)}{d-2}-\frac{4pH_d(p)}{d-2}\left(1+\frac{H_d(p)}{p}-\frac{d}{2p}\right)}\right]
\end{equation}
where for notational purposes we have defined
\begin{equation}
H_d(p)=\frac{p^{d/2}e^{-p}}{G_d(p)\Gamma\left(d/2\right)}
\end{equation}
We can see that at large distances, being $G_d(p)\to 0$ and $H_d(p)\to 0$, the heat capacity is negative and
approaches the commutative result $\sim -2\pi x^{d-1}$, while at $x=x_{min}$ the heat capacity vanishes, because
$\partial m /\partial x =0$ in (\ref{extraC}).  Further we note that the sign of $C$ is uniquely determined by
$\partial T/\partial x$, which switches from negative value for $x>x_{max}$  during the Schwarzschild phase to
positive ones for $x_{min}<x<x_{max}$ in the SCRAM phase. Such change of sign is accompanied by the appearance of
a polar singularity of $C$, which occurs when $\partial T/\partial x$ vanishes at $x=x_{max}$, i.e. where the
temperature reaches the maximum.  The Fig.~\ref{ExtraHeat} is illuminating about the stability of black hole
relics and the behavior of $C$ for different dimensions.

\begin{figure}[ht!]
\begin{center}
\includegraphics[width=5.5cm,angle=270]{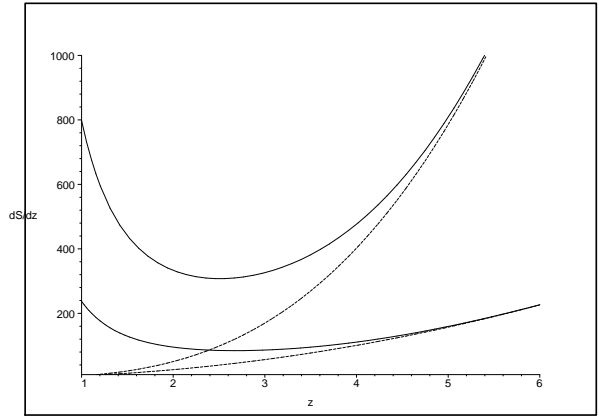}
\caption{ \label{ExtraEntropy} \textit{The noncommutative extradimensional Schwarzschild solution.} The derivative
of the black hole entropy $dS/dz$ as a function of $z$. The dotted curve correspond to the commutative case, while
the upper solid curve regards $d=5$ while the lower $d=4$. In both case, at large distances, $z\sim 5$, the
entropy is essentially coincident with its commutative equivalent. In spite of the singularity in the heat
capacity and the final zero temperature state, the entropy is always a finite function.}
\end{center}
\end{figure}

The thermal history of the black hole can be analyzed also in terms of the entropy which can be defined via
\begin{equation}
S=\int dx T^{-1}\frac{\partial m}{\partial x}=2\pi\int_{x_{min}}^xt^{d-2}G_d^{-1}(t^2/4y^2). \label{extraS}
\end{equation}
In the above definition, we have made the natural choice that $S=0$ at $x_{min}$, where the mass $m=m_{min}$. We
note that, in the commutative limit, $y\to 0$, the function $G_d\to 1$, while the lower limit of integration
$x_{min}= z_0 y$ vanishes. Thus one recovers the conventional result
\begin{equation}
S_{comm}=2\pi\ \frac{x^{d-1}}{d-1}
\end{equation}
corresponding to the large distance behavior of the entropy.  The integration of (\ref{extraS}) leads to the extra
dimensional generalization of (\ref{Entropy})
\begin{equation}
S= 2\pi \,\frac{x^{d-1} - x_{min}^{d-1} }{(d-1)\ G_d\left(z\right)} +\delta S
\end{equation}
where
\begin{equation}
\delta S\approx \frac{1}{(2y)^{m}}\left(x_{min}^{2m-3} e^{-x_{min}^2/4y^2} -x^{2m-3} e^{-x^2/4y^2}\right)
\end{equation}
is an exponentially suppressed correcting term. If we take into account that the area of the event horizon is $ A=
2 \pi^{d/2} x^{d-1}/\Gamma\left(d/2\right)$, we obtain the generalization of the four dimensional entropy-area law
to higher dimensions
\begin{equation}
S=\Gamma\left( d/2\right)\pi^{1-\frac{d}{2}}\ \frac{\left(\, A_+ -A_e\,\right)}{(d-1)\ G_d\left(\, z\,\right)}
\,+\delta S.
\end{equation}
 As further note, looking at
\begin{equation}
S=\int dx\ \frac{C}{T}\ \frac{\partial T}{\partial x}
\end{equation}
one finds that the entropy is always finite at any critical point of both $C$ and $T$ to get a reliable black hole
thermodynamics (see Fig.~\ref{ExtraEntropy}).

Finally, also the free energy can be extended to higher dimensional cases, combining the previous thermodynamic
quantities
\begin{equation}
F=m-TS
\end{equation}
which coincides with the commutative value, $F=x^{d-2}/2(d-1)=m/(d-1)$ at large distances. On the other hand, at
$x=x_{min}$, we have $F=m$ for all value of $y$. At intermediate value of $x$, the free energy decreases, reaching
a minimum where the temperature is maximized, namely at $x_{max}$ where $\partial T/\partial x =0$. After that,
the free energy matches onto the asymptotic $\sim x^{d-2}$ behavior.  Again the value $x_{max}$ marks the frontier
between an unstable Schwarzschild phase and a stable SCRAM phase, in which a loss of mass implies a decrease of
total energy and thermal emission.

As final comment, the present analysis confirms our conjectures: we can expect that black holes be produced with
large cross sections and studied in detail at the LHC. In particular, the extradimensional model confirms the four
dimensional prevision about the existence of stable remnants, whose mass and radius are fixed by the
noncommutative scale and the number of dimensions and whose detection turns out to be vital for the understanding
of Quantum Gravity.

\subsection{The noncommutative extradimensional charged solution}

The final step of this review regards the recently determined noncommutative higher dimensional charged solution
of Einstein's equation\cite{Spallucci:2008ez}.  The starting point of any noncommutative quasi classical
coordinates is the smearing which fatally affects source terms of equations. To this purpose it has been studied,
as preliminary case, the possibility for Standard Model fields to propagate into the bulk rather than being
constrained to a four dimensional brane. This strong hypothesis is in agreement with recent arguments in
literature within the framework on the Universal Extra Dimensions, that are expected to be compactified at a scale
$1/R$ above few TeV\cite{Appelquist:2000nn}\cdash\cite{Rai:2005vy}. For this reason, one can determine, in the
case of spherical symmetry, the quasi classical source terms of both the gravitational and the electromagnetic
field, in $D=d+1$ dimensions as
\begin{eqnarray}
 \rho_{matt}\left(\,r\,\right)&=&
\frac{{\cal M}}{\left(\,4\pi\theta\,\right)^{d/2}}\, \exp\left(-r^2/4\theta\,\right)
  \\
 \rho_{el}\left(\,r \,\right)&=&
\frac{{\cal Q}}{\left(\,4\pi\theta\,\right)^{d/2}}\, \exp\left(-r^2/4\theta\,\right).\label{due}
  \end{eqnarray}
Such sources are going to describe the noncommutative behavior of the electro-gravitational system within the
framework of the quasi classical set of field equations
\begin{eqnarray}
&& R^M_{N}-\frac{1}{2}\, \delta^M_N\, {\cal R} = \frac{8\pi}{M_\ast^{d-1}} \,\left(\, T^M_N\vert_{matt.} +
T^M_N\vert_{el.} \,\right)
\\
&& \frac{1}{\sqrt{-g}}\,\ \partial_M\,\left(\, \sqrt{-g}\, F^{MN}\, \right)= J^M
\end{eqnarray}
with $M,N=0,..., d$. The tensors $T^M_N\vert_{matt.}$ and $T^M_N\vert_{el.}$ are the higher dimensional equivalent
of the energy momentum tensors, describing the matter and the electromagnetic content, in the four dimensional
solution\cite{Ansoldi:2006vg}.

\begin{figure}[ht!]
\begin{center}
\includegraphics[width=5.5cm,angle=270]{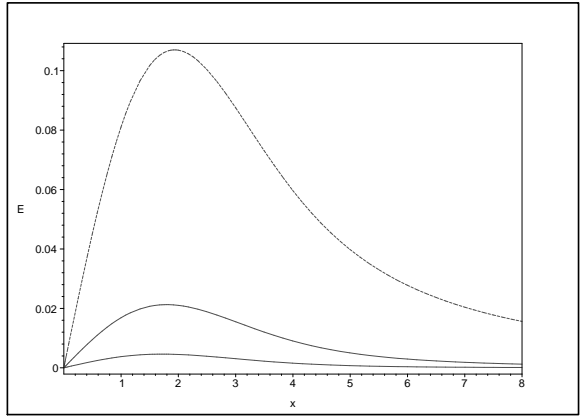}
\caption{\label{coulomb} \textit{The noncommutative extradimensional charged solution.} The electric field
$E\left(x \right)$ as a  function of $x=r M_\ast$ for $e=1\times M_\ast^{(d-3)/2}$ and $y=M_\ast\sqrt{\theta}=1$.
The electric field reaches its maximum intensity near $x\approx 2$, and then drops to zero for any $d$. An
increase of $d$ determines a weaker field, since the spacetime turns out to be larger. Regarding the long distance
behavior, we find that of ordinary Coulomb field, $1/x^{d-1}$.}
\end{center}
\end{figure}

The above Maxwell field can be obtained as
\begin{equation}
F^{MN}=\delta^{0[\, M\,\vert}  \delta^{r\,\vert\, N \,]}\, E\left(\, r\,\right) 
\end{equation}
where $E(r)$ is the electric field produced by the charge distribution $\rho_{el}(r)$
\begin{equation}
E\left(\, r\,\right)=\frac{1}{\pi^{(d-2)/2}} \frac{2{\cal Q}}{r^{d-1} } \ \gamma\left(\, \frac{d}{2}\ ,
\frac{r^2}{4\theta} \,\right),\label{c1}
\end{equation}
while the corresponding current density is
\begin{equation}
 J^M\left(\,x\,\right)= 4\pi\, \rho_{el}\left(\, r \,\right)\,\delta^M_0.
\end{equation}
We note that the electric field results regular and vanishing at the origin for any $d$, while at large distances
it exhibits the conventional $1/r^{d-1}$ behavior. This is a further check of consistency of the quasi coordinates
approach to noncommutative geometry. For the above conditions, we assume a metric of the following form
\begin{equation}
ds^2_{(d+1)}= -f(r)\, dt^2 + f^{-1}(r)\, dr^2 + r^{2} d\Omega^2_{d-1} \label{extrareisds}
\end{equation}
with the usual demand that $f(r)\to 1$ as $r\to \infty$, while $d\Omega_{d-1}$ is still the surface element of
$d-1$-dimensional unit sphere. Plugging the above line element into the Einstein-Maxwell system, we can determine
the function $f(r)$
\begin{equation}
f(r)= 1 -\frac{1}{M_\ast^{d-1}}\frac{4 {\cal M}}{\pi^{(d-2)/2}\, r^{d-2}}\ \gamma\left(\, \frac{d}{2}\ ,
\frac{r^2}{4\theta}\,\right)
 +\frac{d-2}{M_\ast^{d-1}}\ \frac{4{\cal Q}^2 \, \,F_d\left(\, r\,\right)}{\pi^{d-3}\, r^{2d-4}}
\end{equation}
where
\begin{equation}
F_d(r)\equiv  \gamma ^2\left(\,\frac{d}{2}-1\ , \frac{r^2}{4\theta}\,\right)
 -  \frac{2^{(8-3d)/2}\,r^{d-2}}{(d-2)\theta^{(d-2)/2}}\,
  \gamma\left(\, \frac{d}{2}-1\ , \frac{r^2}{2\theta} \,\right).
\end{equation}
In analogy of what we have already seen in the four dimensional case, the resulting manifold is curvature
singularity free. Indeed the short distance geometry is governed by a deSitter line element, whose effective
cosmological constant coincides with what found in the neutral extradimensional case
\begin{equation}
\Lambda= \frac{1}{M^{d-1}}\frac{4\,{\cal M} }{d\, 2^{d-1}\pi^{(d-2)/2 } \theta^{d/2}}.
\end{equation}
This feature enlightens the validity of the present method and represents a matter to reject all the other
noncommutative geometry approaches. Introducing the total mass energy, due to both matter and electromagnetic
field energy contributions
\begin{equation}
  M\left(\, r\,\right) = \frac{2{\cal M} }{\pi^{ \frac{d-2}{2}}}\
  \gamma\left(\, \frac{d}{2}\ , \frac{r^2}{4\theta}\,\right)
  +\frac{2^{\frac{8-3d}{2}}}{\pi^{d-3}}\frac{2{\cal Q}^2}{\theta^{ \frac{d}{2}-1 }}
  \gamma\left(\, \frac{d}{2}-1\ , \frac{r^2}{2\theta}\,\right)
\end{equation}
and its asymptotic limit $M$, namely the total mass energy as measured by an asymptotic observer
\begin{equation}
  M = \lim_{r\to\infty}  M\left(\, r\,\right)=
  \frac{2{\cal M}}{\pi^{\frac{d-2}{2}}}\,\Gamma\left(\, \frac{d}{2}\,\right)
  + \frac{2^{\frac{8-3d}{2}} \, \ 2{\cal Q}^2}{
  \pi^{d-3}\, \theta^{\frac{d-2}{2}}}\,
\Gamma\left(\,\frac{d}{2}-1\,\right),
  \end{equation}
 one can write the above line element as
\begin{eqnarray}
  f(r)&&= 1
-\frac{1}{M_\ast^{d-1}}\frac{2M}{r^{d-2} \Gamma\left(\,\frac{d}{2}\,\right)}\, \
\gamma\left(\,\frac{d}{2}\ , \frac{r^2}{4\theta}\,\right)\nonumber\\
&&+ \frac{4\,{\cal Q}^2\ (d-2)}{M_\ast^{d-1}\pi^{d-3}\, r^{2d-4}}\, \left[\, F_d\left(\, r\,\right)
 + c_d\, \left(\frac{r}{\sqrt{\theta}}\right)^{d-2}\, \gamma\left(\,  \frac{d}{2}\ ,  \frac{r^2}{4\theta}\, \right)\,\right]
 \label{extrancrn}
   \end{eqnarray}
where the coefficient $c_d$ is given by
\begin{equation}
  c_d \equiv \frac{2^{\frac{8-3d}{2}} }{d-2}
  \, \,
  \frac{\Gamma\left(\, \frac{d}{2}-1\,\right)}
  {\Gamma\left(\, \frac{d}{2}\,\right)}\\.\label{met}
\end{equation}
This line element nicely behaves at large distances too: indeed for $r\gg\sqrt{\theta}$ we have
\begin{equation}
f(r)\sim 1 -\frac{1}{M_\ast^{d-1}}\frac{2M}{r^{d-2}}\, + \frac{4\,{\cal Q}^2\ (d-2)}{M_\ast^{d-1}\pi^{d-3}\,
r^{2d-4}}\,\Gamma^2\left(\, \frac{d}{2}-1\,\right),
\end{equation}
namely the line element (\ref{extrancrn}) asymptotically reproduces the ordinary Reissner-Nordstr\"{o}m geometry.

\subsubsection{The horizon equation}

\begin{figure}[ht!]
\begin{center}
\includegraphics[width=5.5cm,angle=270]{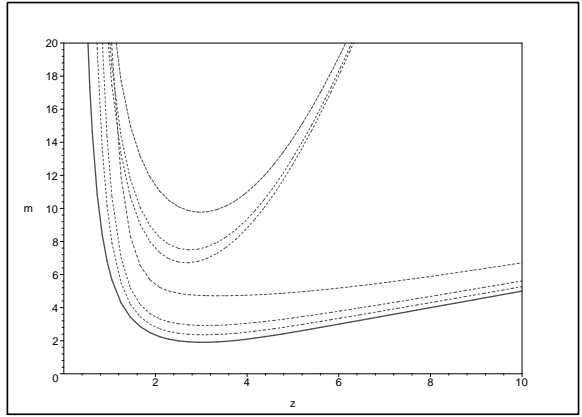}
\caption{\label{massaq} \textit{The noncommutative extradimensional charged solution.}  The mass function $m$
versus the horizon radius $z$ for different $q$'s and $d$'s and $y=1$. The solid curve represent the neutral three
dimensional case, while the the dashed curves, from the bottom to the top, are respectively for the cases $d=3$,
$q=2$; $d=3$, $q=3$; $d=3$, $q=5$; $d=4$, $q=1$; $d=4$, $q=10$; $d=4$, $q=20$.  At first glance we can conclude
that minimal masses increase with the charge $q$ and the number of dimensions $d$, while their radii decreases
with $d$ and increase with $q$. }
\end{center}
\end{figure}

At this point, we can study the existence of the horizon solving the equation $f(r)=0$.  For sake of notational
clearness, it is worthwhile to introduce dimensionless quantities, such as $m=M/M_\ast$, $x=r_HM_\ast $,
$y=M_\ast\sqrt{\theta}$ and $q={\cal Q}M_\ast^{(d-3)/2}$ and write the horizon equation as
\begin{equation}
2m\ G_d(z)=x^{d-2}+a_d\ q^2 \ x^{2-d}\left[F_d(z) + b_d\ z^{d-2}\ G_d(p)\right]
\end{equation}
where we recall the definition of effective Newton constant
\begin{equation}
G_d(p)=\frac{1}{\Gamma(d/2)}\ \ \gamma(d/2, p)
\end{equation}
with $z=x/y$, $p=z^2/4$, while
\begin{equation}
a_d=\frac{4(d-2)}{\pi^{d-3}}
\end{equation}
and $b_d=\Gamma\left(\, d/2\,\right)\ c_d$. The function $F_d$ can be written in a more convenient form as
\begin{equation}
F_d(z)=\Gamma^2\left(\, d/2-1\,\right)\ G^2_{d-2}(p)-b_d\ z^{d-2}\ G_{d-2}(p)
\end{equation}
where $p=z^2/2$.

 \begin{figure}[ht!]
 \begin{center}
\includegraphics[width=9.5cm,angle=0]{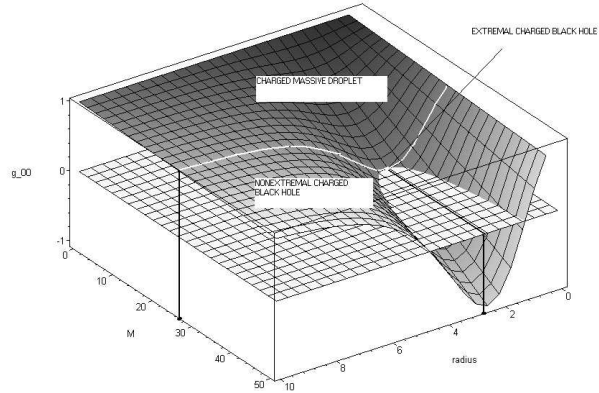}
\caption{\label{extraRNf}\textit{The noncommutative extradimensional charged solution.} The function $f$ versus
$m$ and $z$, for a charge, $q=1$ and $d=4$. With respect to the $d=3$ case, the intersection with the $f=0$ plane
(light grey) occurs at larger value of both $m$ and $z$. }
\end{center}
\end{figure}

We can see that the total mass energy diverges both at short distances $m\sim z^{-2}$ and at large distances
$m\sim z^{d-2}$. Therefore we do expect the existence of a minimum value of $m$ at some intermediate value of $x$,
in agreement with Figs.~\ref{massaq} and \ref{extraRNf}. It is also evident that for any $d$ there can be two
horizons when $m>m_{min}$, one degenerate horizon for $m=m_{min}$, or no horizon in $m<m_{min}$.  Again, from the
study of the temperature, the mass  $m_{min}$ of the extremal black hole, is expected to be the end of the Hawking
evaporation process. The analysis of the neutral case showed that black holes are too massive to be produced at
LHC for $y$ about $1$. In the charged case the situation is even worse. In Fig.~\ref{massaq}, families of curves
labelled by the value of $d$ and assigned value of the electric charge $q$. The horizons, and the corresponding
values of $m$, can be read as the intersections of the grid lines, with each curve in Fig.(\ref{extraRNf}). The
minimum of the curves defines the extremal black hole configurations in different dimensions.  We see that the
effect of both extradimensions and charges is to lift upward the curves, increasing the value of the mass for a
given radius of the horizon, including the degenerate case. Therefore, we conclude that the evolution of the black
hole towards its extremal configuration is qualitatively the same as described in four dimensional
case\cite{Nicolini:2005vd}. The increase of the mass $m_{min}$ and its eventual experimental verification could
indicate the number of extra-dimensions. On the other hand, our current experimental ability would demand lower
value for the masses and therefore  $y\sim 0.1$. Further on smaller $y$ would lead to the commutative limit, in
which the horizon equation becomes
\begin{equation}
2m=x^{d-2}+a_d\ q^2\ x^{2-d}\ \Gamma^2\left(\, d/2-1\,\right)
\end{equation}
whose roots are given by
\begin{equation}
x^{d-2}_\pm=m\pm\left(\ m^2-a_d\ q^2\ \Gamma^2\left(\, d/2-1\,\right)\ \right)^{1/2}.
\end{equation}

\begin{figure}[ht!]
\begin{center}
\includegraphics[width=5.5cm,angle=270]{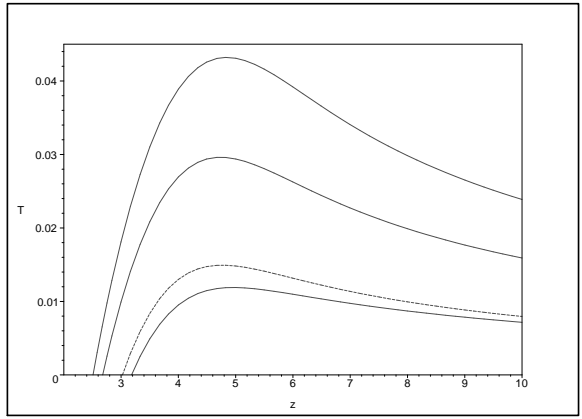}
\caption{ \label{ExtraRNTemp}\textit{The noncommutative extradimensional charged solution.} The dimensionless
Hawking temperature $T$  as a function of $z$, for different values of $q$ and $d$. The dashed curve regards the
case $d=3$, $q=0$, while the solid curves from the bottom to the top regard the cases $d=3$ $q=1$, $d=4$ $q=1$ and
$d=5$ $q=1$.  In all such cases $T$ reaches a maximum whose value increases with $d$ before dropping to zero. The
role of the charge is just to lower down the temperature at any $z$. }
\end{center}
\end{figure}

\subsubsection{Higher dimensional charged black hole thermodynamics}

On the thermodynamics side, the temperature is going to reveal the behavior of the black hole in the so called
balding phase. The four dimensional analysis showed that very quickly the black hole reaches the Schwarzschild
phase and thus subsequent SCRAM basically occurs in a neutral regime.  In other words, Schwinger pair production
and Hawking emission behave like two temporally distinct processes.  At this point, it is imperative to know
whether this four dimensional scenario could be altered by the presence of extradimensions. The starting point is
the explicit form of the Hawking temperature
\begin{eqnarray}
T_H &&= \frac{d-2}{4\pi r_+}\,\left[\,1 -\frac{ r_+}{d-2}
 \frac{\gamma^\prime\left(\, d/2\ ,r_ +^2 /4\theta\, \right)}
 {\gamma\left(\, d/2\ , r_ +^2 /4\theta\,\right)}\,\right]+\\
 &&-\frac{16\ \ {\cal Q}^2\  (d-2) }
{M_\ast^{d-1}\ \pi ^{d - 3} r_+^{2d - 3}} \left[\, \frac{1}{d-2}\ \gamma ^2\left(\, d/2\ , r_ +^2
/4\theta\,\right)
 +\, F\left(\, r_+ \,\right)\,\frac{r_+}{4}\,
 \frac{\gamma^\prime\left(\, d/2\ ,r_ +^2 /4\theta\,
 \right)}{\gamma \left(\, d/2\ , r_ +^2 /4\theta\, \right)}\right]\nonumber
\label{extrarntemp}
\end{eqnarray}
where, we replaced $M$ with $r_+$ by using the horizon equation.  The dimensionless form of the above formula
reads
\begin{equation}
T=\frac{d-2}{4\pi x}\left[1-\frac{2}{d-2}\ H_d(p)\right]-\frac{16 q^2  (d-2) } {\pi ^{d - 3} x^{2d - 3}} \left[\,
\frac{\Gamma ^2\left( d/2\right)\ G^2_d(p) }{d-2}
 +\, \frac{1}{2}\ F_d\left(p\right)\,H_d(p)\right]
\end{equation}
where $H_d=p^{1/2}e^{-p}/G_d(p)\Gamma(d/2)$. It is clear that the additional term due to the charge may become
important for small $x$ only. Indeed from Fig.~\ref{ExtraRNTemp}, one learns that the temperature behavior is
equivalent to that encountered in the neutral case. The effect of the charge is just to lower down the already
regular curve of the temperature. Also the entropy resembles that encountered in the previous cases. Both from
\begin{equation}
dm = T\, dS + \frac{\partial m}{\partial q}\, dq \label{RNM1}
\end{equation}
and
\begin{equation}
dm= \frac{\partial m}{\partial z }\ d z + \frac{\partial m}{\partial q}\ dq \label{RNM2}
\end{equation}
one gets
\begin{equation}
dS= \frac{1}{T}\, \left(\,\frac{\partial m}{\partial z}\,\right) dz
\end{equation}
Therefore, calculating $\partial m/\partial z$ and integrating from $x_{min}$ to a generic $x$, one again finds
\begin{equation}
S=\Gamma\left( d/2\right)\pi^{1-\frac{d}{2}}\ \frac{\left(\, A_+ -A_e\,\right)}{(d-1)\ G_d\left(\, z\,\right)}
\,+\delta S,
\end{equation}
in analogy to what already seen for the four dimensional charged case and the extradimensional neutral case. In
the present case, the event horizon areas $A= 2\, \pi^{d/2}\, z^{d-1}/\Gamma\left(\, d/2\,\right) $ do depend on
both $d$ and $q$ through $z$.  Again, one can find that the correcting term $\delta S$ is negligible and the area
entropy law still holds up to exponentially suppressed corrections.

\subsubsection{Schwinger mechanism and dyadosphere}

 At this point, we still have to study the
discharge time of the black hole.  First of all, the analysis in the four dimensional case about the development
of the dyadosphere can be extended to the extradimensional case and leads to the same conclusions. The condition
to have a pair producing field implies
\begin{equation}
  \frac{2{\cal Q}}{\pi^{\frac{d-2}{2}}\, r^{d-1}}\, \
\gamma\left(\, \frac{d}{2}\ , \frac{r^2}{4\theta}\,\right) \ge  \frac{m^2_e}{e}
\end{equation}
where $m_e$ and $e$ are the mass and the charge of each particle of the pair. Supposing that the black hole total
charge is ${\cal Q}=Ze$, a multiple of the elementary charge $e$, we have $q={\cal Q }M_\ast^{(d-3)/2}=Z$ and we
can conveniently write the above relation in term of dimensionless quantities as
\begin{equation}
\frac{1}{y^{d-1}}\frac{2Z}{\pi^{(d-2)/2} z^{d-1}} \ \gamma\left(\, \frac{d}{2}\ , \frac{z^2}{4}\,\right)\ge
\left(\frac{m_e}{M_\ast}\right)\sim 2.5\times 10^{-13}
\end{equation}
Therefore, for any $y\le 10$ and for any $d=3-10$, we can conclude that it is sufficient a single elementary
charge, $Z=1$, to have Schwinger pair creation at the black hole  event  horizon. In other words it is very likely
that dyadospheres, the spherical surface where the above condition is satisfied, have radii even larger than
$r_+$. To this purpose it is crucial to know whether such dyadosphere can be formed. According to
\refcite{Page:2006cm}--\refcite{Page:2006zk}, dyadospheres develop only if
\begin{equation}
\frac{E}{E_{th}}=\frac{e}{r^{d-1}\ m_e^2}\frac{{\cal Q}}{\Gamma(d/2)}\ \gamma\left(\, \frac{d}{2}\ ,
\frac{r^2}{4\theta} \,\right)\ll\frac{m_p \ M G}{r^{d-1}\ m_e^{2}}\sim \left(\frac{m_p}{m_e}\right)\frac{M}{m_e}
\sim 4\times 10^9
\end{equation}
where $E_{th}\equiv m^2_e/e$ is the Schwinger threshold field. Here we see that the above condition which is
always met in any dimension, being $m_p$ the proton mass.  This means that the electric field may be strong enough
to produce pairs without the electrostatic repulsion overcoming the gravitational attraction among collapsing
hadronic partons. At first order, we find that the dyadosphere radius $r_{ds}$ is
\begin{equation}
 r_{ds}^{d-1}\simeq \frac{2 e{\cal Q}}{\pi^{(d-2)/2}\,  m^2_e}\,
\gamma\left(\,\frac{d}{2}\ ,\frac{r_0^2 }{4\theta}\,\right)\ ,\\
 \label{dyado}
\end{equation}
where $r_0$ is the $0^{th}$ order result, given by
\begin{equation}
 r_0^{d-1}= \frac{2 e{\cal Q}}{\pi^{(d-2)/2}\,  m^2_e}\,
\Gamma\left(\,\frac{d}{2}\,\right).
\end{equation}
At this point, we can calculate the black hole discharge time. To this purpose, let us introduce the surface
charge density
\begin{equation}
\sigma\left(\,r\,\right)=\frac{1}{A}\ \frac{{\cal Q}}{\Gamma(d/2)}\ \gamma\left(\, \frac{d}{2}\ ,
\frac{r^2}{4\theta} \,\right)= \frac{{\cal Q}}{ 2\pi^{d/2}r^{d-1}}\ \gamma\left(\, \frac{d}{2}\ ,
\frac{r^2}{4\theta} \,\right)
\end{equation}
and divide the dyadosphere in thin shells, whose width is the electron Compton wave length $\lambda_e=1/m_e$.
Thus, the resulting electric field $E(r)=4\pi\sigma(r)$ can be considered constant within each of such shells and
we are allowed to apply Schwinger formula
\begin{equation}
W=\frac{(e\,E)^{(d+1)/2}}{2^{d+1}\pi^d}\, \exp\left(-\frac{\pi\, m^2_e}{eE}\,\right)
\end{equation}
where $W$ is the pair production rate per volume.  As a consequence, the total number of pairs produced in the
time interval $\Delta \tau$ turns out to be
\begin{eqnarray}
\Delta N&&\equiv \lambda_e\, A\left(\,r\,\right)\,
  W\ \Delta \tau\\
&&=\frac{  \lambda_e \, r^{d-1}  m_e^{d+1} }{2^d \, \pi^{d/2}\, \Gamma\left(\,\frac{d}{2}\,\right)
}\left(\,\frac{\sigma}
 {\sigma_c}\, \right)^{(d+1)/2}
\exp\left(-\pi\,\frac{\sigma_c}{\sigma}\right)\ \Delta \tau \nonumber
\end{eqnarray}
where the threshold density $\sigma_{th}= m_e^2/4\pi\, e$ is obtained when $E=E_{th}$. The integration of the
above formula has to be done assuming that the discharge process takes place till $\sigma$ reaches the threshold
value $\sigma_{th}$.  As a result we have
\begin{equation}
 \Delta \tau = \frac{\theta\, m_e}{\alpha_{em}}\,
\left(\frac{2\pi}{m_e c\sqrt{\theta}}\right)^{d-1} \frac{s -1}{s^{(d+1)/2}} \exp\left(\,\frac{\pi}{s}\,\right)
\label{time}
\end{equation}
where $s=\sigma/\sigma_{th}$ and $\alpha_{em}=1/137 $ is the fine structure constant in four dimensions. The
dimensionless version of the above relation reads
\begin{equation}
 \Delta t= \frac{1}{y^{d-3}}\frac{m_e/M_\ast}{\alpha_{em}}\,
\left(\frac{2\pi}{m_e/M_\ast }\right)^{d-1} \frac{s -1}{s^{(d+1)/2}} \exp\left(\,\frac{\pi}{s}\,\right)
\label{dimetime}
\end{equation}
where $\Delta t=\Delta \tau M_\ast$, $m_e/M_\ast\approx 5\times 10^{-7}$ and, as usual,  $y=\sqrt{\theta}M_\ast$.
The Eqs.~ (\ref{time}) and (\ref{dimetime}) give the discharge time assuming that the process occurs in the $d+1$
dimensional bulk spacetime. To this purpose, it is interesting to compare the extradimensional time $\Delta \tau$
to $\Delta \tau_{d=3}$, the corresponding one for $d=3$. In particular, being $\Delta \tau_{d=3}\le 1.76\times
10^{-19}$ s, we can conclude that
\begin{equation}
 \Delta \tau\le \frac{1}{y^{d-3}} \left(\,\frac{2\pi}{m_e/M_\ast}\,\right)^{d-3}
 1.76\times 10^{-19}\ \mathrm{s} \label{tau}.
 \end{equation}
As a consequence, if the pair production takes place in the bulk, the discharge time significantly increases with
$d$, while, according to the brane universe scenario in which only gravity can probe transverse higher dimensions,
$\Delta \tau=\Delta \tau_{d=3}\le 1.76\times 10^{-19}$ s. In the latter case, the noncommutative charged solution
can be efficiently employed in brane models: indeed quickly the black hole turns into the noncommutative
Schwarzschild one, whose event horizon radius is small with respect to the size of the extradimensions and let us
adopt the spherical symmetry condition to describe the subsequent thermal emission, even in the presence of a
$D$-brane.

 \begin{table}[pht!]
 \tbl{  Schwinger discharge time upper bounds, for different $d$ and $y=1$}
 {\begin{tabular}{@{}ccccccccc@{}}  \toprule d
&  3  & 4  & 5  & 6  & 7  & 8  & 9  & 10 \\
 \colrule $\Delta \tau  $ (seconds)
&  $10^{-19}$  & $10^{-12}$  & $10^{-5}$  & $10^{2}$  & $10^{9}$  &
$10^{16}$  & $10^{23}$  & $10^{30}$ \\
 \botrule
\end{tabular} \label{ta4}}
\end{table}

\begin{table}[pht!]
 \tbl{  Schwinger discharge time upper bounds, for different $d$ and $y=10$}
 {\begin{tabular}{@{}ccccccccc@{}}  \toprule d
&  3  & 4  & 5  & 6  & 7  & 8  & 9  & 10 \\
 \colrule $\Delta \tau  $ (seconds)
&  $10^{-19}$  & $10^{-13}$  & $10^{-7}$  & $10^{-1}$  & $10^{5}$  &
$10^{11}$  & $10^{17}$  & $10^{23}$ \\
 \botrule
\end{tabular} \label{ta5}}
\end{table}

From Eq.~\ref{tau} and Tables \ref{ta4}, \ref{ta5}, it emerges that Schwinger mechanism is relevant for larger
$y$, while becomes negligible in the commutative regime, namely for $y\ll 1$. In particular, the Schwinger
mechanism dominates the Hawking emission in the early stages of the black hole life for $d\le 5$, $d=1$ and $d\le
6$, $y=10$.

\subsubsection{Phenomenological implications}

As final comment, we showed that the noncommutative extradimensional charged solution basically reproduces all the
main feature of the previous noncommutative solution. Anyway, the presence of the charge is important because
since we do expect the production of charged black holes that might be surrounded by clouds of opposite charges.
At hadron colliders, such as LHC, generally electrons are more easily seen among other hadronic particles.
Therefore, the presence of many electrons near an object among the hadronic fragments, could provide a further
signal in support to the creation of a mini (charged) black hole, even if, in case of quick discharge time, its
remnant is expected to be neutral.

\section{Future Perspectives}

This review represents both a summary of the state of the art and a starting
point towards the exciting arena of new physical phenomena that may be observed
in the next few months at the CERN Large Hadron Collider and in subsequent
experiments. We have merged together some basic ingredients including Hawking
radiation phenomenology, General Relativity, Particle Physics, Noncommutative
Geometry and TeV Quantum Gravity in order to make a reliable picture of the
nature of mini black holes when noncommutative effects are taken into account.
Indeed, black hole production at the LHC may be one of the early signatures of
the validity of the models which we have been reviewing and, more in general,
of Noncommutative Geometry and TeV Quantum Gravity.

In particular, we have provided a detailed analysis of all of the existing black hole models within a
noncommutative background spacetime. For doing this, we introduced the most popular formulations of Noncommutative
Geometry, with particular emphasis on those which lead to mathematically tractable field equations for gravity and
hence to physically meaningful solutions. We presented all of the different kinds of line element ever proposed in
the noncommutative literature and proceeded by following those approaches which, by an efficient introduction of a
natural cutoff in spacetime, have successfully reached the goal of removing the singular short distance behavior
of black hole solutions in General Relativity. This has really been an unavoidable requirement for two reasons:
first, at phenomenological level, we need a tool that lets us describe the final stages of black hole evaporation
and predict finite, experimentally testable values for temperatures, masses, cross sections and lifetimes; second,
there is no reason to justify the effort in formulating and adopting a noncommutative background spacetime if the
Planck phase of the evaporation is plagued by the same divergences and singularities as in the conventional method
based on the usual differential spacetime manifold. With this in mind, we showed that the approach based on
quasi-classical noncommutative coordinates is able to implement the nonlocal character of noncommutativity for any
field theory propagating on the noncommutative manifold, including gravity. This approach let us spread out any
point-like source in the field equations, determining the exact form of the noncommutative smearing in terms of
Gaussian distributions, whose width is governed by the minimum length $\sqrt{\theta}$, induced by
 \begin{equation}
\left[\, \mathbf{x}^i\ , \mathbf{x}^j\, \right]= i \, \theta^{ij},
\end{equation}
 the fundamental relation of Noncommutative Geometry. Furthermore, we have seen
that the quasi-coordinates approach leads to field equations which formally
resemble the conventional classical equations: for this reason, our analyses of
black hole solutions in four dimensions, for both the neutral and charged
cases, could appear, at first sight, as new nonsingular line elements of
ordinary General Relativity. By itself, this could already be considered as
being a nontrivial result, but there is much more. After a deeper inspection,
new features arise that can only be explained in terms of quantum geometrical
effects. Indeed, the fate of the evaporating black hole is drastically modified
by Noncommutative Geometry and in place of the conventional Planck phase, a new
final phase emerges, the SCRAM phase, during which the black hole, after a
temperature peak, cools down to a stable zero-temperature remnant
configuration. A conclusion of this kind for the black hole's life would not
only provide a concrete answer to the long-standing problem of information
loss, but would also be the signature for the experimental evidence of
mini-black hole production. In particular, once we abandon the conventional
four-dimensional spacetime, ruled by the unreachable Planck scale $M_P$, the
possibility of producing mini-black holes and therefore of finding their
remnants at $M_\ast\sim 1$ TeV, becomes more than a wild conjecture.
Experimentally, this implies that we should detect a suppression of di-jet
events above the black hole production threshold energy, because the Hawking
radiation is expected to mask the standard scattering in the energy range of
the emission\cite{Stocker:2007zz}. Of course this consideration has to be
reviewed in the light of the noncommutative solutions, which imply lower upper
bounds for the peak of the temperature and therefore a modified scenario for
the decay, taking for granted that black holes can be produced in hadronic
collisions. Indeed, all that we know about the particles emitted during the
evaporation is based on the conventional Schwarzschild
solution\cite{Harris:2003eg}. An analysis of this kind can still be applied
before the SCRAM phase, namely until the Hawking temperature reaches
$T_H\lesssim T_H^{max}$, where the temperature peak $T_H^{max}$ is around
$30-100$ GeV depending on the number of extra dimensions. In particular, it is
believed that a $D$-dimensional black hole can emit energy and angular momentum
both in the $(3+1)$ brane universe and in the $D$-dimensional bulk, where the
brane is embedded. Some estimates suggest that the black hole radiates mainly
on the brane, even if almost half of the total energy is lost in the
bulk\cite{Emparan:2000rs}\cdash\cite{Kanti:2002ge} where direct observations
are not possible. The brane emission is governed by
 \begin{equation}
\left<N\right>_{\omega s}=\frac{\left|A\right|^2}{e^{\omega/T_H}-(-1)^{2s}}
\end{equation}
 where $\left<N\right>_{\omega s}$ is the average number of particles with
energy $\omega$ and spin $s$, emitted by a Schwarzschild black hole with temperature $T_H$, while
$\left|A\right|^2$ is the greybody factor taking into account the gravitational potential barrier in the vicinity
of the horizon. Integration of the above formula lets us determine the contribution of spin $s$ Standard Model
particles to the power and flux emitted on the brane. It has been shown that $\sim 75 \%$ of the decay products
are quarks, anti-quarks and gluons, while only $\sim 12 \%$ are charged leptons and photons\cite{Casanova:2005id}.
Because of this, since the dominant parton contribution cannot be directly observed, one must take into account
all of the QED/QCD interactions, including hadronization, in order to understand the kind of spectra that we are
going to detect. In particular, a lot of attention has been devoted to the presence of quark-gluon plasma around
the black hole before hadronization takes place\cite{Heckler:1995qq}. Indeed the parton density near to the event
horizon grows like $T_H^3$: if the temperature is high enough, namely higher than a critical value $T_c$, one
expects interactions among quarks and gluons through bremsstrahlung and pair production processes. These $2\to 3$
processes can increase the number of quarks and gluons, determining the onset of a quark/gluon plasma appearing in
the vicinity of the horizon. Therefore propagation through this plasma is affected by a loss of energy which leads
to the recombination of partons into hadrons. Also photons, electrons and positrons are equally subject to these
processes when QED is considered. For this reason it is worthwhile to define two regions surrounding the black
hole, the {\it photosphere} where QED mechanisms lead to a $e^{\pm},\gamma$ plasma, and the {\it chromosphere},
where QCD mechanisms lead to a quark-gluon plasma. Both regions have been studied numerically by solving the
Boltzmann equation for the distribution function of the interacting particles, assuming that scattering processes
become important only beyond some critical temperature, $T_c^{QED}\sim 50$ GeV for the photosphere and
$T_c^{QCD}\sim 175$ MeV for the chromosphere. On the other hand, these considerations hold only in the vicinity of
the horizon: when interactions of the emitted particles are properly considered, the emission spectrum at the
photo/chromosphere edge is modified and the black hole behaves like a black body at a lower temperature. It is
clear that the above analysis of photosphere development has to be revised when the black hole has a maximal
temperature, as in the noncommutative case. However the considerations about the chromosphere seem to still be
valid in the noncommutative scenario and it is reasonable to have $T_H\gtrsim\Lambda_{QCD}\gtrsim T_c^{QCD}$
leading to a strongly coupled quark-gluon plasma and final spectra dominated by hadrons. In connection with this,
we should mention that the case of direct hadronization in the absence of a photosphere or chromosphere is already
the subject of investigations using PYTHIA 6.2 with the CHARYBDIS event generator, leading again to a hadron
dominated spectrum\cite{Harris:2003db}. Other contributions on the numerical side include the TRUENOIR\cite{Dimopoulos:2001en} black hole event generator and, more recently, CATFISH\cite{Cavaglia:2006uk} and BlackMax\cite{Dai:2007ki} which employ a richer set of external parameters and allows a certain flexibility among different theoretical models.

In addition to all of this, the detection of stable black hole
remnants would provide further interesting new signatures favoring
the Noncommutative Geometry inspired models. Electrically charged
remnants would leave an ionizing track in the detector, allowing
their identification and direct measurement of their masses. On
the other hand, neutral remnants would not be visible in the
detector but could be identified by modifications of the $p_T$
distribution, multiplicities and angular correlations of Standard
Model particles emitted as Hawking evaporation. The remnants would
also carry a fraction of the total energy and therefore a search
for events with $\sim$ TeV missing energy could give a signature
for the presence of a minimum length related to $\sqrt{\theta}$.
It is generally believed that the large acceptance of the
detectors at the LHC will enable a complete event reconstruction
to be made so that the missing energy can be determined. The
effects of the formation of these remnants on black hole
evaporation have been investigated in Refs.
\refcite{Hossenfelder:2005ku,Casadio:2008qy}.

A lot of work is still needed in order to get a definitive picture
of the phenomenology of black hole production at the LHC. Even if
we have found encouraging cross sections $\sim 20-400$ pb, a still
open problem, at least within the framework of the models
presented, regards the threshold energy required to prime the
production, since for $M_\theta\sim M_\ast\sim 1$ TeV and for
$d\sim 10$ this could be around $10^4$ TeV. The difficulty caused
by this could be overcome in the case of black hole events in the
Earth's atmosphere originated by ultra high energy cosmic
rays\cite{Ahn:2005bi} for which a scale of around $\sim$ PeV or
even beyond is considered reasonable. Therefore it is very likely
that in the near future we could be able to obtain some signatures
of the production of tiny black holes in particle collisions, so
as to address long standing questions about the quantum nature of
gravity. Given the strong hypothesis about the presence of
additional dimensions in order to have an experimentally testable
scale to work with, any upcoming observations with particle
accelerators, at the Pierre Auger
Observatory\cite{Stojkovic:2005fx,Auger} or with other
experiments, can really be considered as being both the first and
the final appeal to Quantum Gravity.

\section*{Acknowledgments}

I would like to warmly thanks E.~Spallucci for a long-time and successful collaboration on the majority of the
topics included in this review. I am also grateful to J.~C.~Miller for illuminating discussions and careful
reading of the draft and  A.~Smailagic for the collaboration on this area. I would like to express my gratitude
also to S.~Ansoldi, R.~Balbinot, R.~Casadio, M.~Cobal, M.~Cavagli\`a, G.~Giannini, A.~Gruppuso, S.~D.~H.~Hsu,
D.~A.~Singleton, M.~Tessarotto for fruitful discussions and/or comments and references. Finally I want to thank
E.~T.~Akhmedov, H.~Balasin, S.~Carlip, V.~Cardoso, B.~Harms, R.~Jackiw, D.~Grumiller, H.~Nicolai, A.~C.~Ottewill,
W.~Pezzaglia and T.~G.~Rizzo for all their support, interest and valuable hints since the early stages of this
developing research area.

This work has been partially supported by a CSU Fresno
International activity grant.

\appendix
\section{Mathematical Formulas}

Definitions of incomplete lower $\gamma$ and upper $\Gamma$  functions
\begin{eqnarray}
\gamma\left( \frac{m}{2}+1,\  x^2\right) &\equiv&
\int_0^{x^2} dt\, t^{m/2} \ e^{-t}\nonumber\\
\Gamma\left( \frac{m}{2}+1,\ x^2\right) &\equiv& \int_{x^2}^{\infty} dt \,t^{m/2}\  e^{-t}=\Gamma\left(\,
\frac{m}{2} +1\,\right)-\gamma\left(\, \frac{m}{2}+1,\ x^2\,\right).
\end{eqnarray}
Integral and differential properties of incomplete $\gamma$ function
\begin{eqnarray}
 &&\int_0^r  \frac{dx}{x^{m + 1}  }\ \gamma \left(\, \frac{m}{2}
 +1\ , x^2 \,\right)  =  -
\frac{1}{2}\frac{\gamma \left(\, \frac{m}{2} \ , r^2\,\right)}{r^m }
\nonumber\\
 &&\oint_{sphere} {d^m } x\ \rho _\theta  (x^2 ) =
 M\frac{\gamma \left(\,\frac{m}{2}\ ,R^2\,\right)}{\Gamma
 \left(\, \frac{m}{2}\,\right)} \nonumber\\
 && \gamma^{\prime}\left(\, \frac{m}{2},\
 x^2\, \right)=2\, x^{m-1}\, e^{-x^2}\nonumber\\
 && \gamma\left( \frac{m}{2}+1,\ \frac{r^2}{4\theta}\,\right)=
 \frac{m}{2}\ \gamma\left( \frac{m}{2},
\frac{r^{2}}{4\theta }\right)-\left(\, \frac{r}{2\sqrt{\theta}}\, \right)^m\, e^{-r^{2}/4\theta }.
\end{eqnarray}
 Long and short distance behavior of incomplete $\gamma$ functions
 \begin{eqnarray}
 && \left.\gamma \left(\, \frac{m}{2},\ x^2 \,\right)\right|_{x>>1}
 =\frac{2}{m}\ x^m e^{ - x^2} \left[ \, 1 - \frac{2}{{m + 2}}x^2  + \frac{2}{{m +
2}}\frac{2}{{m + 4}}x^4  + \dots \right] \nonumber\\
&&  \left.\gamma \left(\, \frac{3}{2}\ ,\frac{r^2}{4\theta}
  \,\right)\right|_{\frac{r^2}{4\theta}<<1}
\approx  \frac{r^3}{12\sqrt{\theta^3}} \left(1 -
\frac{7}{20}\frac{r^2}{\theta}\right)\nonumber\\
&& \left.\Gamma\left(\, \frac{m}{2}, \ x^2\, \right)\right|_{x>>1} = x^{m-2}\
e^{-x^2}\left[1+\left(\frac{m}{2}-1\right)\frac{1}{x^2}+\left(\frac{m}{2}-1\, \right)
 \left(\frac{m}{2}-2\right)\frac{1}{x^4}+\dots\right]\nonumber\\
 &&\left.\gamma\left( \, \frac{3}{2}\ ,
\frac{r^2}{4\theta}\, \right)\right|_{\frac{r^2}{4\theta}>>1}
 =\frac{\sqrt{\pi}}{2}-\Gamma\left( \frac{3}{2}
\frac{r^2}{4\theta}\right)\vert_{\frac{r^2}{4\theta}>> 1}\approx\frac{\sqrt{\pi}}{2}
 +\frac{1}{2}\frac{r}{\sqrt{\theta}}\,e^{-\frac{r^2}{4\theta}}.
\end{eqnarray}

\end{document}